\documentclass[draft]{agujournal2019}
\usepackage{url} 
\usepackage{lineno}
\usepackage[inline]{trackchanges} 
\usepackage{soul}
\usepackage{amsmath}
\usepackage{graphicx}
\usepackage[percent]{overpic}
\draftfalse

\journalname{JGR: Space Physics}

\begin{document}

%
%


\title{Reconstructing coronal hole areas with EUHFORIA and adapted WSA model: optimising the model parameters}

%
%




\authors{E. Asvestari\affil{1,2}, S. G. Heinemann\affil{1}, M. Temmer\affil{1}, J. Pomoell\affil{2}, E. Kilpua\affil{2}, J. Magdalenic\affil{3}, S. Poedts\affil{4}}

\affiliation{1}{Institute of Physics, University of Graz, Universit{\"a}tsplatz 5, 8010 Graz, Austria}
\affiliation{2}{Department of Physics, University of Helsinki, P.O. Box 64, 00014 Helsinki, Finland}
\affiliation{3}{Solar-Terrestrial Centre of Excellence - SIDC, Royal Observatory of Belgium, 1180 Brussels, Belgium}
\affiliation{4}{Centre for mathematical Plasma Astrophysics (CmPA), KU Leuven, 3001 Leuven,
Belgium}

\correspondingauthor{Eleanna Asvestari}{eleanna.asvestari@helsinki.fi}




\begin{keypoints}
\item We assess the capability of the PFSS+SCS models adopted by EUHFORIA in reconstructing CH areas.
\item The heights of the source surface and the inner boundary of the SCS model impact the results.
\item Lower values for these two heights improved the results.
\end{keypoints}

%
%

\begin{abstract}
The adopted WSA model embedded in EUHFORIA (EUropean Heliospheric FORecasting Information Asset) is compared to EUV observations. According to the standard paradigm coronal holes are sources of open flux thus we use remote sensing EUV observations and \textsc{catch} (Collection of Analysis Tools for Coronal Holes) to extract CH areas and compare them to the open flux areas modelled by EUHFORIA. From the adopted WSA model we employ only the Potential Field Source Surface (PFSS) model for the inner corona and the Schatten Current Sheet (SCS) model for the outer (PFSS+SCS). The height, $R_{\rm ss}$, of the outer boundary of the PFSS, known as the source surface, and the height, $R_{\rm i}$, of the inner boundary of the SCS are important parameters affecting the modelled CH areas. We investigate the impact the two model parameters can have in the modelled results. We vary $R_{\rm ss}$ within the interval [1.4, 3.2]$R_{\rm \odot}$ with a step of 0.1$R_{\rm \odot}$, and $R_{\rm i}$ within the interval [1.3, 2.8]$R_{\rm \odot}$ with the same step, and the condition that $R_{\rm i}<R_{\rm ss}$. This way we have a set of 184 initial parameters to the model and we assess the model results for all these possible height pairs. We conclude that the default heights used so far fail in modelling accurately CH areas and lower heights need to be considered.
\end{abstract}


%
%

\section{Introduction}
\label{sec:intro}

The ambient Solar Wind (SW) and the frozen-in open magnetic field sculpt the medium through which Coronal Mass Ejections (CMEs) and Solar Energetic Particles (SEPs) propagate. Consequently, modelling the solar wind with high accuracy is a significant component towards more reliable space weather forecasts. Fast SW and open magnetic field are primarily associated with Coronal Holes (CHs) \cite{levine_solar_1977,wang_magnetic_1996, schwenn_solar_2006, schwenn_space_2006}, which in X-ray and extreme-ultraviolet (EUV) imaging observations can be identified as dark structures in the solar corona.  Their dark appearance at these wavelengths is a result of having significantly lower density and temperature compared to their surroundings. Although a plethora of both empirical and intricate magnetohydrodynamic (MHD) models exist, due to its simplicity and comparability to computationally expensive MHD ones \cite{riley_comparison_2006,macneice_assessing_2018}, the semi-empirical Wang-Sheeley-Arge (WSA) model \cite{arge_improvement_2000} is the model most commonly employed for estimating the solar wind flow close to the Sun. It's magnetic field model comprises the Potential Field Source Surface (PFSS) \cite{altschuler_magnetic_1969, schatten_model_1969} and the Schatten Current Sheet (SCS) \cite{schatten_current_1971} models to extract the solar wind speed and the magnetic field in the corona. And then by using a 1D kinematic scheme attempts to predict the solar wind speed at Earth.

\begin{figure}[h]
    \centering
    \begin{minipage}{1\textwidth}
    \begin{minipage}{0.55\textwidth}
        \centering
        \begin{overpic}[width = 0.8\textwidth]{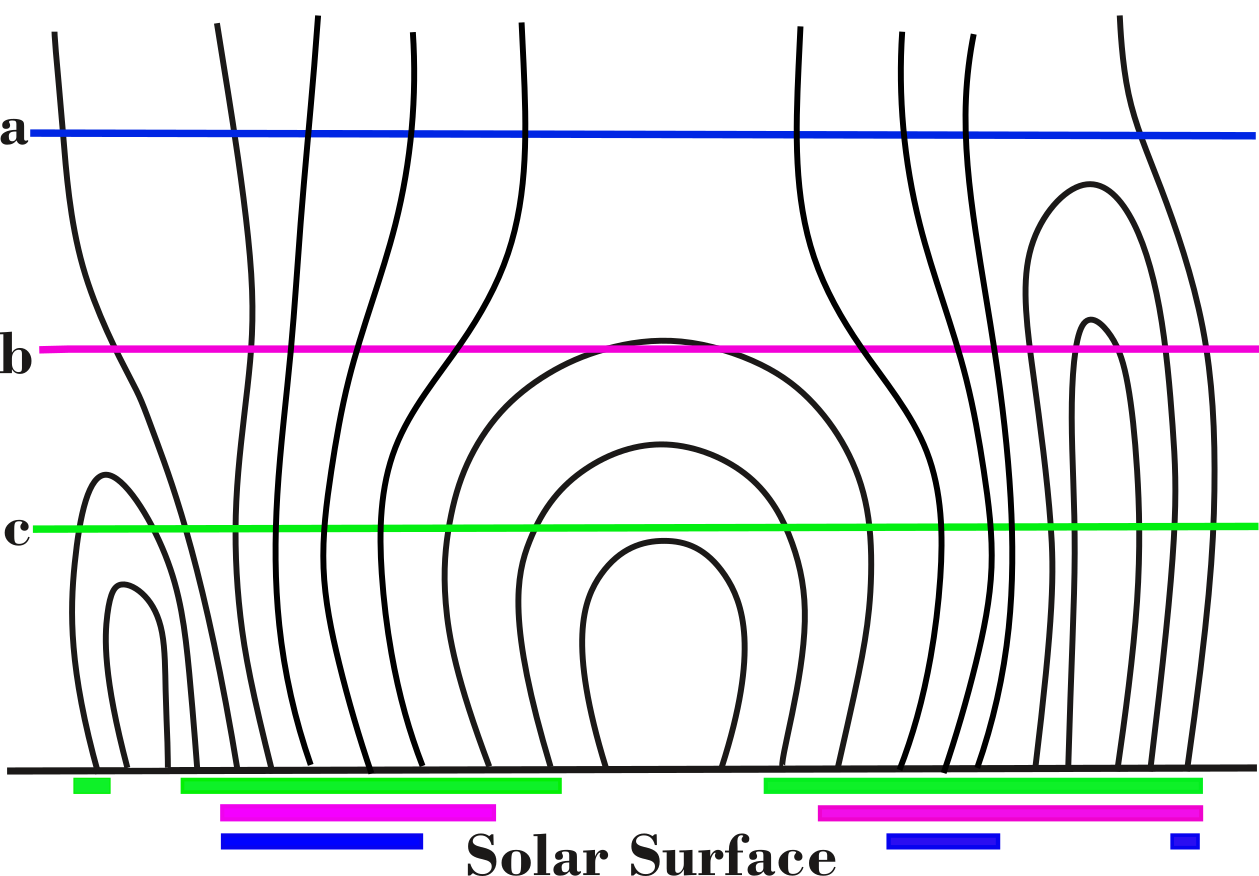}
        \put(49,-13){(a)}
        \end{overpic}
    \end{minipage}%
    \begin{minipage}{0.41\textwidth}
        \centering
        \begin{overpic}[width = 1\textwidth]{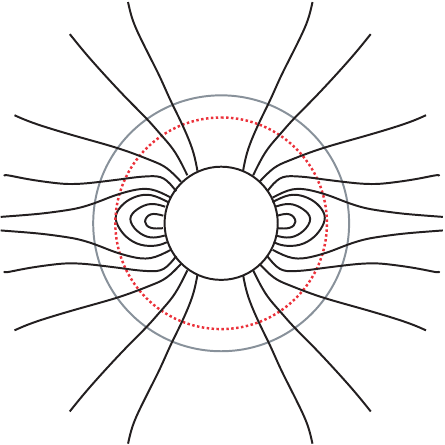}
        \put(45,0){(b)}
        \end{overpic}
    \end{minipage}
    \end{minipage}
    \caption{(a) Schematic representation of the impact of the source surface height in the open and closed field line assignment. Lines a (thin blue line), b (thin magenta line) and c (thin green line) illustrate three possible positions for the source surface. The lower it is placed the more field lines are considered as open. To visualise this we added the thicker lines at the lower part of the image, with representative colours, which show how the area the open field lines map down to grows the lower the source surface is placed. (b) An illustration depicting the coronal domain divided into the inner and outer corona by the source surface (grey circle). The red dotted circle represent the inner boundary of the SCS model which approximates the magnetic field in the outer corona.}
 \label{theory_imag}
 \end{figure}

Being an empirical model, the WSA consists of a parametrisation including several free parameters, which throughout the decades have been fitted and re-fitted to conform in--situ observations of the solar wind plasma at 1AU. One such parameter is the height of the outer boundary of the PFSS model, known as the source surface. In the traditional WSA model it serves as the inner boundary to the SCS model \cite{wang_solar_1990, wang_potential_1992, arge_improvement_2000}. The source surface and its distance from the Sun, R$_{\rm ss}$, divides the corona into two zones; the inner and outer. Field lines piercing through the source surface are considered as open, while those forming closed loops below it are identified as belonging to closed magnetic structures. Open field lines are stretched out and away in the Heliosphere by the solar wind forming the Interplanetary Magnetic Field (IMF) \cite<e.g.,>[]{wang_topological_2003}. The PFSS does not account for dynamic effects that are especially important at the top of streamers \cite<e.g., the high closed loops>[]{arge_two_2002}. Due to the association of open field to CHs, the areas on the source surface that are defined by concentration of open field lines map down to the low corona to CHs, i.e. each field line that is open is rooted in the low corona to (one or more) CHs. The location/height of $R_{\rm ss}$ determines strongly the total area of open field regions rooted in the low corona. Lowering $R_{\rm ss}$ will result in more field lines being open, and subsequently in larger areas of CHs. On the contrary, increasing it will reduce the number of open field lines and thus shrinking CH areas. This is visualised in panel (a) of Figure \ref{theory_imag}, where changing the source surface from height \textit{a} to \textit{b} and \textit{c} expands the areas of open flux. Qualitatively some field lines that were forming closed loops for the case of a source surface at height \textit{a} are crossing through the source surface placed at height \textit{b} and therefore are not considered closed anymore. Decreasing the height even lower to height \textit{c} will result in increasing the number of closed loops that are becoming open, and therefore, in further growing of the open flux areas. 

Since fast solar wind and open magnetic field are related to CHs, for the WSA model to successfully reconstruct the in-situ measured solar wind properties and the open magnetic flux, it should model the size and geometry of CHs as accurately as possible. The dependence of the latter on the source surface height makes the choice of $R_{\rm ss}$ a definitive element in the model's success.

Selecting the optimal source surface height has been a long-standing debate. Earlier studies \cite<e.g.,>[]{altschuler_magnetic_1969} indicated that the source surface is best placed at 2.5$R_{\rm \odot}$ ($R_{\rm \odot}$ = 1 solar radius). This value has been adopted by many as the canonical height; however, a range of values from 1.6$R_{\rm \odot}$ to 3.25$R_{\rm \odot}$ have been used \cite<e.g.,>[]{arden_breathing_2014}. Criteria for the choice of these values have been (i) the sector structure of the Heliospheric Current Sheet (HCS) \cite{hoeksema_structure_1982,hoeksema_structure_1983,hoeksema_atlas_1986}, (ii) in-situ and remote observations of the HCS by Ulysses \cite{phillips_ulysses_1994, wang_solar_1995} and during total solar eclipses \cite{schatten_model_1969}, (iii) the interplanetary magnetic field polarity \cite{hoeksema_structure_1983}, (iv) the coronal hole boundaries and (v) the Interplanetary Magnetic Field (IMF) strength \cite<e.g.,>[]{lee_coronal_2011}. While plenty of studies were restricted to and/or concluded the use of a singular value for $R_{\rm ss}$ \cite{ altschuler_magnetic_1969, hoeksema_structure_1983,  arge_improvement_2000, mcgregor_analysis_2008, wang_coronal_2009, riley_role_2015, linker_open_2017}, others have been suggesting a solar cycle dependant source surface height \cite<e.g.,>[]{lee_coronal_2011, arden_breathing_2014}, a concept known as the "breathing" source surface \cite<termed by>[]{arden_breathing_2014}. In addition to that concept, the idea of a non-spherically shaped source surface has also been proposed \cite<e.g.,>[]{levine_solar_1977,schulz_non-spherical_1997, schulz_non-spherical_2008}.

Aiming to eliminate a known discontinuity in the WSA model, namely kinks in the field lines occurring at the transition from the low--coronal PFSS to the high--coronal SCS domain, \citeA{mcgregor_analysis_2008} re-positioned the inner boundary of the SCS model at a radius $R_{\rm i}$ lower than the source surface ($R_{\rm i}<R_{\rm ss}$). This concept is illustrated in Figure \ref{theory_imag} panel b, where the red dashed curve encircling the Sun represents the new inner boundary of the SCS model, and the grey solid curve is the source surface placed at height $R_{\rm ss}$. In establishing the optimal pair of [$R_{\rm i}, R_{\rm ss}$] heights for their new adaptation of the WSA model \citeA{mcgregor_analysis_2008} aimed for conservation of the absolute total magnetic flux between the one from the original WSA model and that from their updated version. The purpose of that was to facilitate the comparison between the two models, original and improved WSA. The pair they deduced in their study, is [2.3, 2.6]$R_{\rm \odot}$ for $R_{\rm i}$ and $R_{\rm ss}$ respectively. 

EUHFORIA (EUropean Heliospheric FORecasting Information Asset) is a novel heliospheric wind and CME evolution model \cite{pomoell_euhforia:_2018} consisting of two main building blocks, the coronal model and the heliospheric full MHD model. The coronal model aims to reconstruct the solar wind plasma and the open flux conditions in the upper corona at 21.5$R_{\rm \odot}$, which serve as boundary conditions to the heliospheric model. It implements for that purpose the WSA model, with the \citeA{mcgregor_analysis_2008} improvements \cite<for more details read>[]{pomoell_euhforia:_2018}. The default [$R_{\rm i}, R_{\rm ss}$] pair of heights is the one proposed by \citeA{mcgregor_analysis_2008}. In a comprehensive study of modelling the background solar wind with EUHFORIA using the adopted WSA model and default heights \cite[in preparation]{hinterreiter_solar_wind_2019} it became evident that the capability of the model to predict a high-speed stream at Earth is often quite low, with predicted high speed streams either shifted in time, having lower than expected amplitude in the solar wind speed profiles or being overall absent. Similar results have been reported in other physics-based reconstructions of the ambient solar wind \cite{owens_metrics_2008,lee_solar_2009, jian_comparison_2011, gressl_comparative_2014}. The novelty of our study is the assessment of the PFSS+SCS capability to reconstruct CH areas with accuracy, knowing the relation between CHS and high speed streams.

In the model the source surface and the SCS inner boundary positions are adjustable parameters. Considering the freedom in the choice of the source surface height, already brought forward by other studies discussed above, we focused firstly in qualitatively assessing the default values of [$R_{\rm i}, R_{\rm ss}$], and subsequently in investigating the possibility of other height choices. To do that, we selected a sample of 15 CHs located around the central meridian of the solar disk as viewed from Earth, but at different latitudes (see section \ref{subsec:CH_selection}). Using \textsc{catch} (Collection of Analysis Tools for Coronal Holes) \cite<>[in preparation]{Heinemann_CATCH_2019} we extracted the boundaries of these CHs based on their appearance in EUV filtergrams at 193~{\AA} (section \ref{subsec:CATCH_tool}). We reconstructed their areas with EUHFORIA, by varying the paired values for the heights of the source surface and the inner boundary of the SCS model, in order to find optimal values that produce a better match between modelled and observed CH areas (section \ref{subsec:euhforia}). A quantitative analysis of the results was performed by defining two parameters, the coverage and the connected pixels of open flux associated to the CH (both parameters are analytically described in section \ref{subsec:results}).

\clearpage
\begin{figure}[b!]
    \centering
    \begin{minipage}{0.8\textwidth}
    \begin{minipage}{0.3\textwidth}
        \centering
        \begin{overpic}[width = 1\textwidth]{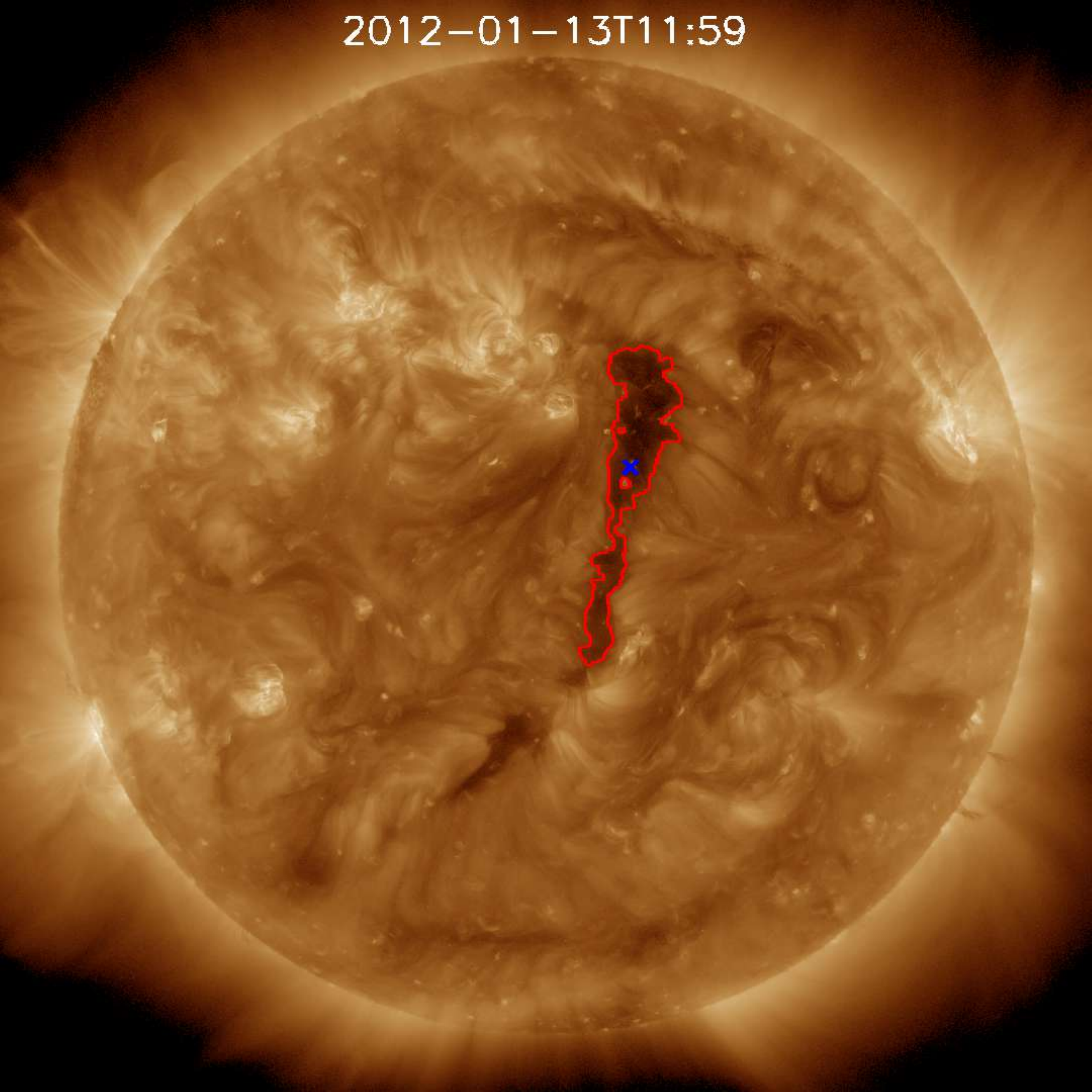}
        \end{overpic}
        \begin{overpic}[width = 1\textwidth]{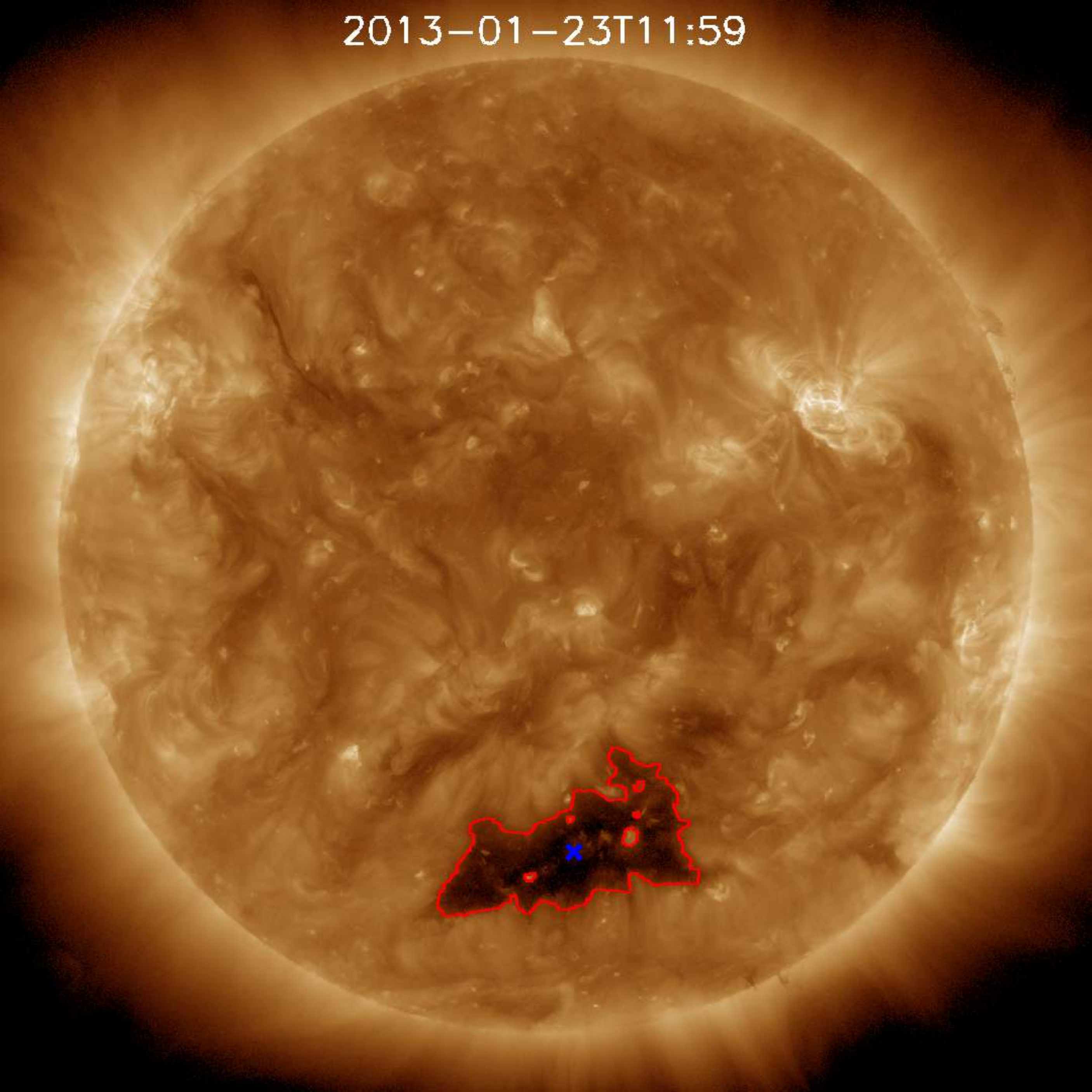}
        \end{overpic}
        \begin{overpic}[width = 1\textwidth]{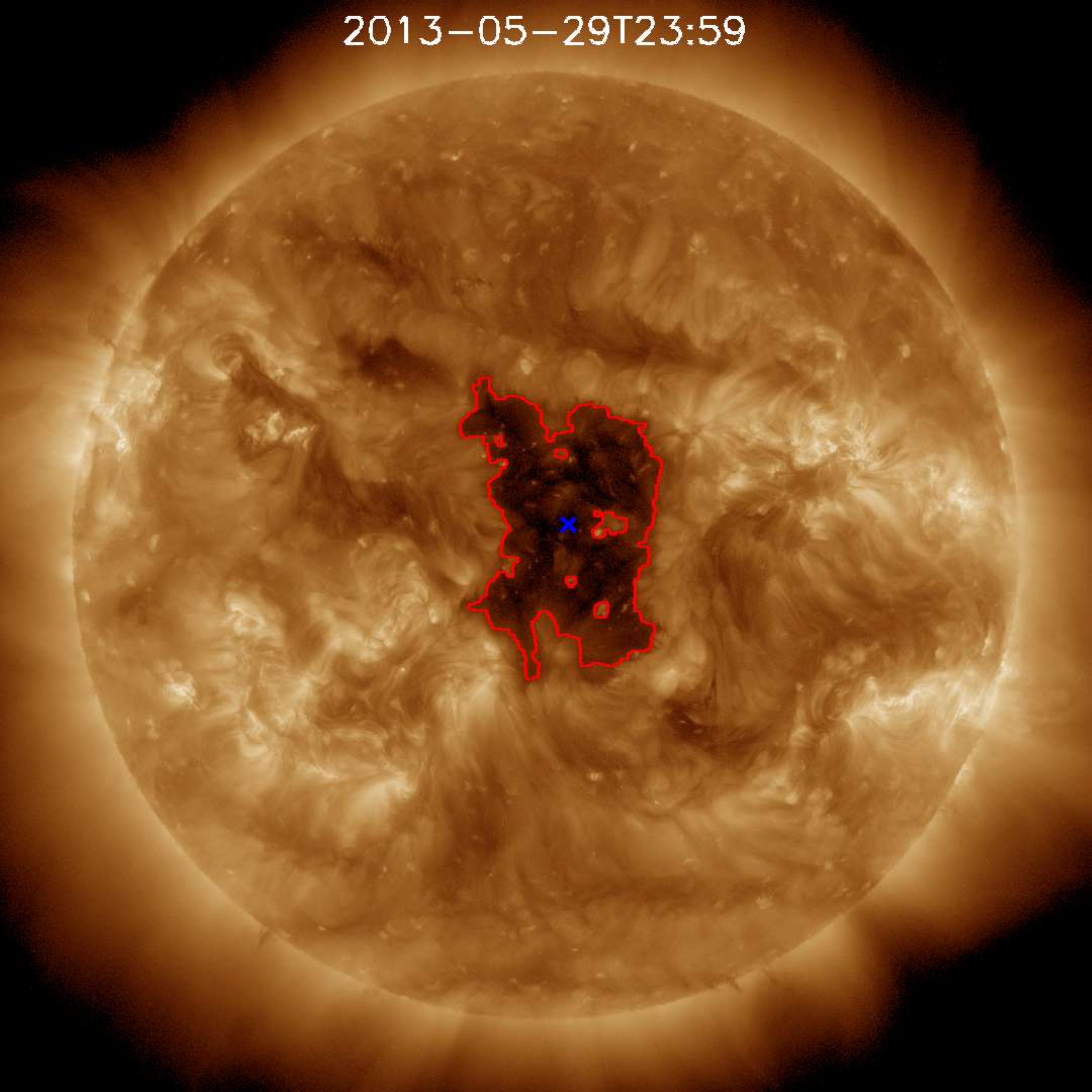}
        \end{overpic}
        \begin{overpic}[width = 1\textwidth]{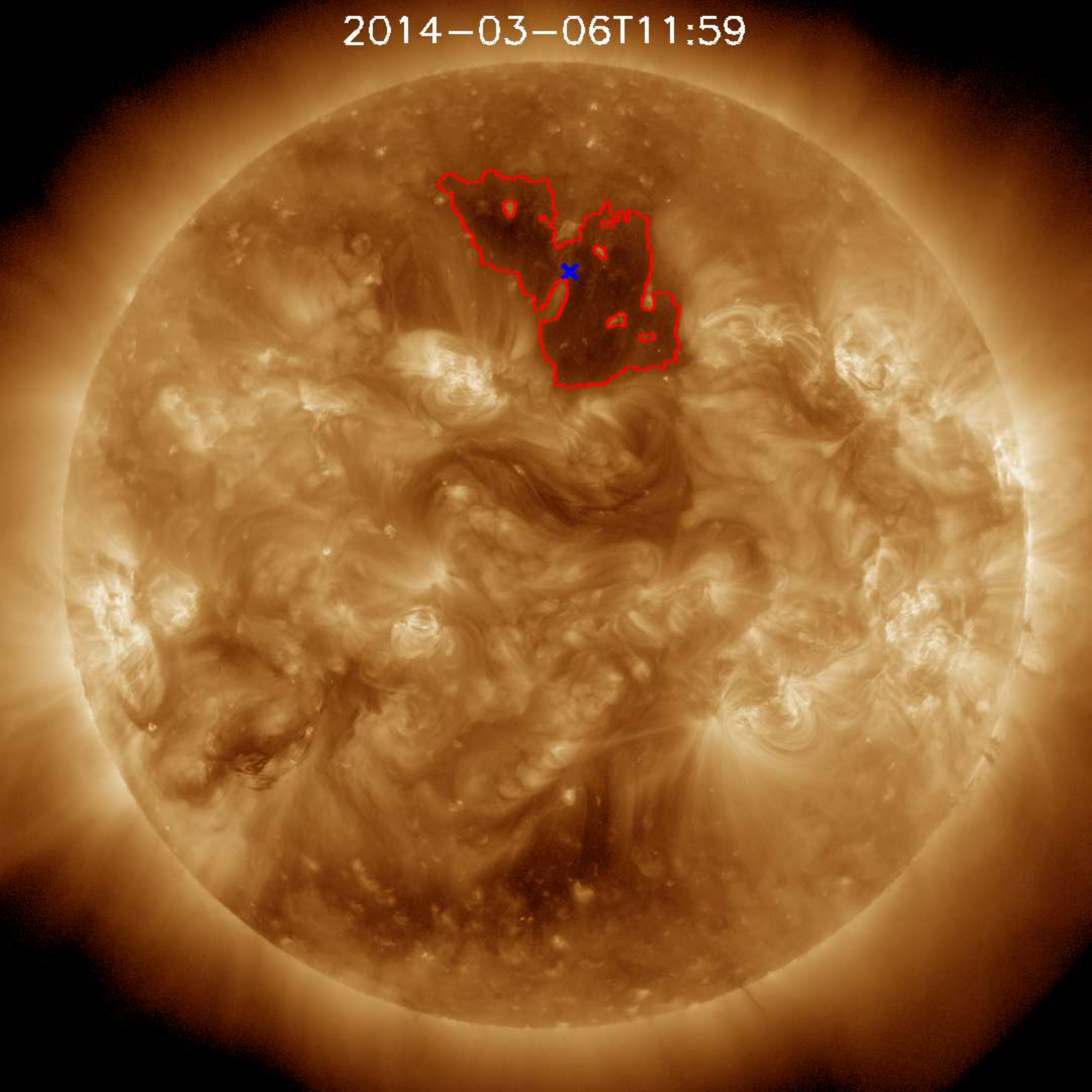}
        \end{overpic}
        \begin{overpic}[width = 1\textwidth]{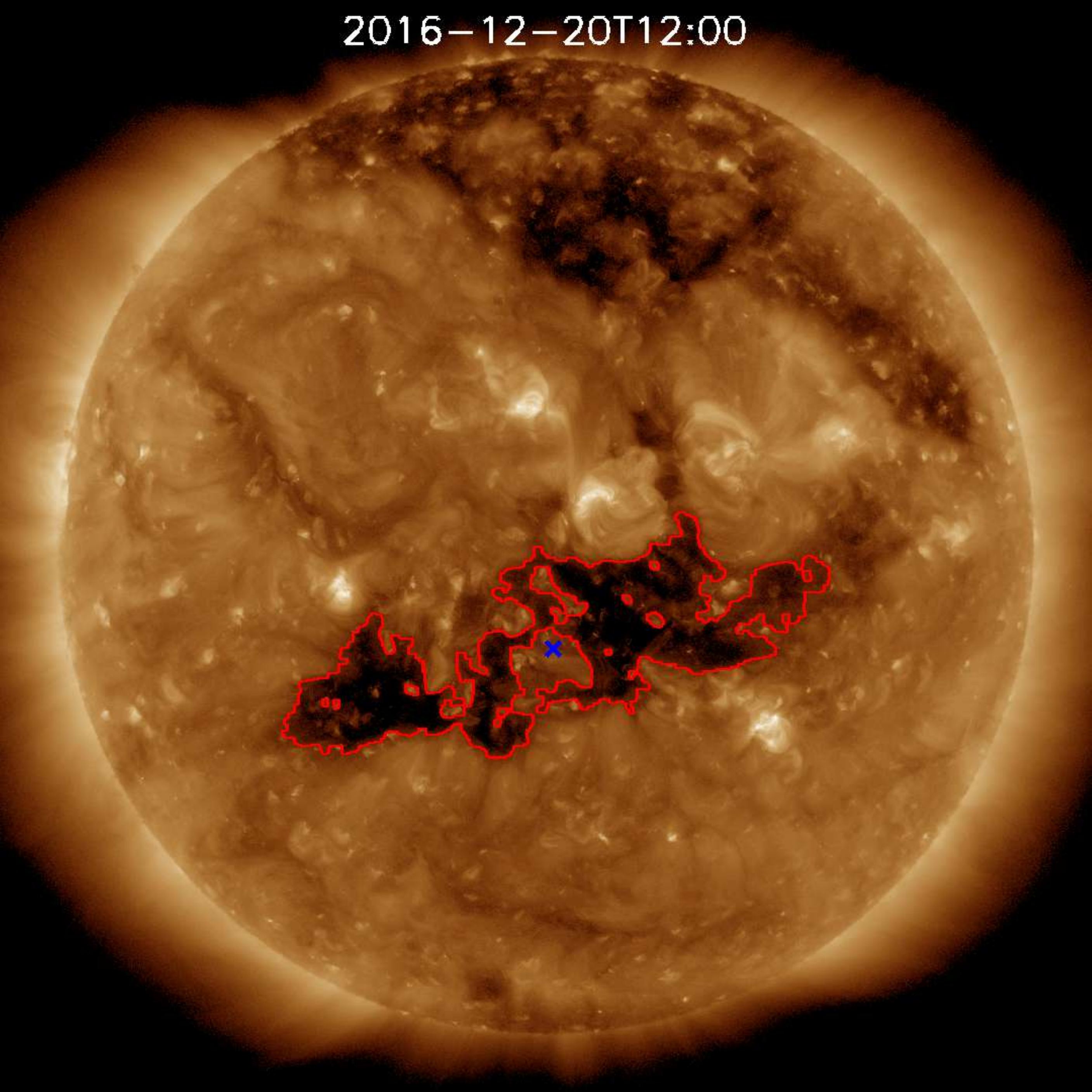}
        \end{overpic}
    \end{minipage}
    \begin{minipage}{0.3\textwidth}
        \centering
        \begin{overpic}[width = 1\textwidth]{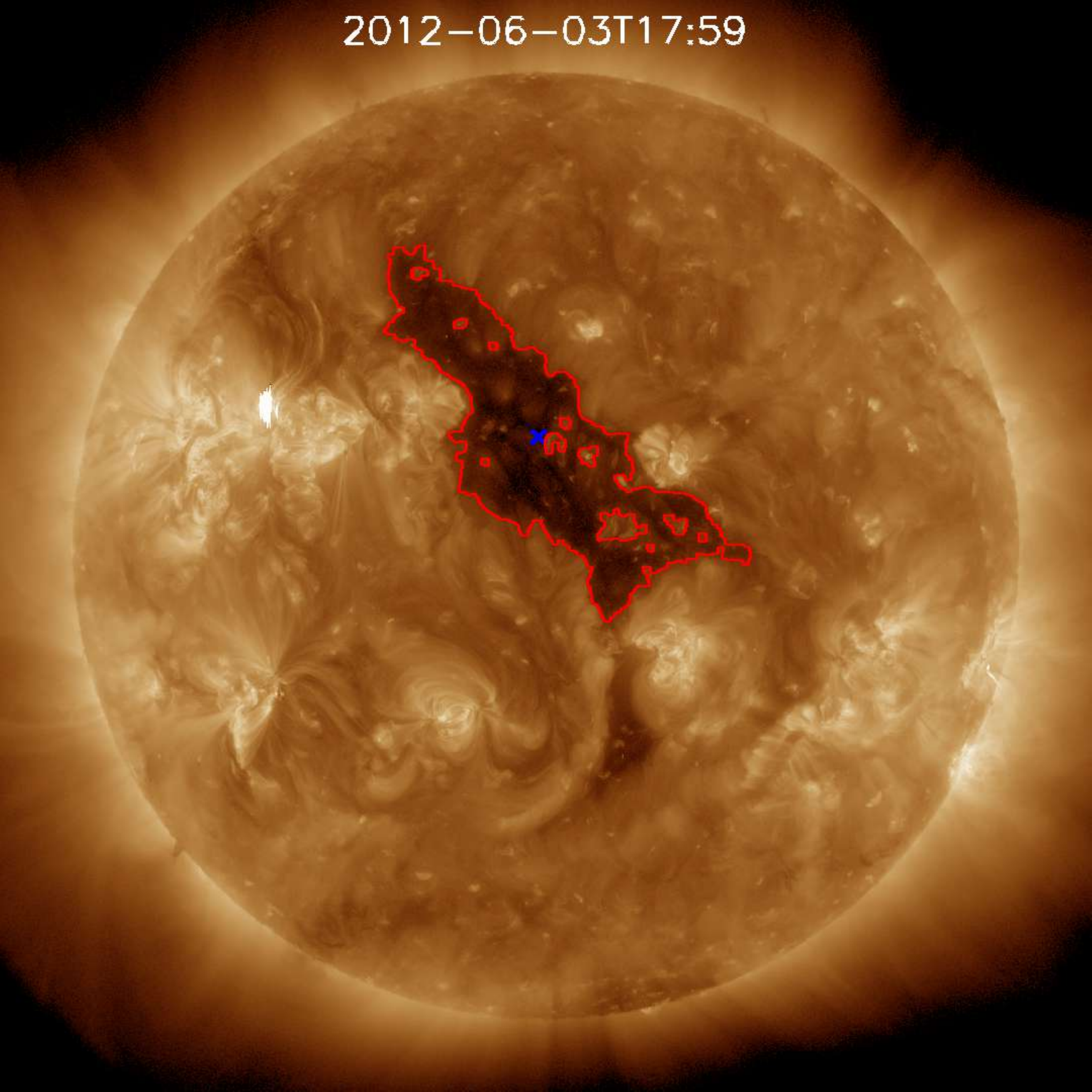}
        \end{overpic}
        \begin{overpic}[width = 1\textwidth]{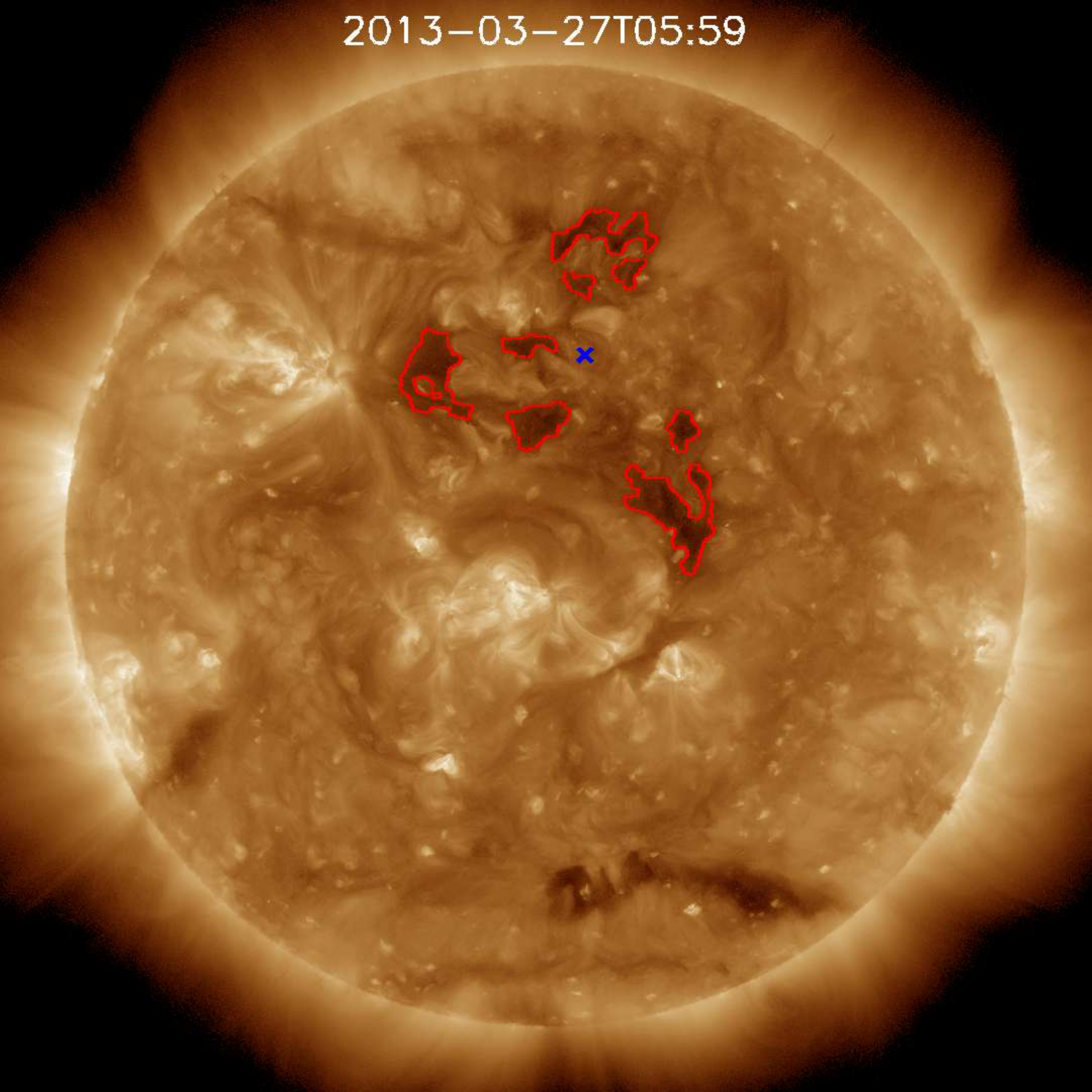}
        \end{overpic}
        \begin{overpic}[width = 1\textwidth]{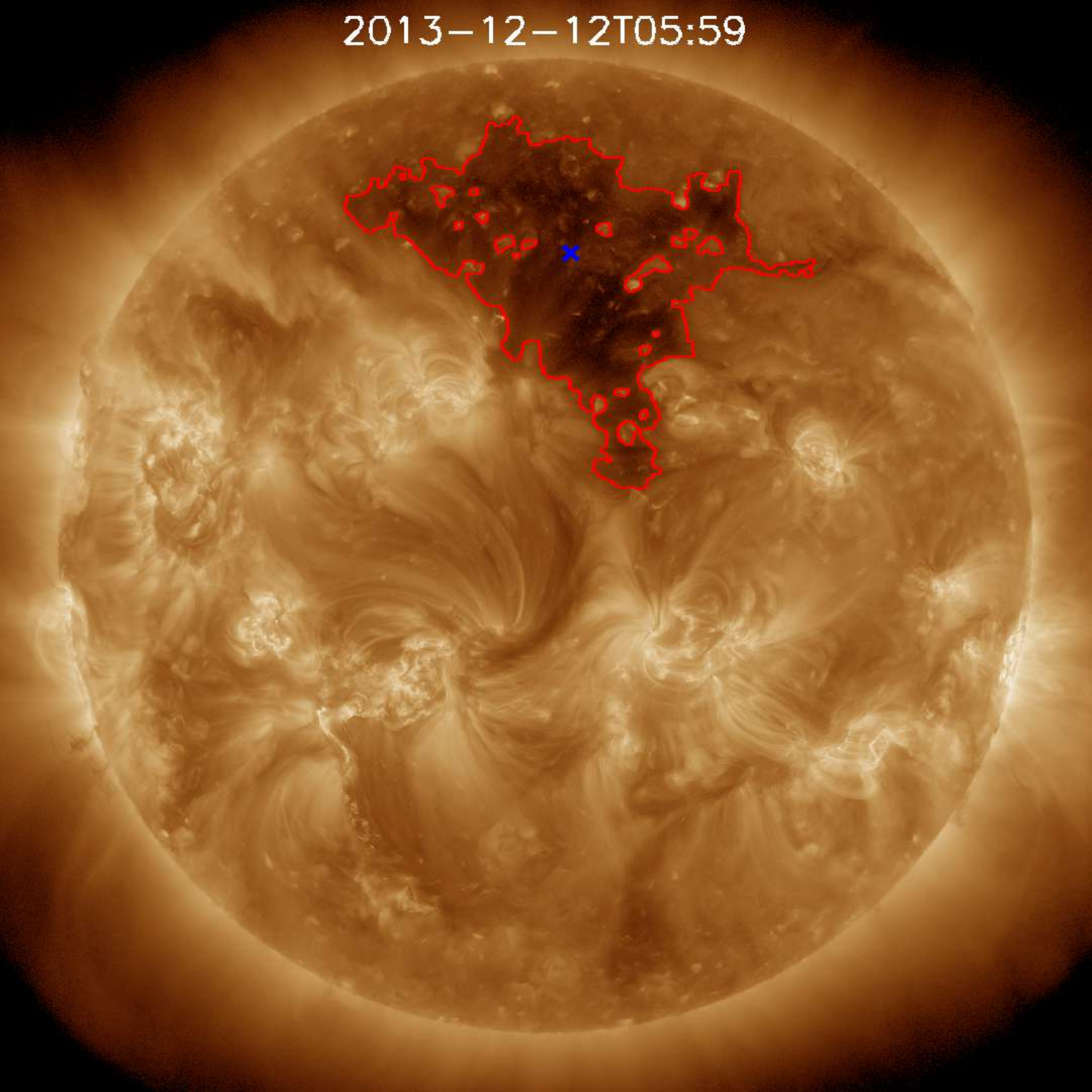}
        \end{overpic}
        \begin{overpic}[width = 1\textwidth]{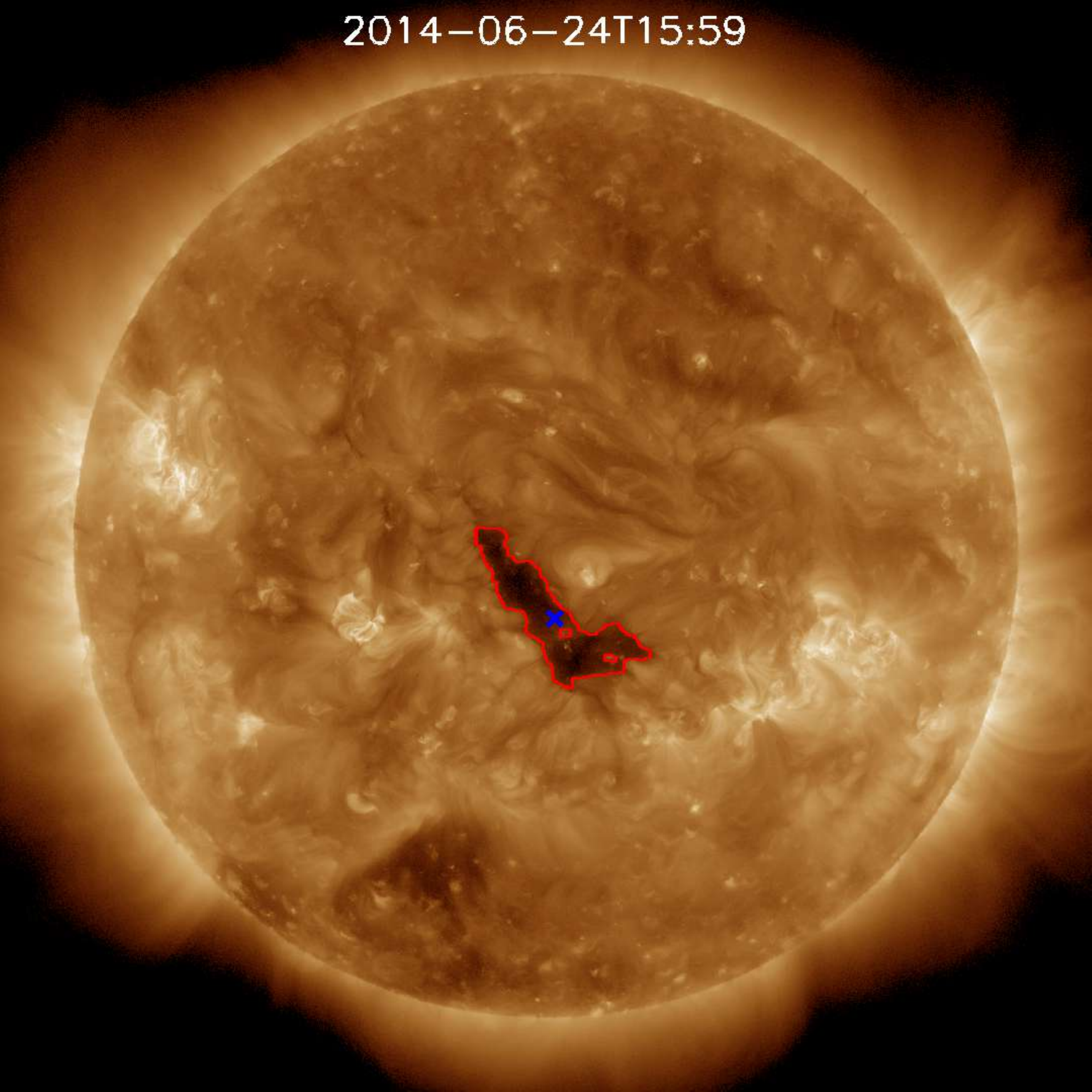}
        \end{overpic}
        \begin{overpic}[width = 1\textwidth]{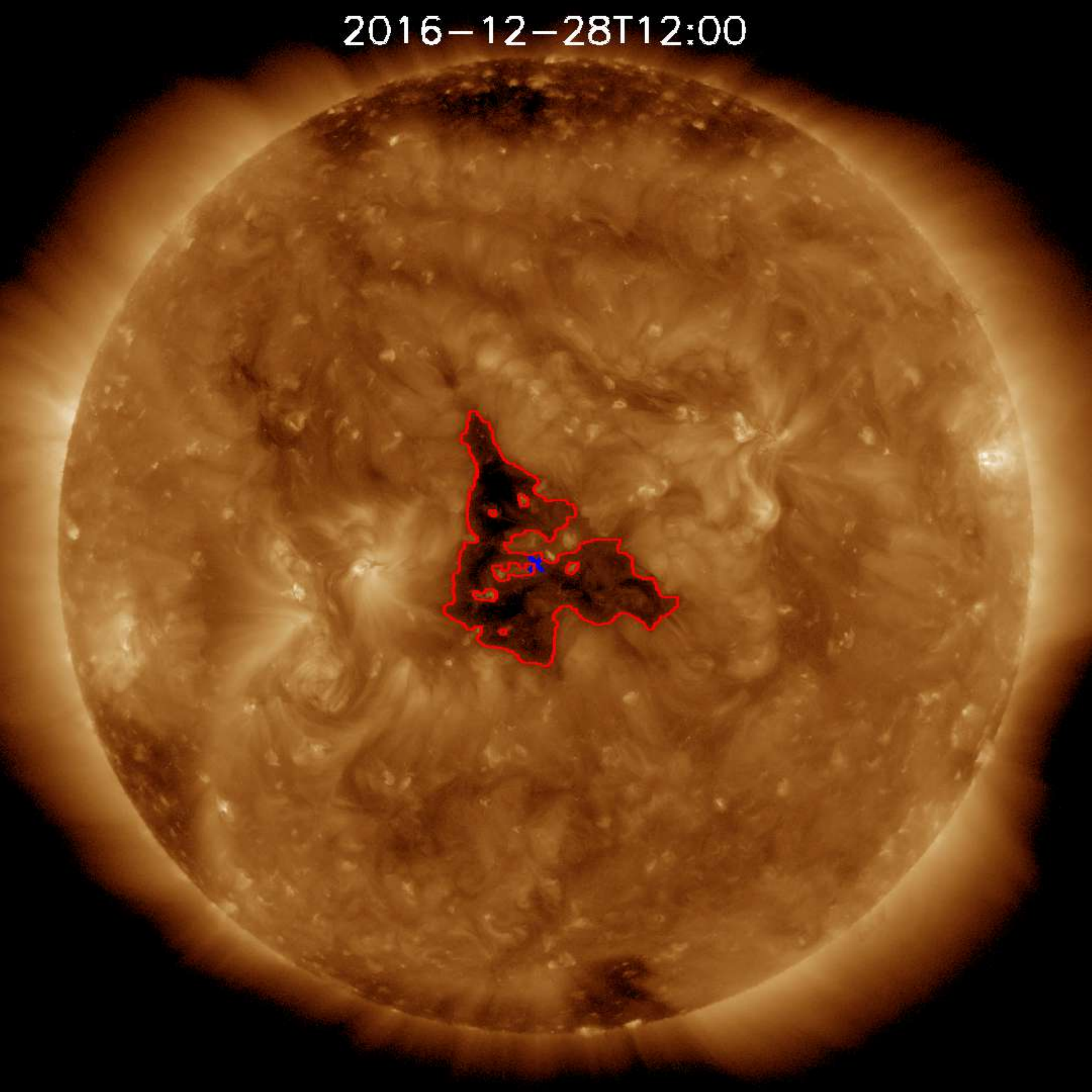}
        \end{overpic}
    \end{minipage}
    \begin{minipage}{0.3\textwidth}
        \centering
        \begin{overpic}[width = 1\textwidth]{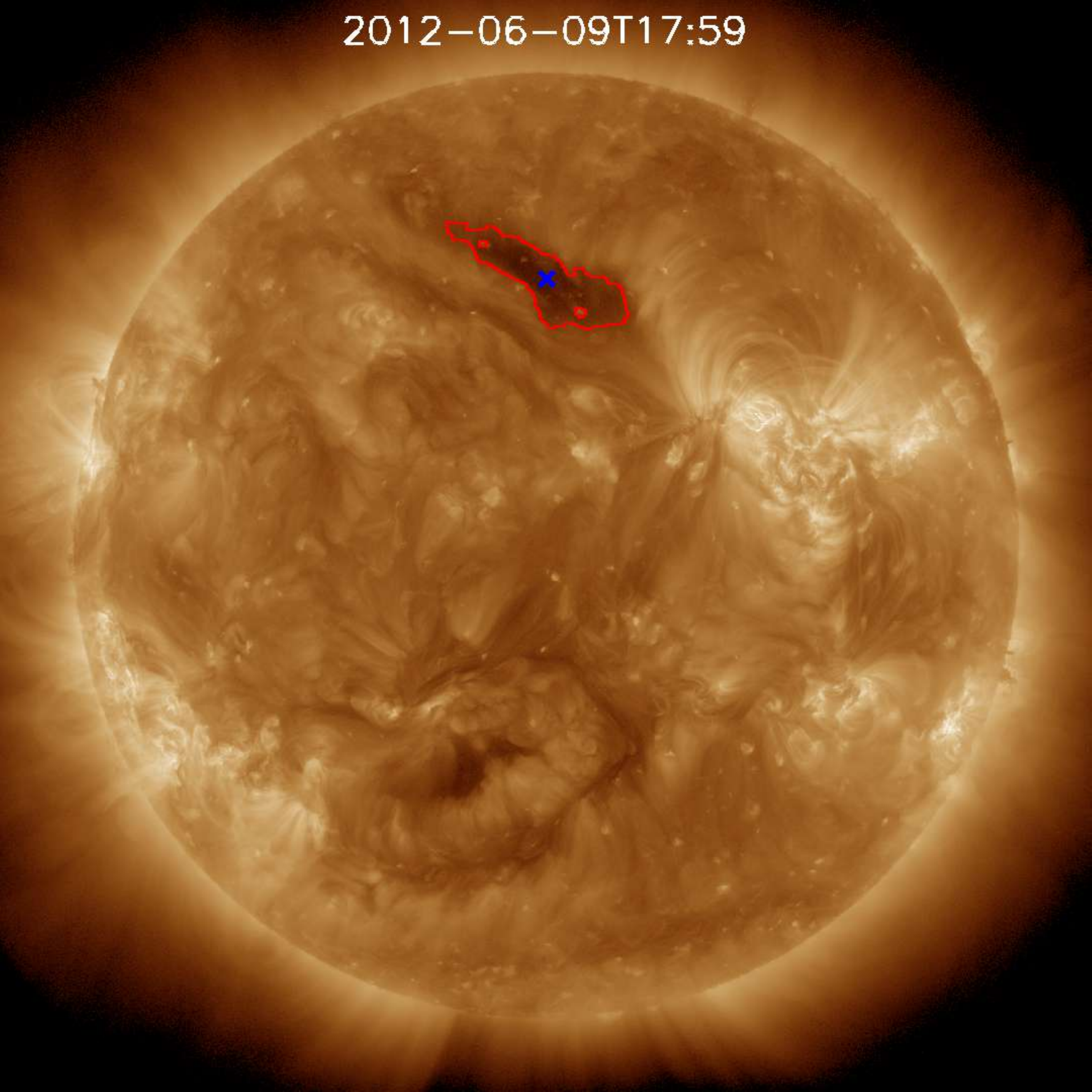}
        \end{overpic}
        \begin{overpic}[width = 1\textwidth]{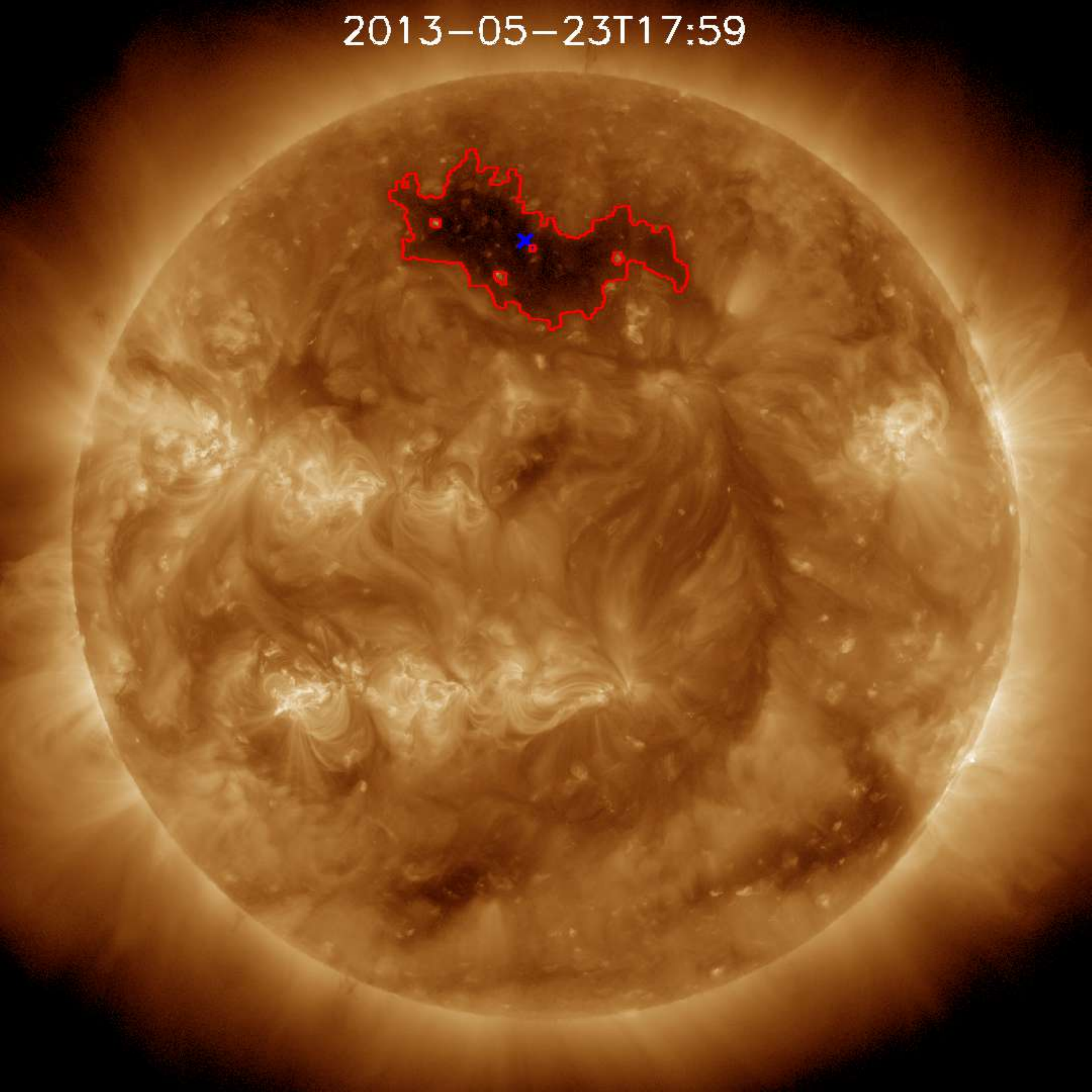}
        \end{overpic}
        \begin{overpic}[width = 1\textwidth]{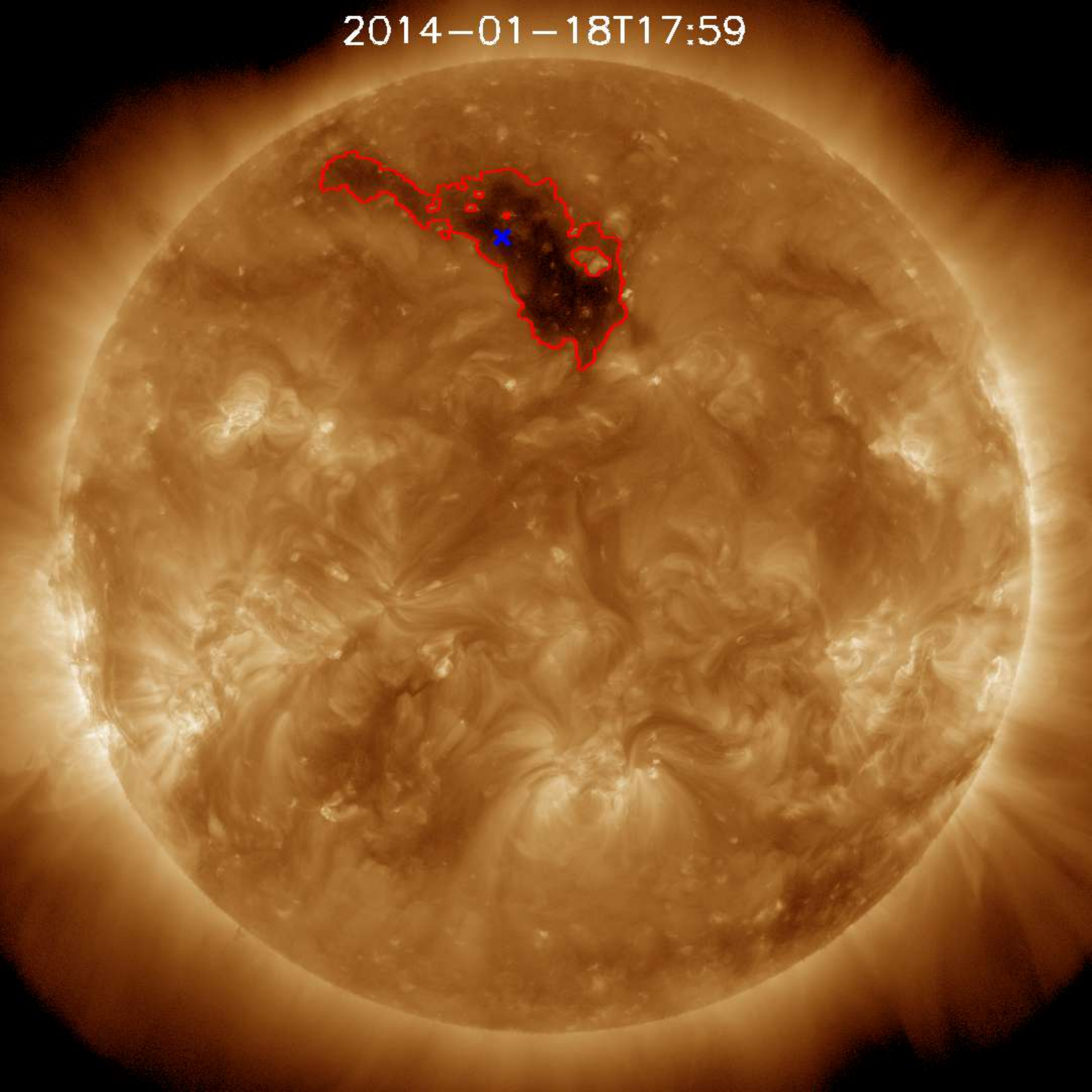}
        \end{overpic}
        \begin{overpic}[width = 1\textwidth]{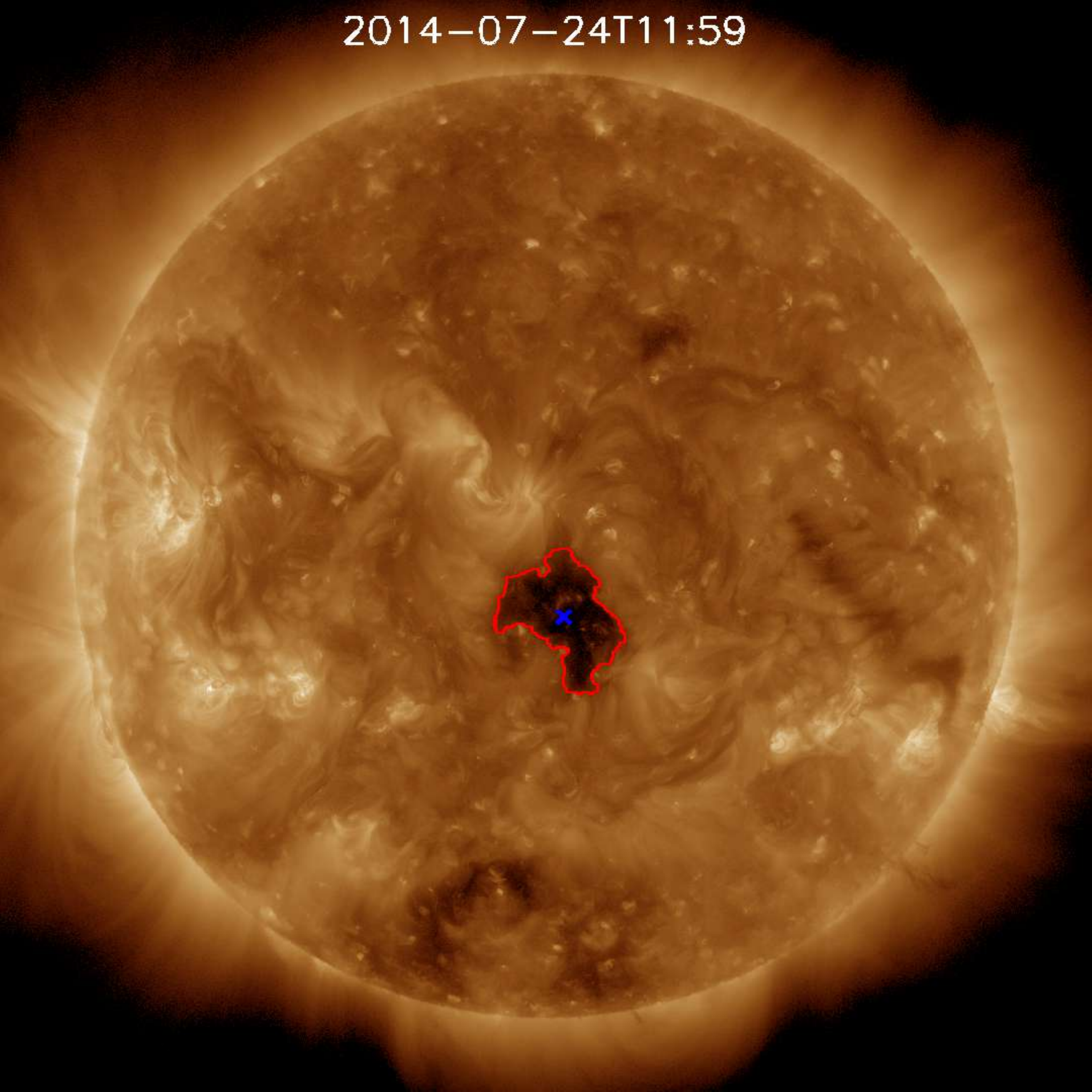}
        \end{overpic}
        \begin{overpic}[width = 1\textwidth]{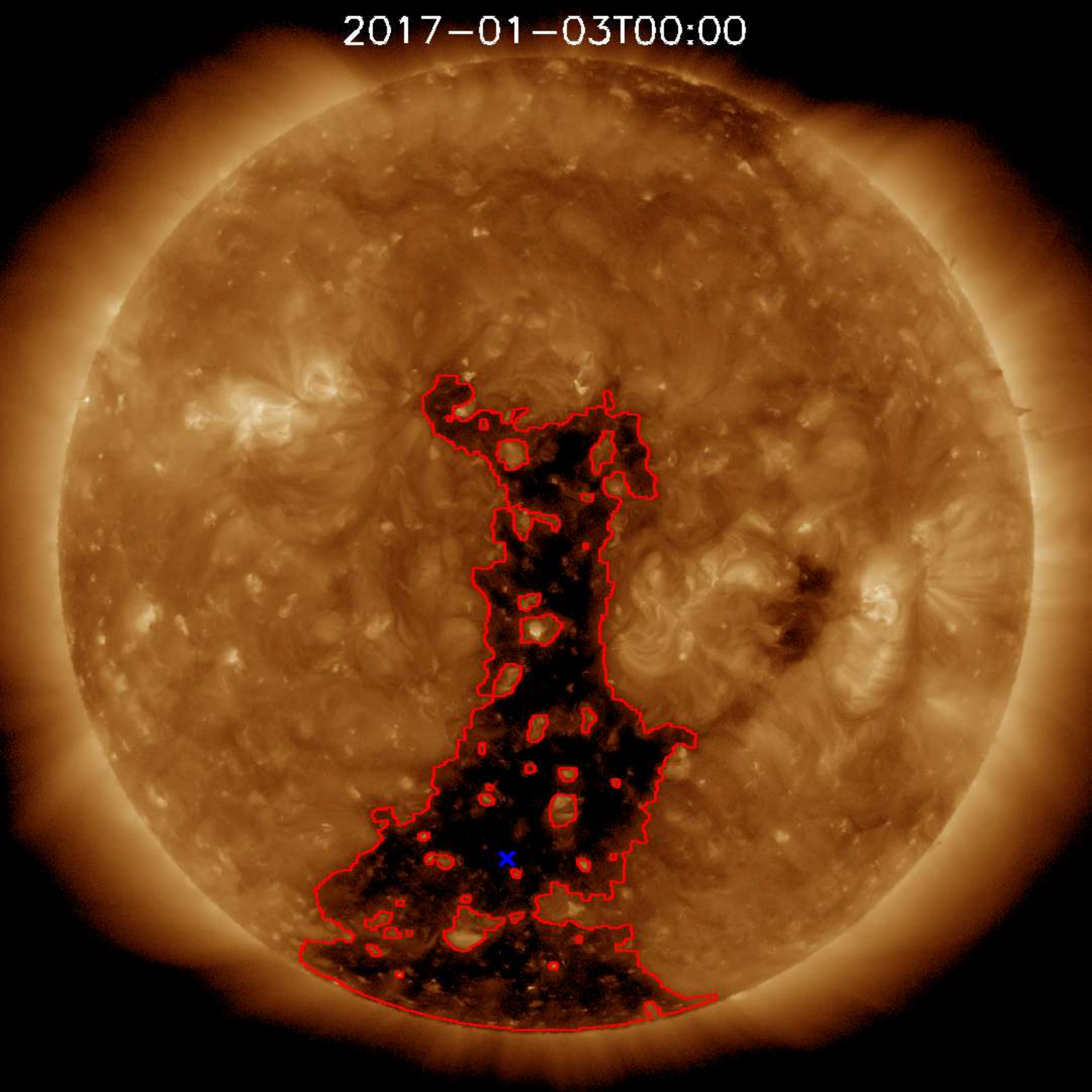}
        \end{overpic}
    \end{minipage}
    \end{minipage}
    \caption{EUV filtergrams at 193~{\AA} depicting the selected CHs in chronological order. The red curves mapping out the boundaries of the CHs are the optimal ones defined using \textsc{catch}.}
 \label{EUV_CH}
 \end{figure}

From this study we conclude that the PFSS+SCS part of the adopted WSA model for the default heights of [2.3, 2.6]$R_{\rm \odot}$ does not properly reconstruct CH areas. Lowering the heights [$R_{\rm i}, R_{\rm ss}$] improved the model results, i.e. open field lines are rooted to areas in the lower corona that better represented the CHs observed in EUV images. However, it also led in the presence of open field lines that mapped down to the corona at areas lying outside the CH boundaries defined via observations. This is an undesirable modelling effect and we attempted to identify optimal height pair candidates by selecting those who lead to a modelling result that has improved CH area reconstruction without expanding the CH area beyond the expected boundaries.

\section{Data and Methodologies}

\subsection{Observations}
\label{subsec:magdbase}
To compute the solar wind and magnetic field in the corona the WSA model requires a magnetogram as input. A variety of magnetograms are available; however, they are obtained at different observatories, using different instruments, and constructed following various distinct techniques. Therefore, they display quantitative differences among each other (Riley et al., 2014). Aiming in a coherent output and to avoid any effects on the modelled results due to disparities between magnetograms from different sources we used exclusively magnetograms provided by the Global Oscillation Network Group (GONG) and developed with the Air Force Data Assimilative Photospheric flux Transport (ADAPT) model. These are synchronic magnetograms, meaning they are developed such as to resemble with high accuracy the magnetic field of the solar surface on a particular time. A detailed description of the ADAPT model can be found in \citeA{arge_air_2010} and \citeA{hickmann_data_2015}, while the GONG-ADAPT database from which magnetograms were acquired is: \url{ftp://gong.nso.edu/adapt/maps/gong/}.   

For extracting the CH boundaries we used EUV observations during 2012 - 2017 from the Atmospheric Imaging Assembly \cite<AIA:>[]{lemen_atmospheric_2012} on board the Solar Dynamic Observatory \cite<SDO:>[]{pesnell_solar_2012}. For this study we utilized the full disk coronal images ($1024\times1024$) at a wavelength of 193 ~{\AA}, which are made available by the Joint Science Operation Center (JSOC). Although, it is argued that CHs appear differently when observed at different wavelengths, i.e. varying boundaries, the 193 ~{\AA}\ filter is most commonly used. The response function of this filter is most sensitive in the temperature range of quiet sun and CH regions resulting in a high contrast. Using these data, we extract the CH area and its boundary in a manner described in section \ref{subsec:CATCH_tool}.

\subsection{Selected Coronal Holes}
\label{subsec:CH_selection}

The 15 selected CHs are shown in chronological order in Figure \ref{EUV_CH}. As mentioned earlier, all CHs have longitudinal position of centre of mass (CoM) located along or within $10^{\circ}$ of the central meridian of the Sun. The latitudinal position of their CoM, however, varies from low to high. Limiting the sample to CHs located only along the central meridional zone provides the opportunity to focus our assessment on the possible impact of the CH latitudinal position in modelling their areas. In addition, the sample covers the extended maximum, declining, and early minimum phases of solar cycle 24, from 2012 to 2017, enabling the investigation of potential solar activity effects in reconstructing CHs and thus in the choice of the source surface height, an idea that has been suggested before \cite{lee_coronal_2011, arden_breathing_2014}.

Out of the 15 CHs, 11 are latitudinally and 3 longitudinally elongated, while one is rather small and bears a more round appearance. Regarding their CoM latitude, 5 and 3 CHs are low latitude ones positioned in the north and south hemisphere respectively, while 5 and 2 CHs are mid latitude ones in north and south hemisphere accordingly. Due to known weaknesses on magnetic field and EUV observations in polar regions (i.e. line-of-site effects) no polar CHs were selected for this study; however, the CH on the 2017-01-03 is clearly connected to a polar CH, and is an exception to our sample. In this particular case we do consider the polar part that is visible in the EUV image, acknowledging the impact this will have on the result. The CH on 2016-12-20 appears to be associated to a polar one as well. A faint channeling between the CH and the polar one is visible in the EUV images. Nevertheless, we assess it individually as its boundaries can be defined and are separated from those of the polar CH. The sample also consists of CHs exhibiting different level of patchiness that is defined by the presence within the CH of large areas with a dipole magnetic field configuration (i.e. 2017-01-03), or due to the CH consisting of a cluster of small dark regions (i.e. 2013-03-27).

\subsection{Extracting CH boundaries with CATCH}
\label{subsec:CATCH_tool}
The CH boundaries were extracted using \textsc{catch} \cite<>[in preparation]{Heinemann_CATCH_2019} employing EUV 193 ~{\AA}\ filtergrams from AIA/SDO. \textsc{catch} uses an intensity threshold method enhanced with a gradient modulation at the CH boundary to find an optimal threshold for the extraction. For better quantifying the uncertainties in the PFSS+SCS computed open magnetic field when comparing to the coronal remote sensing observations, in addition to the optimal CH boundary we also define lower and upper bounds for each CH. The lower and upper bounds present a significantly over- and underestimated threshold for the extraction.

\begin{table}
 \caption{CH characteristics based on remote sensing observations and \textsc{catch} extractions, where CoM is Centre of Mass}
 \centering
 \begin{tabular}{c c c c c c}
 \hline
   \textbf{No} & \textbf{Date-time} & \textbf{CoM$^{\ast}$ latitude} & \textbf{CoM$^{\ast}$ longitude} & \textbf{Area} & \textbf{Mean intensity} \\
& \textbf{[UTC]} & \textbf{[deg]} & \textbf{[deg]} & \boldmath{[$10^{10} km^{2}$]} & \textbf{[DN]} \\
\hline
1 & 20120113T12:00 & 5.01 & 10.07 & 2.34 $\pm$ 0.12​ & 36.31 $\pm$ 0.86 \\
2 & 20120603T18:00 & 13.04 & 0.91 & 9.69 $\pm$ 0.33​ & 31.29 $\pm$ 0.62 \\
3 & 20120609T18:00 & 34.97 & 0.18 & 2.02 $\pm$ 0.16​ & 44.60 $\pm$ 1.36 \\
4 & 20130123T12:00 & -44.45 & 4.70 & 6.03 $\pm$ 0.16​ & 27.79 $\pm$ 0.62 \\
5 & 20130327T06:00 & 16.71 & 4.93 & 3.37 $\pm$ 0.42​ & 38.28 $\pm$ 1.02 \\
6 & 20130523T18:00 & 38.61 & 3.18 & 6.91 $\pm$ 0.36​ & 29.30 $\pm$ 0.81 \\
7 & 20130530T06:01 & 1.69 & 2.76 & 8.07 $\pm$ 0.24​ & 26.68 $\pm$ 0.56 \\
8 & 20131212T05:59 & 36.52 & 3.75 & 17.16 $\pm$ 0.88​ & 38.59 $\pm$ 1.17 \\
9 & 20140118T17:58 & 34.53 & 6.24 & 6.34 $\pm$ 0.34​ & 35.83 $\pm$ 1.17 \\
10 & 20140306T12:00 & 27.46 & 3.23 & 6.71 $\pm$ 0.43​ & 38.80 $\pm$ 1.09 \\
11 & 20140624T18:00 & -6.74 & 1.10 & 1.95 $\pm$ 0.06​ & 29.45 $\pm$ 0.57 \\
12 & 20140724T12:10 & -3.53 & 2.21 & 2.16 $\pm$ 0.05​ & 25.70 $\pm$ 0.46 \\
13 & 20161220T12:10 & -13.86 & 0.92 & 10.55 $\pm$ 0.46​ & 22.31 $\pm$ 0.54 \\
14 & 20161228T12:00 & -4.77 & 1.12 & 4.84 $\pm$ 0.16 & 25.58 $\pm$ 0.45 \\
15 & 20170103T00:00 & -43.35 & 6.21 & 42.18 $\pm$ 0.71 & 19.00 $\pm$ 0.32 \\
 \hline
 \multicolumn{2}{l}{$^{\ast}$ CoM: Centre of Mass}
 \end{tabular}
\label{CH_characteristics}
\end{table}

\subsection{Reconstructing CH areas with EUHFORIA}
\label{subsec:euhforia}

The reconstruction of open flux areas was done using EUHFORIA. As already described in the introduction, EUHFORIA consists of two main building blocks, with the corona model being one of them. It employs the adaptation of the WSA model with the scheme suggested by \citeA{mcgregor_analysis_2008}, which comprises two sub-models, the PFSS and the SCS. The first computes the magnetic field and plasma conditions in the lower corona, from 1$R_{\rm \odot}$ up to the source surface, at height $R_{\rm ss}$. And the latter solves the magnetic field and plasma conditions in the outer corona, from $R_{\rm i}<R_{\rm ss}$ up to 21.5$R_{\rm \odot}$. Although, the PFSS model solution extends to the source surface, it forces the magnetic field there to be purely radial. This configuration is causing unrealistic field line kinks at the source surface in the magnetic field solution of the SCS model when its inner boundary is taken to be the source surface \cite{mcgregor_analysis_2008}. This is the reason why we select as the inner boundary conditions to the SCS model a solution of the PFSS at a surface $R_{\rm i}$ below the source surface where the magnetic field has all components. For each of the selected CHs we run the coronal model of EUHFORIA for a large sample of [$R_{\rm i}$, $R_{\rm ss}$] heights. To define the height pairs we started with $R_{\rm i}$ taking the value of 1.3$R_{\rm \odot}$ up to 2.8$R_{\rm \odot}$ with a step of 0.1$R_{\rm \odot}$, and for each $R_{\rm i}$ we varied $R_{\rm ss}$ from 1.4$R_{\rm \odot}$ to 3.2$R_{\rm \odot}$ with the same step. ADAPT magnetograms, at date--time same as that of the EUV images analysed, were used as input to EUHFORIA which then produced maps of the open and closed magnetic regions for the specified input heights $R_{\rm i}$ and $R_{\rm ss}$.

To create these open - closed flux maps the solar surface is divided in a regular mesh consisting of pixels covering 2x2 degrees per pixel. For each pixel the procedure assigns a field line and traces it from the solar surface upwards towards the source surface. If the tracked field line is curved below the source surface and can be traced back down to the solar surface it is assigned as a closed one. Per contra, if the traced field line pierces through the source surface it is accredited as an open one. Examples of open - closed field areas are given in Figure \ref{default_height_maps}, where dark blue and red colours represent areas of open flux, with positive and negative polarity respectively, while light blue and orange areas are correspondingly closed flux areas. The light green contours are the optimal CH boundaries derived from EUV observations using \textsc{catch}, as described in the previous section. The results presented in this figure are the output of the model run for the default pair of heights, [2.3, 2.6]$R_{\rm \odot}$. To better visualise the modelled CH area and the over-plotted EUV based boundaries, especially in the cases of very small CHs, we only plot in Figure \ref{default_height_maps} the front side of the Sun (Earth view). In addition, for CHs that do not extend to polar regions, high latitudes of northern and southern hemisphere are excluded from the images.

\section{Results}
\label{subsec:results}

\subsection{CH reconstruction using the default model boundary heights}
\label{subsec:results1}

\begin{figure}[b!]
    \centering
    \begin{minipage}{0.32\textwidth}
        \centering
        \begin{overpic}[width = 1\textwidth,clip]{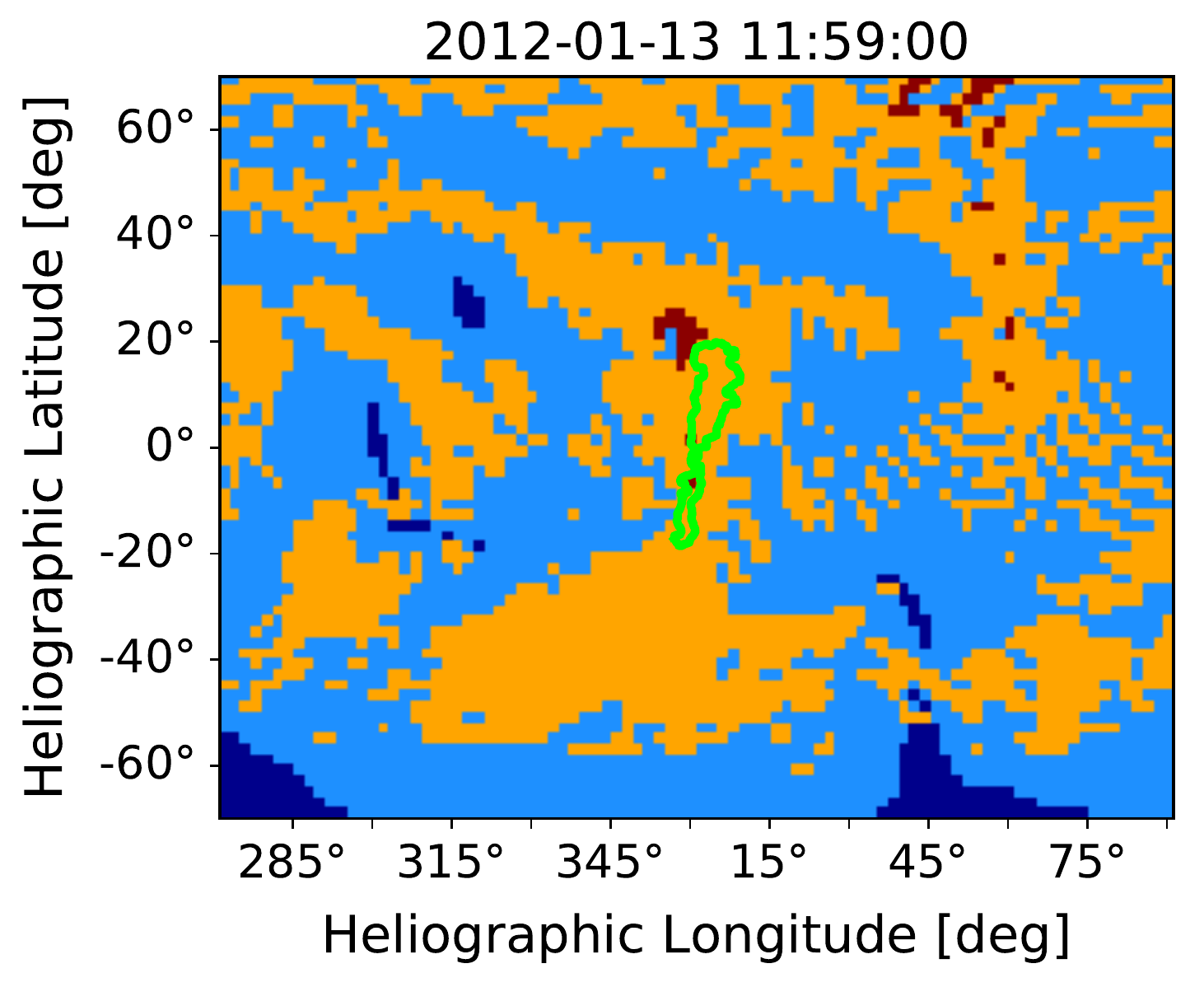}
        \end{overpic}
        \begin{overpic}[width = 1\textwidth]{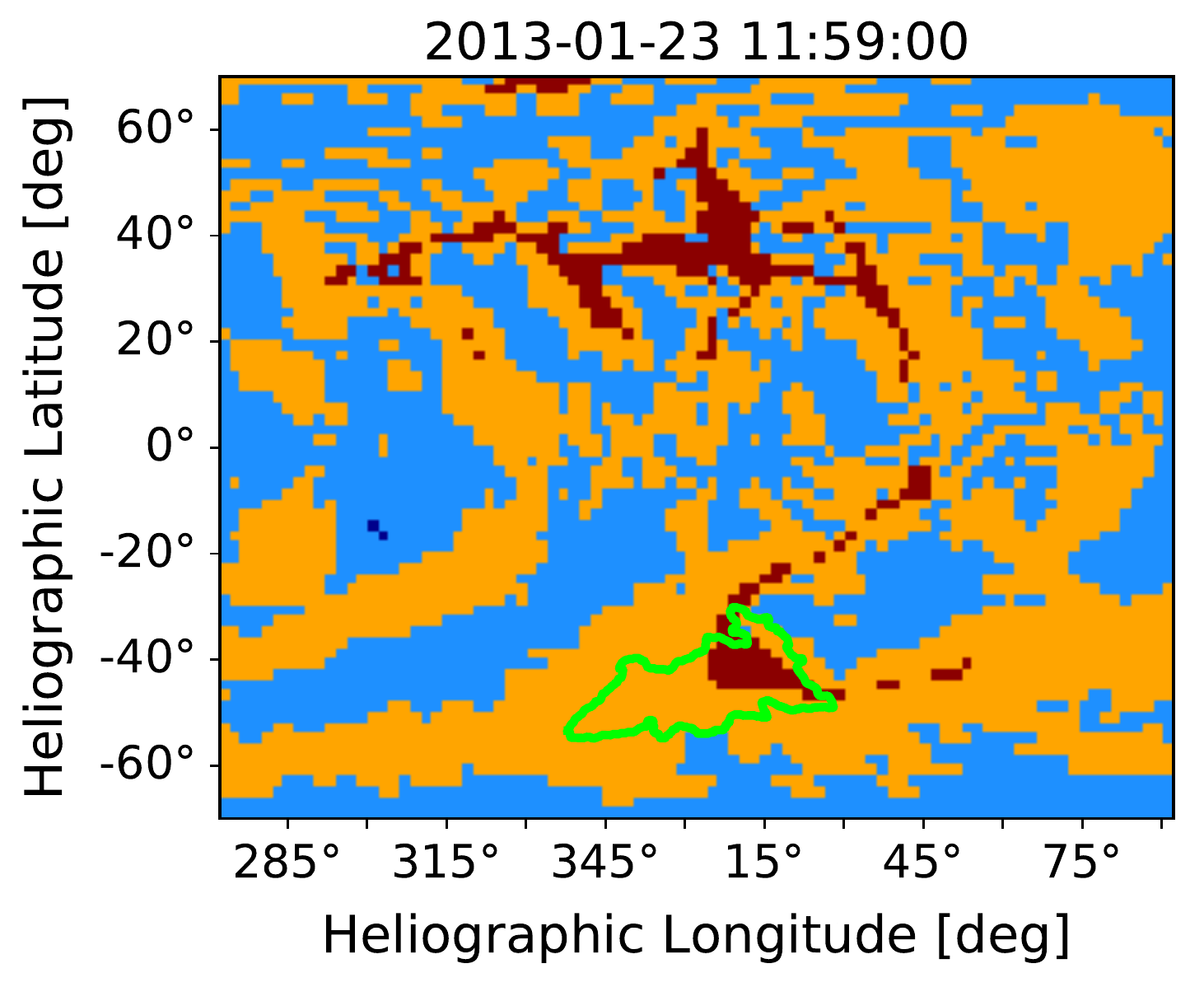}
        \end{overpic}
        \begin{overpic}[width = 1\textwidth]{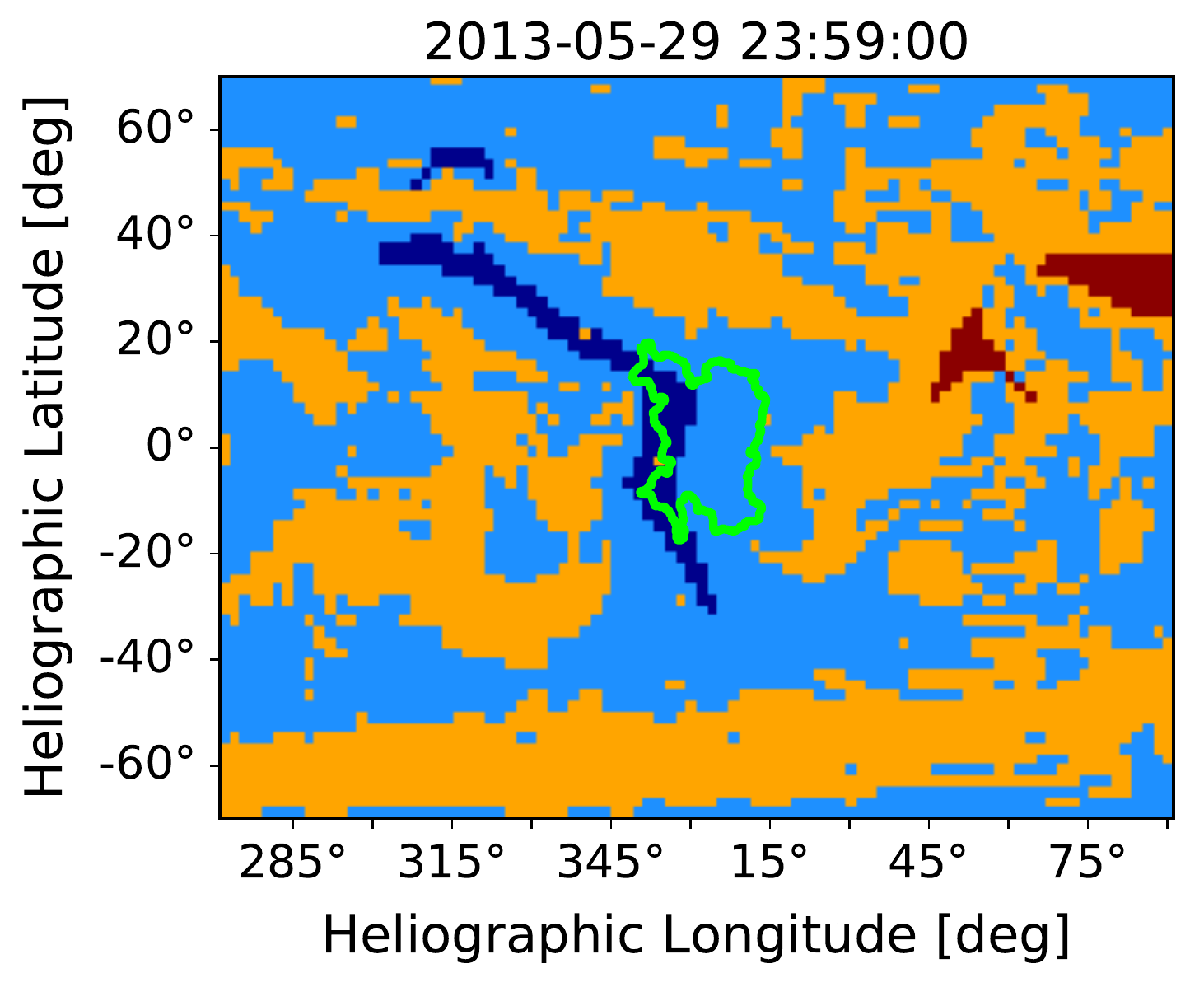}
        \end{overpic}
        \begin{overpic}[width = 1\textwidth]{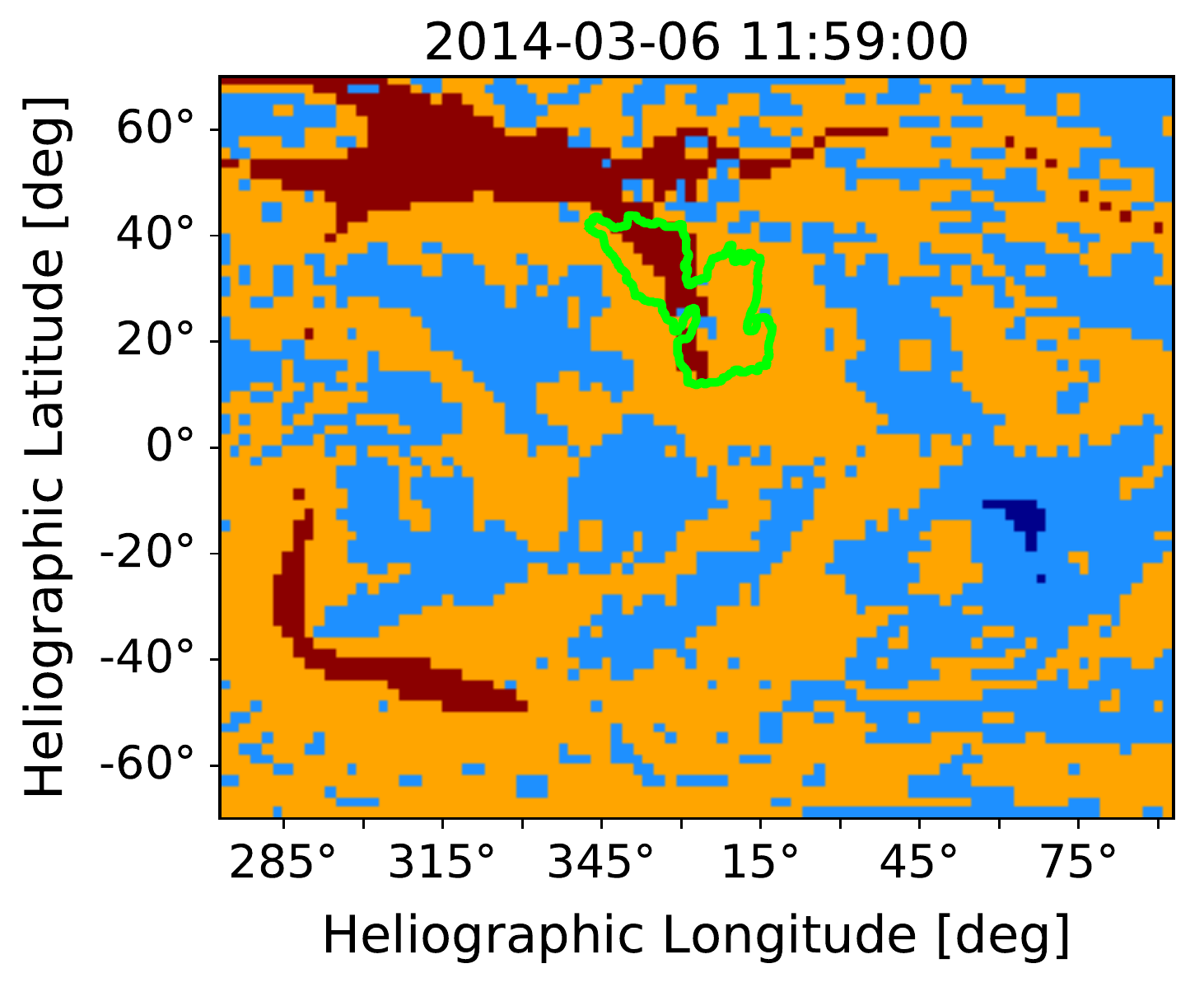}
        \end{overpic}
        \begin{overpic}[width = 1\textwidth]{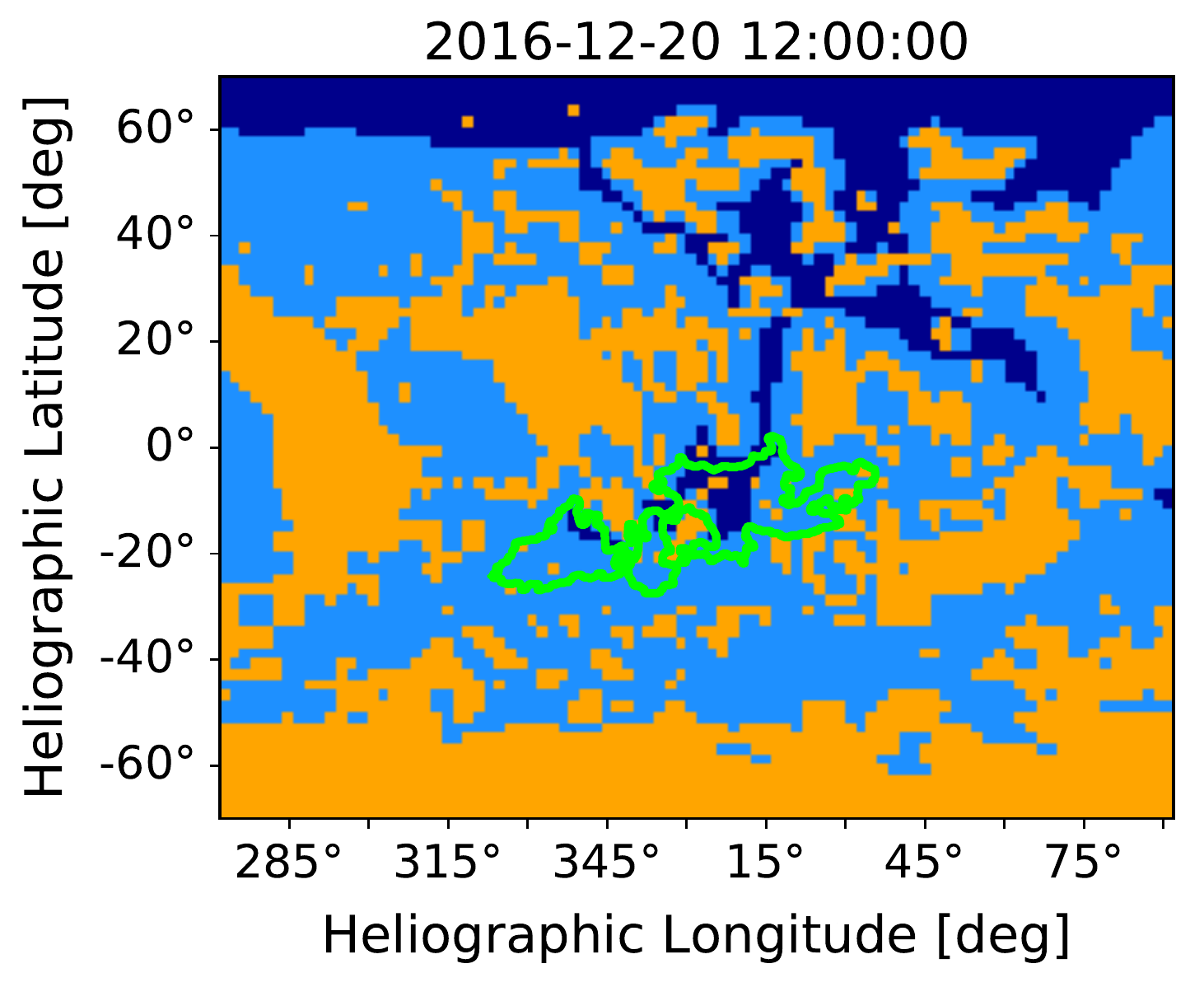}
        \end{overpic}
    \end{minipage}
    \begin{minipage}{0.32\textwidth}
        \centering
        \begin{overpic}[width = 1\textwidth]{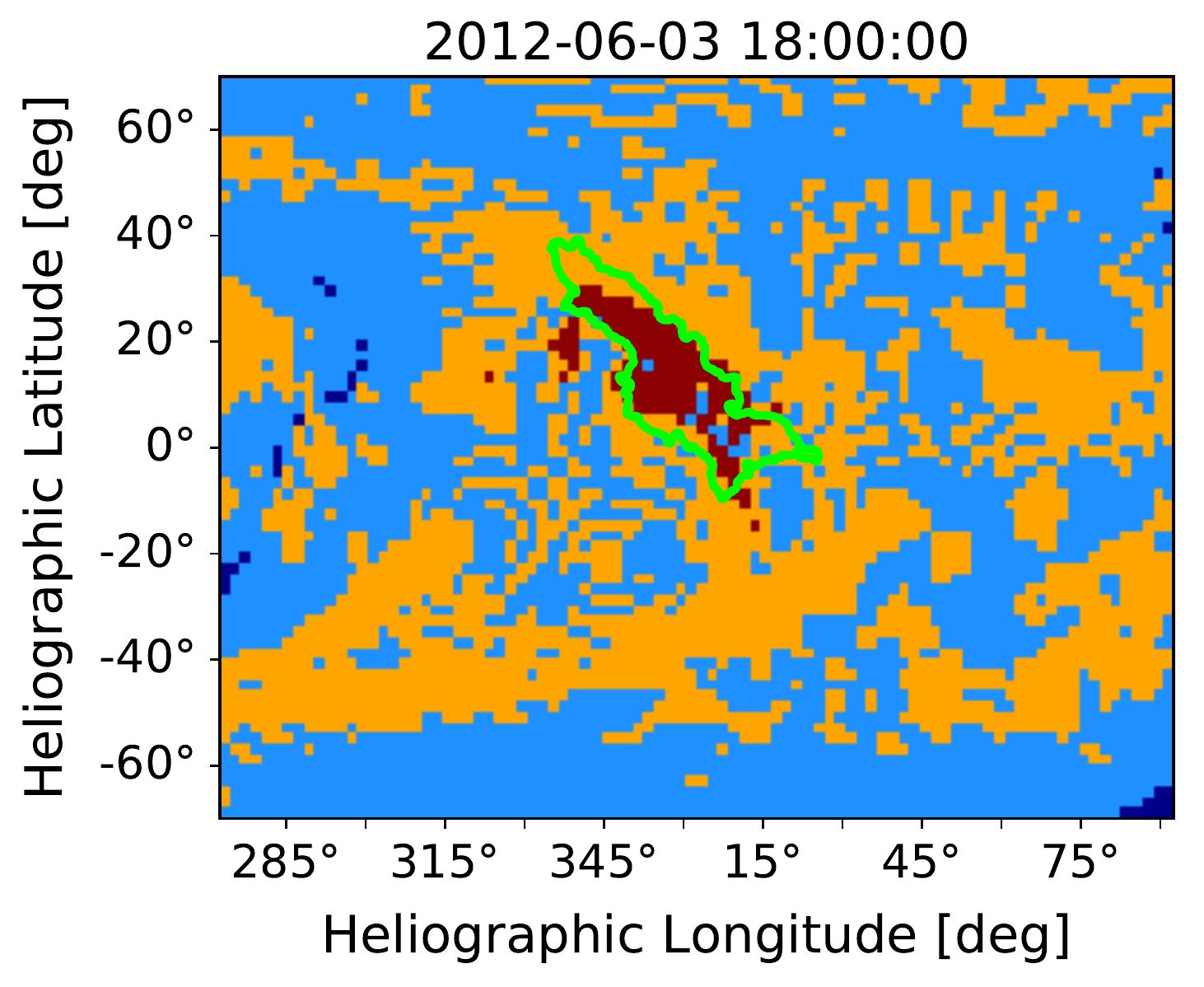}
        \end{overpic}
        \begin{overpic}[width = 1\textwidth]{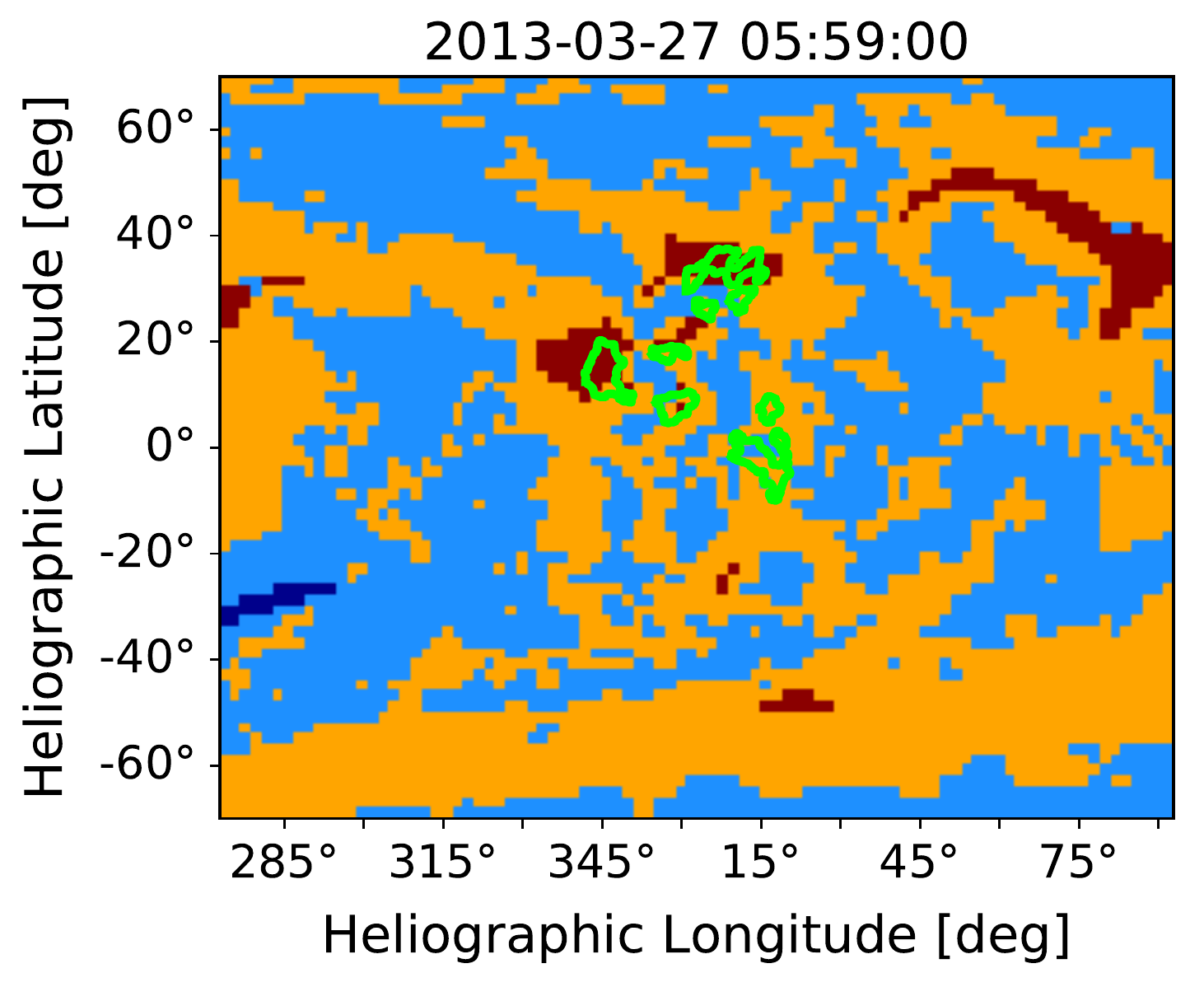}
        \end{overpic}
        \begin{overpic}[width = 1\textwidth]{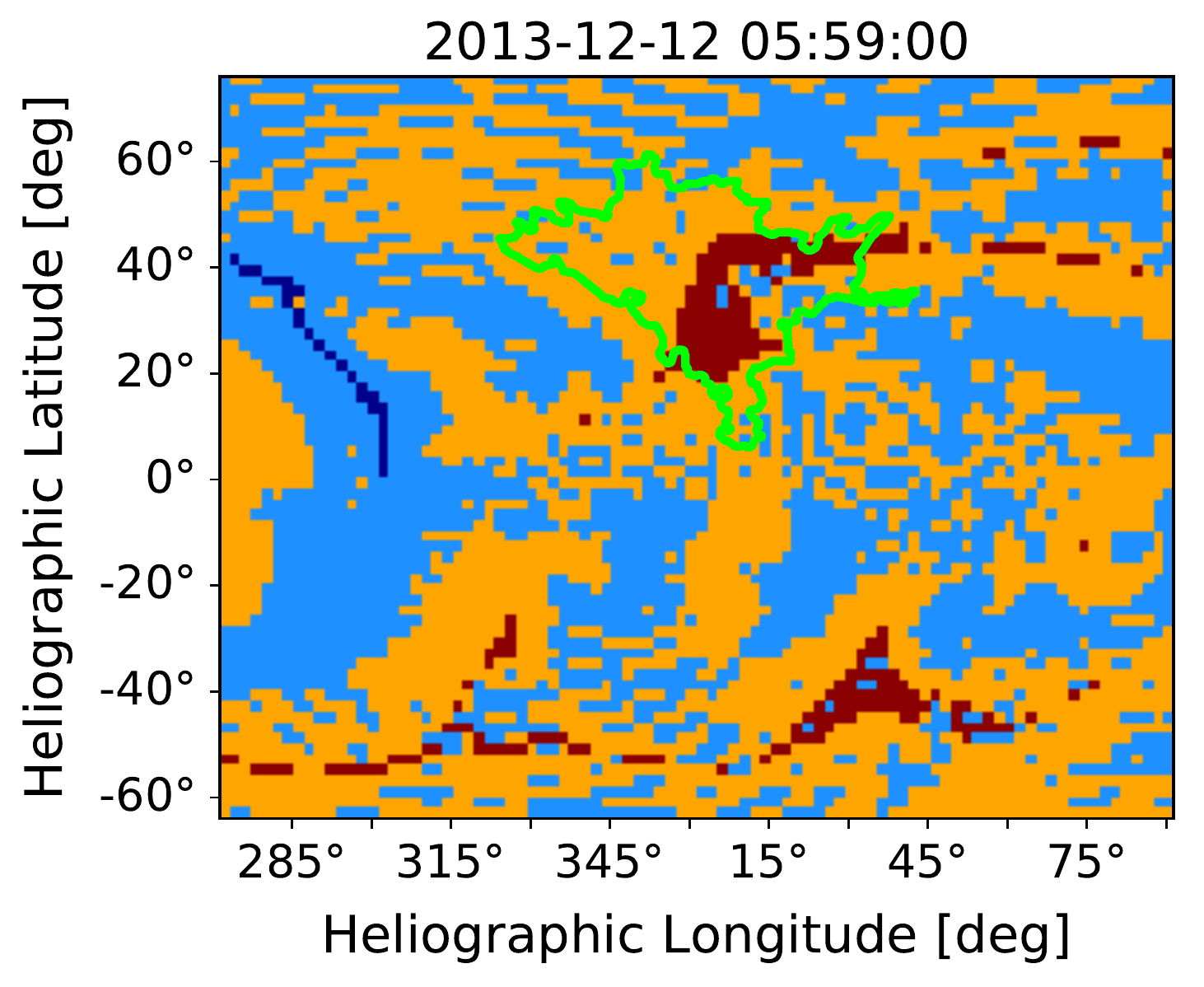}
        \end{overpic}
        \begin{overpic}[width = 1\textwidth]{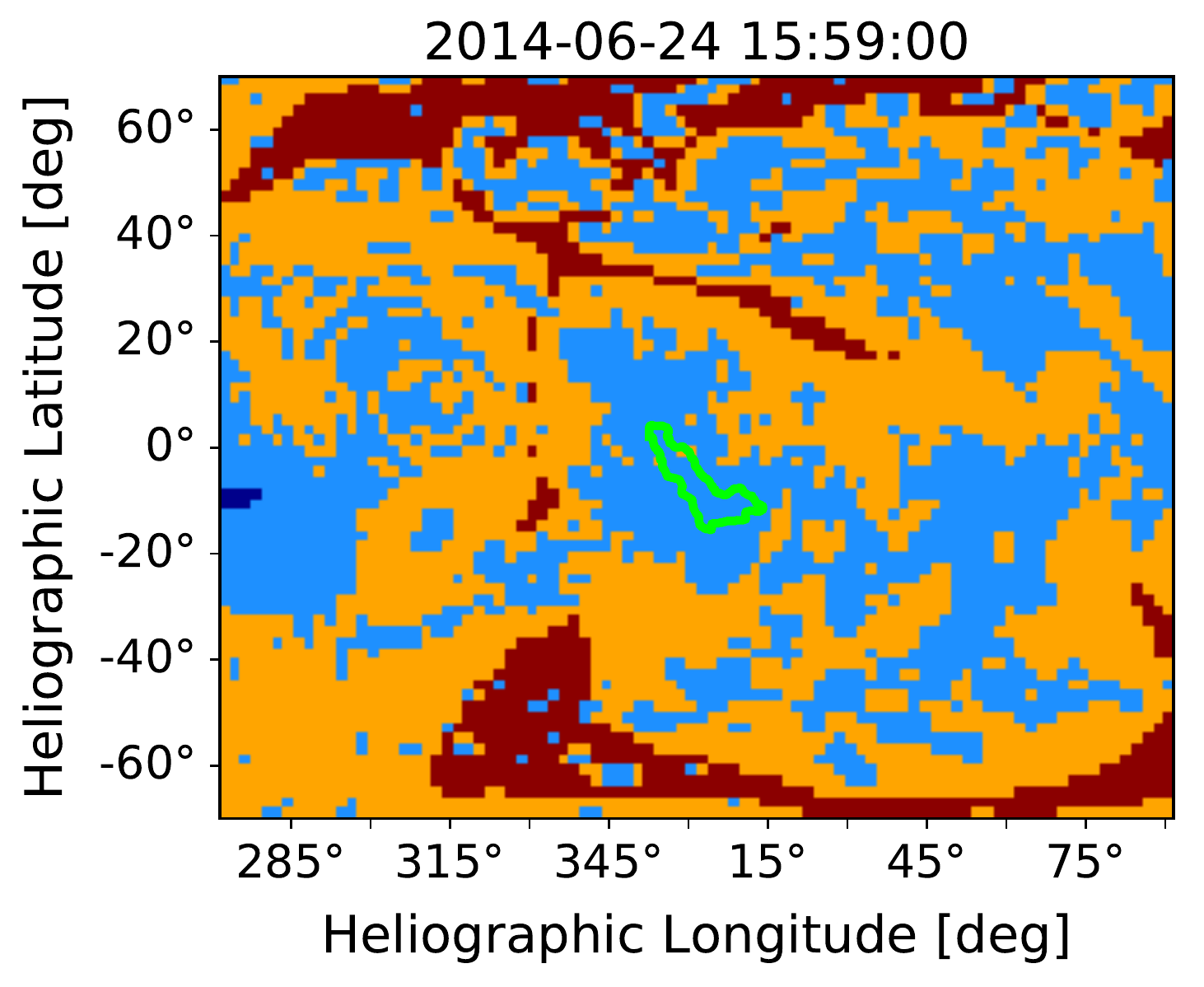}
        \end{overpic}
        \begin{overpic}[width = 1\textwidth]{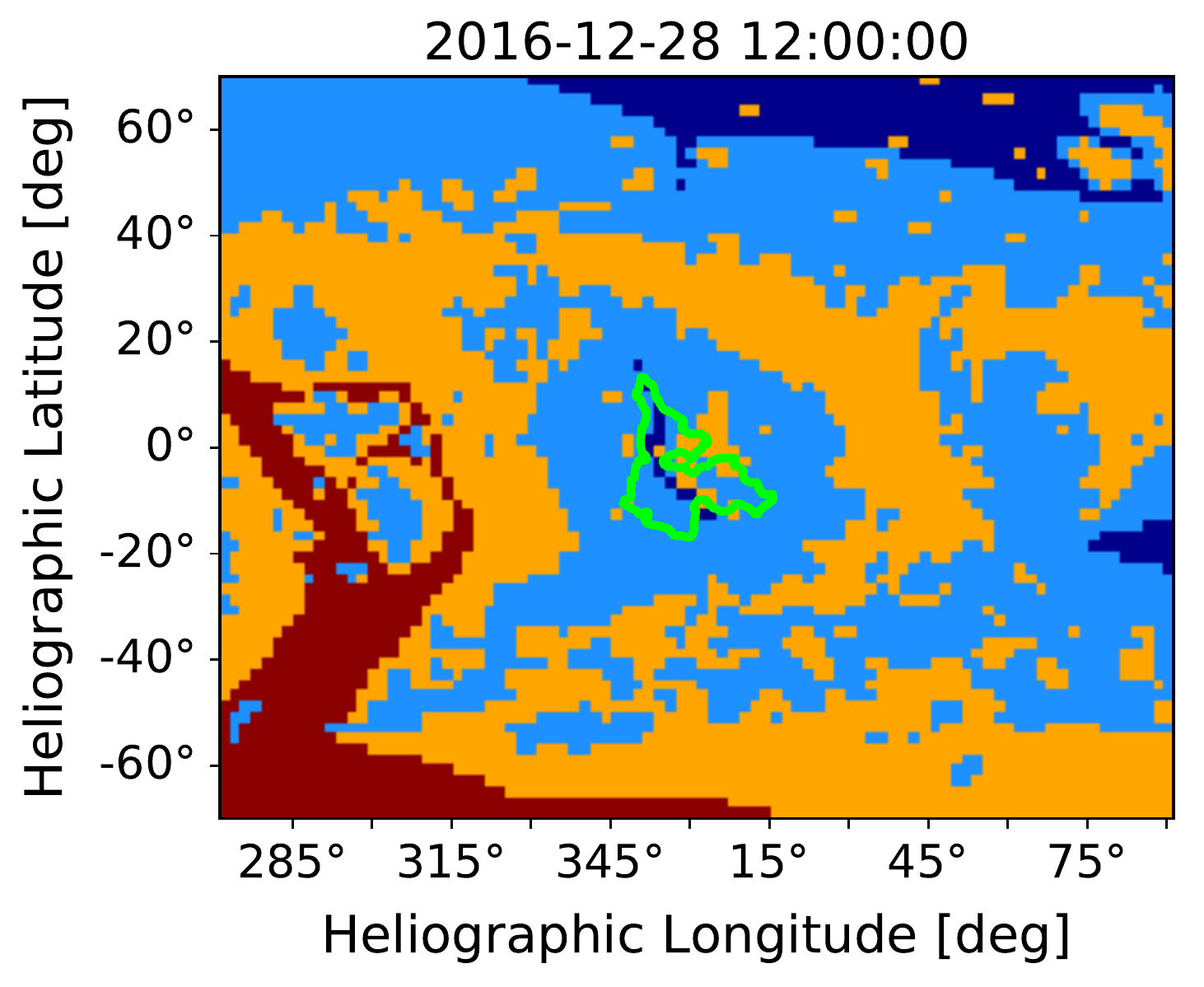}
        \end{overpic}
    \end{minipage}
    \begin{minipage}{0.32\textwidth}
        \centering
        \begin{overpic}[width = 1\textwidth]{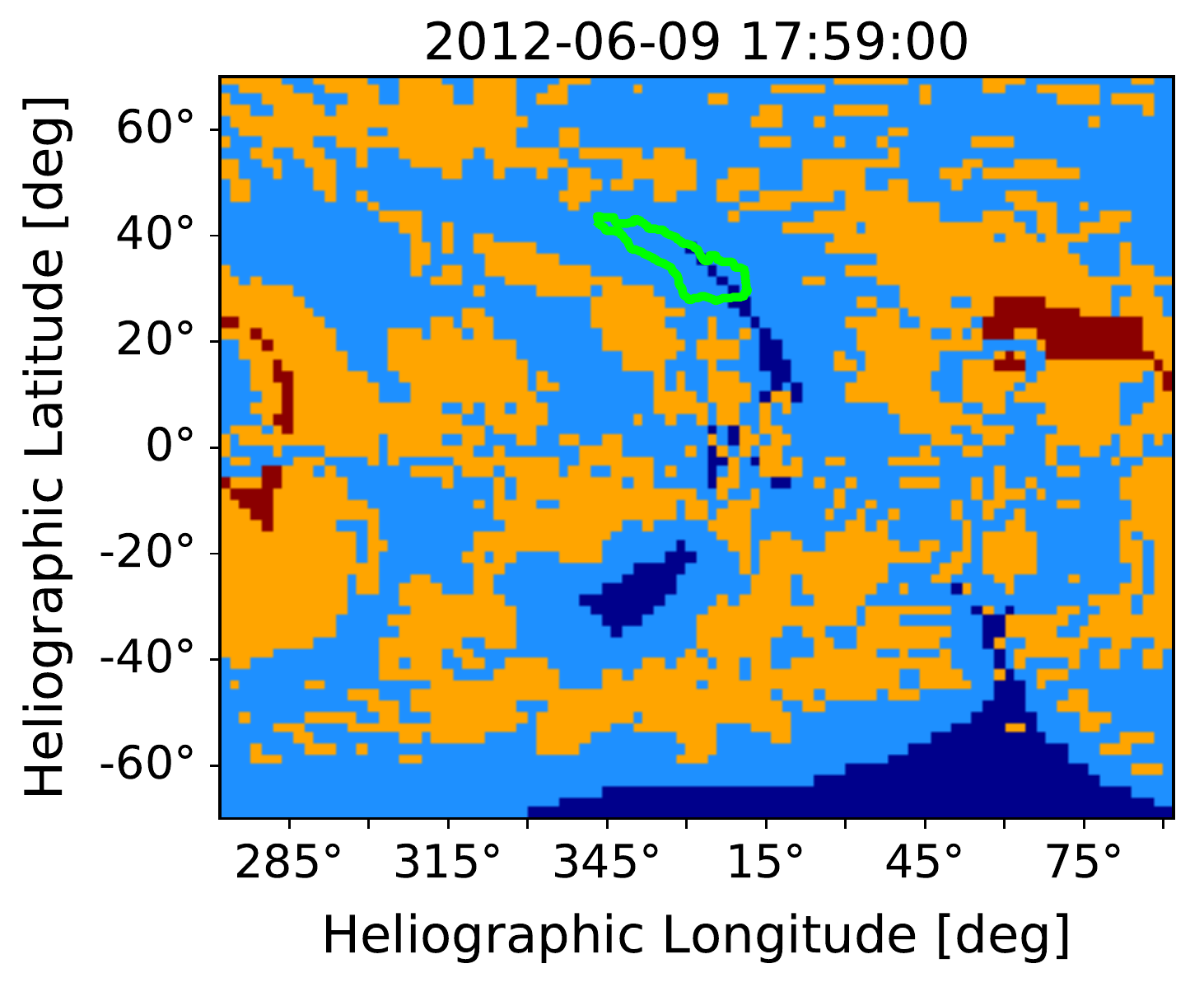}
        \end{overpic}
        \begin{overpic}[width = 1\textwidth]{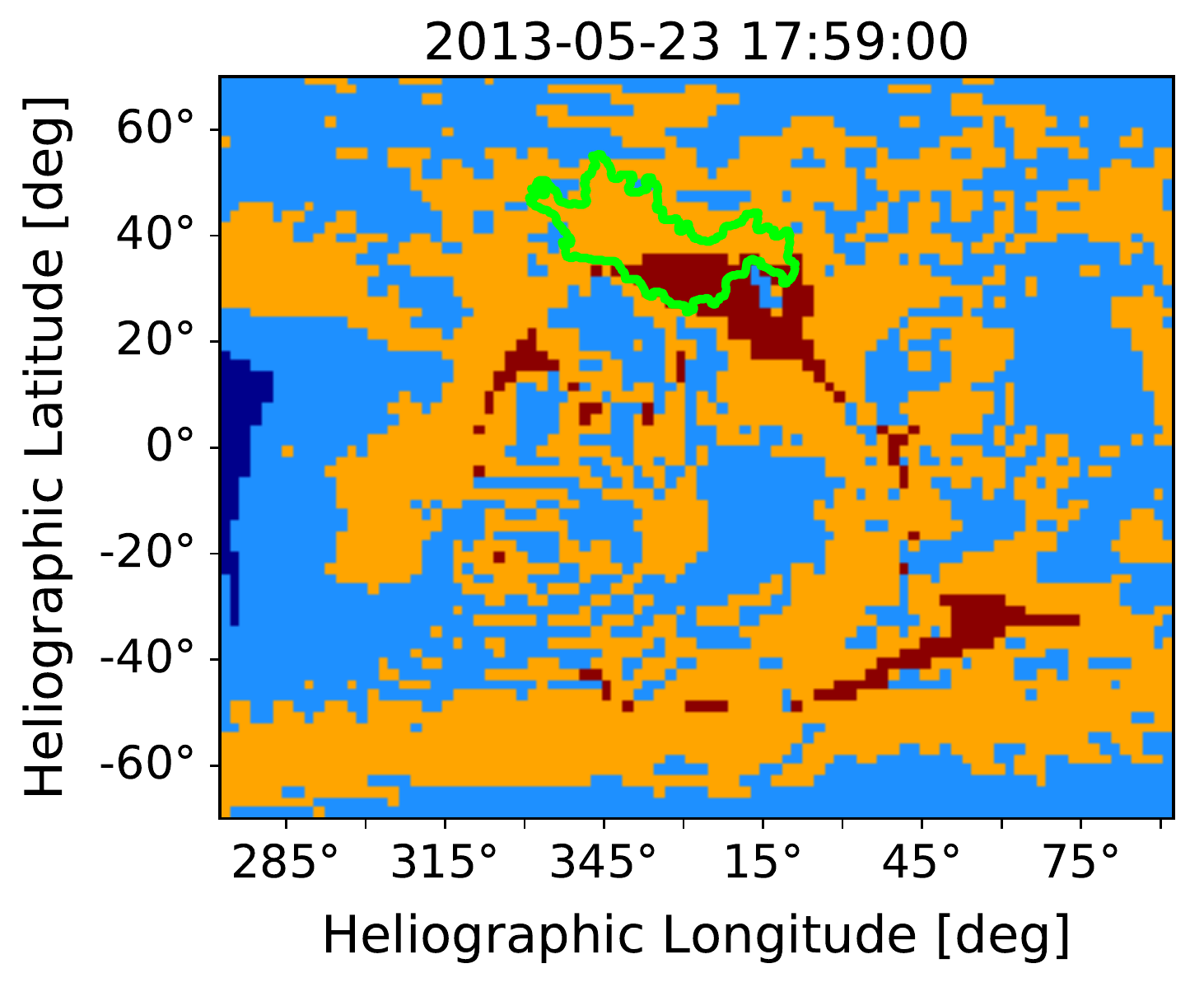}
        \end{overpic}
        \begin{overpic}[width = 1\textwidth]{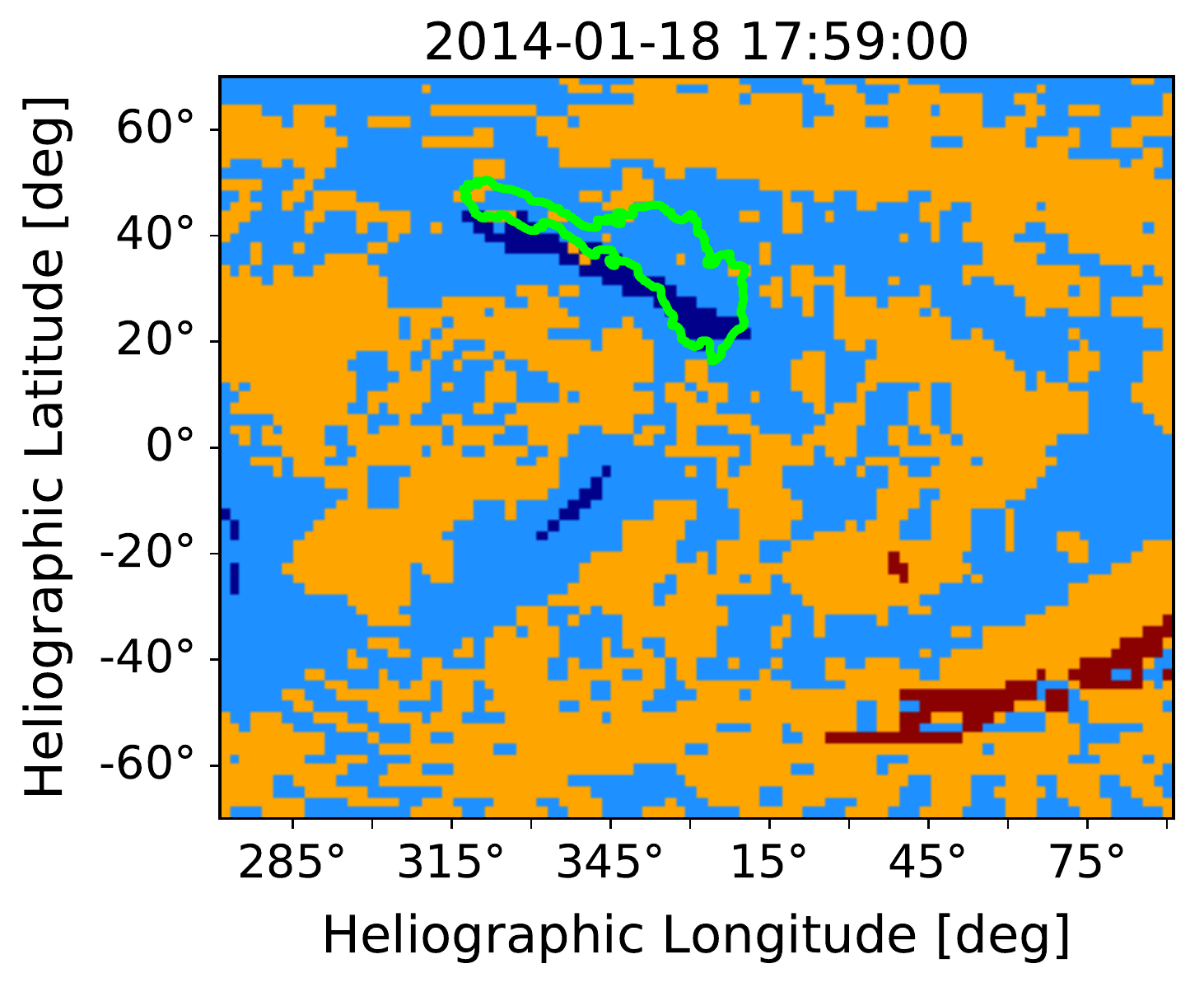}
        \end{overpic}
        \begin{overpic}[width = 1\textwidth]{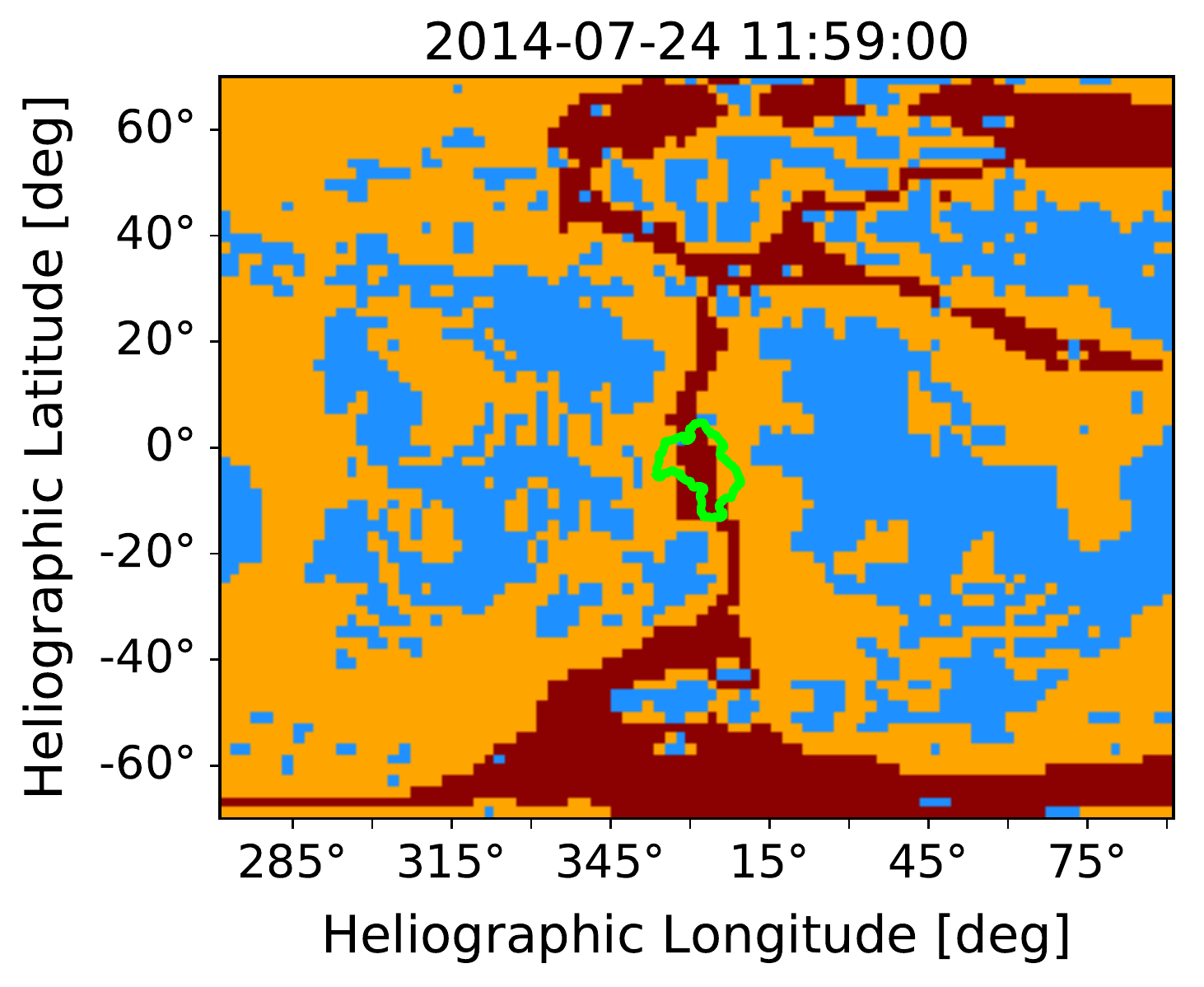}
        \end{overpic}
        \begin{overpic}[width = 1\textwidth]{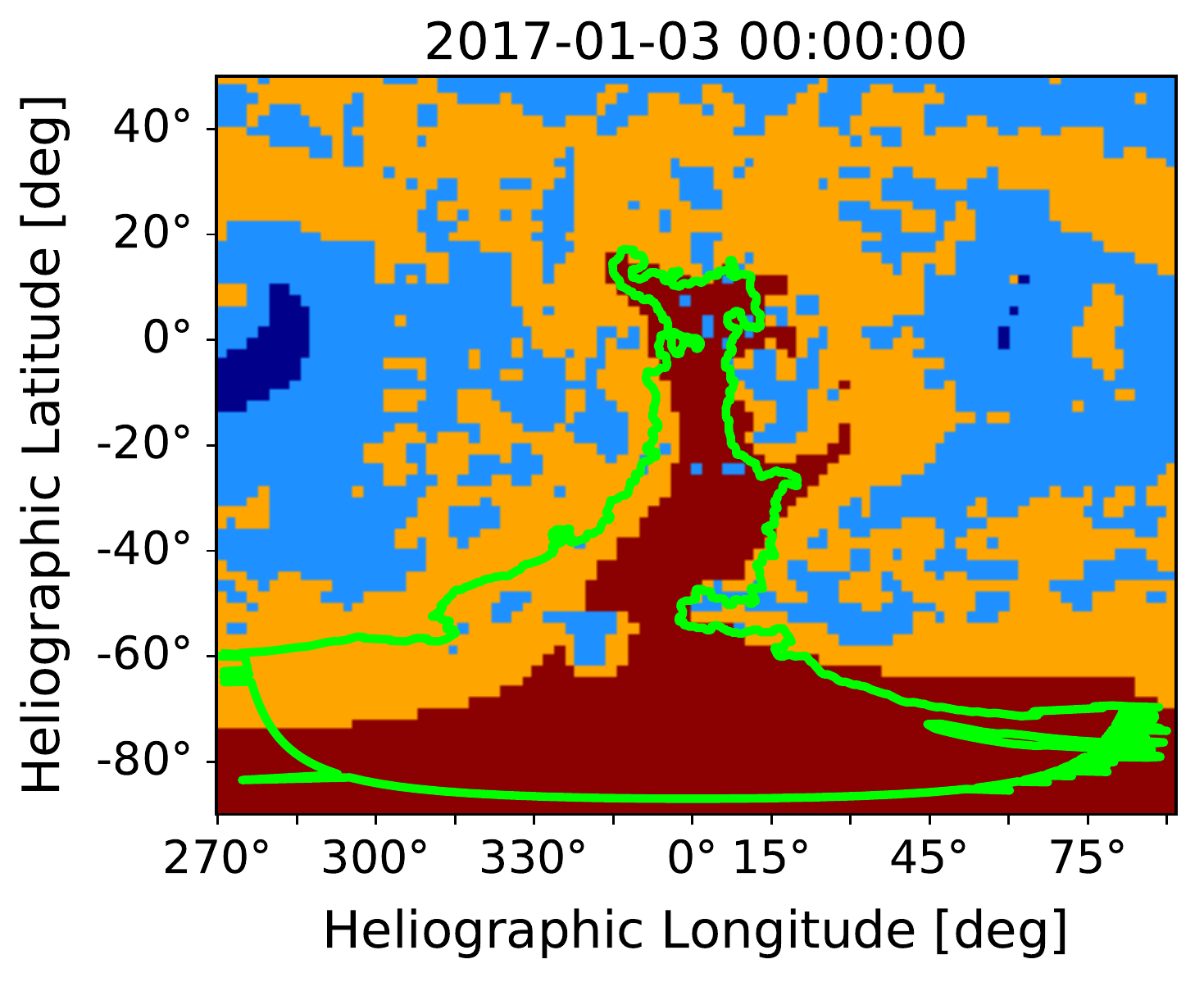}
        \end{overpic}
    \end{minipage}
    \caption{Model generated maps of open - closed field regions based on magnetic field reconstructions by the adopted WSA model in EUHFORIA. Dark blue and red regions represent open field areas of positive and negative polarity respectively, while cyan and orange are accordingly positive and negative closed field areas. The green outlines overplotted on the maps are the EUV defined optimal boundaries of the CHs as extracted with \textsc{catch}. Although, the model generates the map for the entire solar surface we plotted here a fraction of the map aiming to improve the clarity of the CH areas and the overplotted boundaries.}
 \label{default_height_maps}
 \end{figure}

The examples in Figure \ref{default_height_maps} display maps based on the default pairs of heights [2.3, 2.6]$R_{\rm \odot}$ used as default values in EUHFORIA. These values are the ones concluded in \citeA{mcgregor_analysis_2008}. It is clear from this figure that, for the majority of the CHs, the model runs based on the default heights [2.3, 2.6]$R_{\rm \odot}$ fail in reconstructing well the area and geometry as compared to the EUV extraction. Not only are the CHs modelled thinner and overall smaller, but also for some cases (i.e. 2012-06-09, 2013-05-23, 2014-01-18) the CHs appear to be shifted. For the CH of 2014-06-24 the model is unsuccessful in modelling open flux both within the expected boundaries and in the surrounding area. This CH seems to be invisible for the PFSS+SCS models in the adopted WSA model. A CH that appears to be well modelled by EUHFORIA is the one on 2017-01-03, which is a southern polar CH with a large extension that reaches to low northern hemisphere latitudes. It is worth pointing out, however, that the CH boundaries extracted with \textsc{catch} are only for the part of the CH that is visible from Earth's line-of-sight. This explains why also in the EUHFORIA output open/closed field map we do not focus on longitudes beyond $\pm$90 degrees.

\begin{figure}[h]
    \centering
    \begin{minipage}{0.52\textwidth}
        \centering
        \begin{overpic}[width = 1\textwidth]{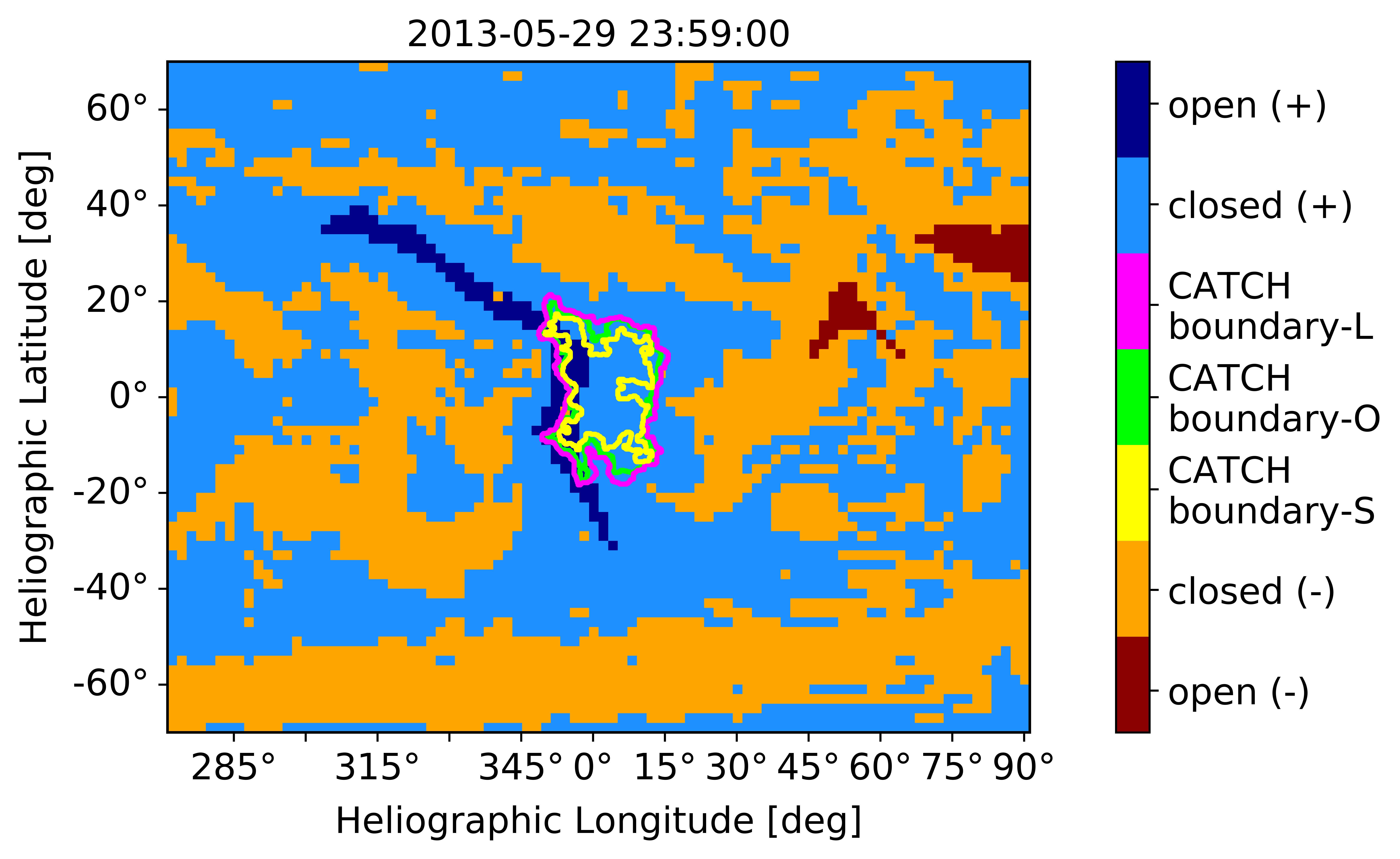}
        \end{overpic}
    \end{minipage}%
    \hfil
    \begin{minipage}{0.45\textwidth}
        \centering
        \begin{overpic}[width = 1\textwidth]{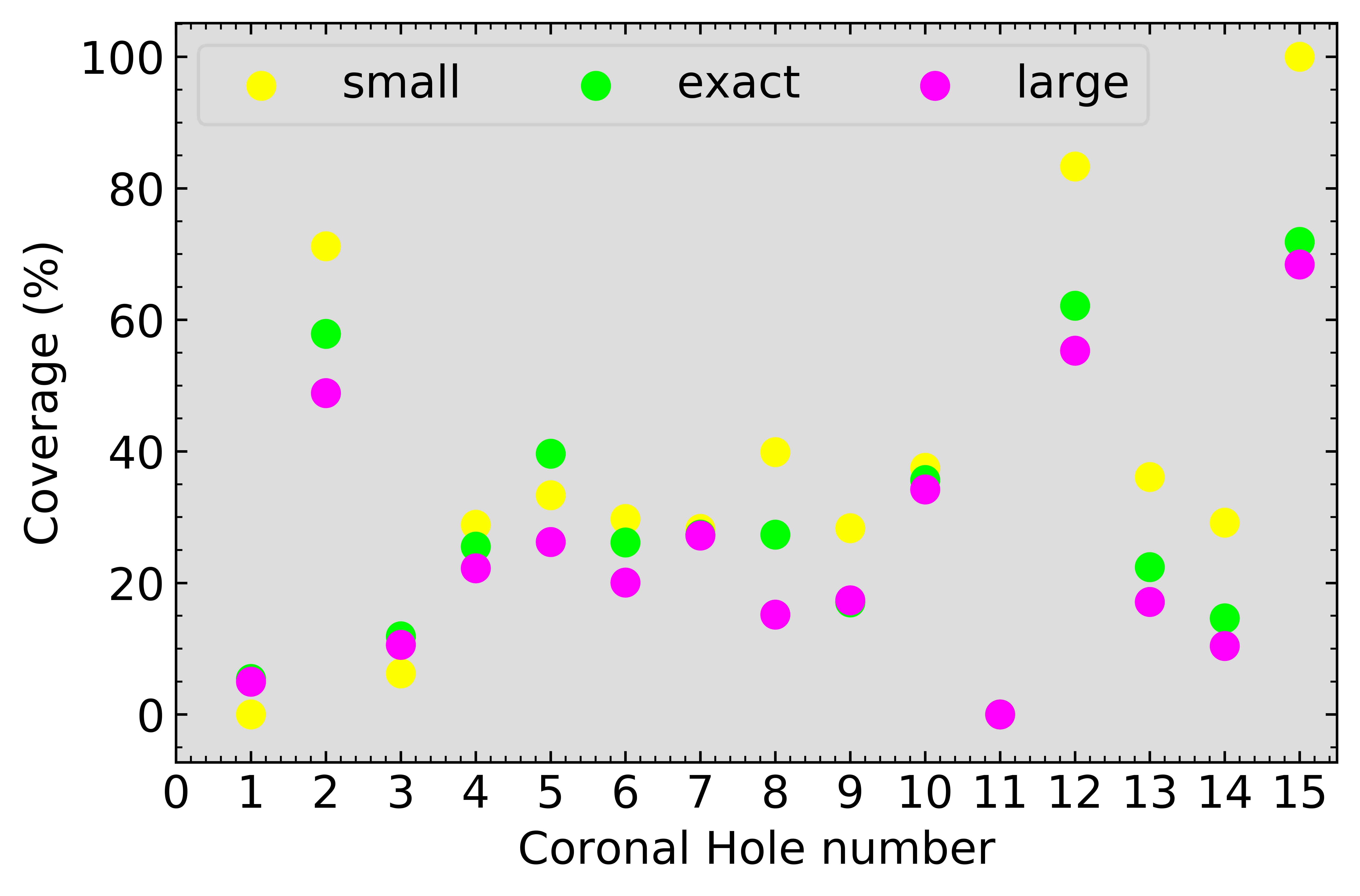}
        \end{overpic}
    \end{minipage}
    \caption{The left panel is an example (CH No7) of how the different boundaries extracted with \textsc{catch} can differ from each other. The colour scheme for open - closed flux of positive and negative polarity is similar to that in Figure \ref{default_height_maps}. With green colour we illustrate the optimal CH boundary extracted with \textsc{catch}, while magenta and yellow coloured boundaries represent the ones obtained using an under- and overestimated thresholds, as described in section \ref{subsec:CATCH_tool}. The right panel shows the coverage achieved by the model when run for the default pair of heights, [2.3, 2.6]$R_{\rm \odot}$, and when the three different boundaries (optimal, smaller and larger) are considered for the analysis. The face colours of the circles represent the colour of the boundary based on which the calculations were carried out.}
 \label{sel}
\end{figure}

In order to quantify how successful the model is in reconstructing coronal hole areas we define the coverage parameter, $cov$, which is given by:
\begin{linenomath*}
\begin{equation}
cov = \frac{N_{open}}{N_{total}}*100 [\%]
\end{equation}
\end{linenomath*}
where $N_{open}$ is the number of open flux pixels contained within the optimal \textsc{catch} boundaries (green contour), and $N_{total}$ is the total number of pixels enclosed by that same boundary (open and closed flux pixels). Low coverage indicates not only that the area size of the CH is not correctly modelled but also that the CH could possibly be modelled shifted in space. High coverage suggests that both area size and location are possibly well captured by the model reconstruction of the CH. The coverage for the model results using the default source surface and SCS inner boundary heights, [2.3, 2.6]$R_{\rm \odot}$, is given in the right panel of Figure \ref{sel} as green coloured circles. The x-axis is the nominal number of the CHs when they are accounted in chronological order (column 1 of table \ref{CH_characteristics}). It can be seen that for 13 out of the 15 CHs we studied the model gives a coverage below 60\%, with the majority of CHs ranking below 30\% coverage. This is a clear indication that the model under the default setup does not properly model the CH areas.

\clearpage

We also investigated the possibility of comparing the modelled CH areas to the ones defined by the smaller and larger area boundaries obtained from the EUV images using \textsc{catch}. An example of how these boundaries differ from each other is shown in the left panel of Figure \ref{sel}. In yellow we show the smaller area boundaries and in magenta the larger area boundaries. For most CHs the differences between the areas defined by the three different boundaries (small, optimal and large) is not that substantial. The coverage estimated for these is shown in corresponding colours in the right panel of the same figure. Although, smaller area boundaries, as expected, improve the coverage percentage for the majority of the CHs, for some this improvement is not significant (i.e. 2013-01-23, 2013-05-23, 2014-03-06, 2014-06-24). For three CHs (i.e. 2012-01-13, 2012-06-09, 2013-03-27) the coverage worsens with the smaller boundaries. An explanation to this is that open flux pixels are lying outside the smaller area boundaries. This concludes that there is no systematic improvement if smaller area boundaries were to be obtained from the EUV observations when a smaller threshold is considered. In a similar manner, larger area boundaries do not worsen the result for all CHs.

\subsection{CH characteristics and their effect}
\label{subsec:results2}

The open closed flux maps given in Figure \ref{default_height_maps} and the coverage shown in Figure \ref{sel} indicate that the model performs better for some CHs comparing to others when the model runs are setup using the default pair of heights. In order to exclude the possibility of the model showing preference in better modelling CHs bearing particular features we assess how the coverage relates to CH characteristics. The elongation, latitudinal position of CoM, patchiness, area size, mean intensity are apparent features of a CH that are of interest to this study. 

The elongation of a CH can have an effect to the modelled results due to the way the selected magnetograms are constructed. Even though dynamical processes are applied to them, synchronic ADAPT magnetograms best represent the magnetic field on the Sun along the central meridian as viewed from Earth. This suggests that CHs which are latitudinally elongated and lie within the central meridional zone can be possibly better modelled when selecting a magnetogram from the date the CH was located there. Thus a longitudinally elongated CH might not be as accurate modelled at its full length. So when studying a CH on a particular moment can also have an impact to the modelled output based on that perspective.

CH mean intensity and area are parameters extracted using \textsc{catch}, as described in the section \ref{subsec:CATCH_tool}. The mean intensity is automatically computed with \textsc{catch} based on the threshold used for extracting the CH boundaries. It is expected that active regions in the vicinity of a CH will have an impact on its configuration, which is also imprinted on magnetograms, and thus has the potential to affect the modelled CH areas. The sample's average mean intensity is 31.3 DN, and only 6 out of the 15 CHs studied have mean intensity above that value. The size of a CH is expected to affect the width/duration of a high speed stream as well as its speed and open flux, and thus is a parameter we investigate \cite{nolte_coronal_1976,vrsnak_coronal_2007, rotter_relation_2012, hofmeister_dependence_2018}. In terms of the area the CH sample consists of both small and large CHs, with the biggest on 2017-01-03 extending from polar to solar equatorial latitudes. The sample mean is $8.7x\times10^{10} ~km^2$ and 4 CHs have area above this value.

\begin{figure}[h]
    \centering
    \begin{minipage}{0.33\textwidth}
        \centering
        \begin{overpic}[width = 1\textwidth]{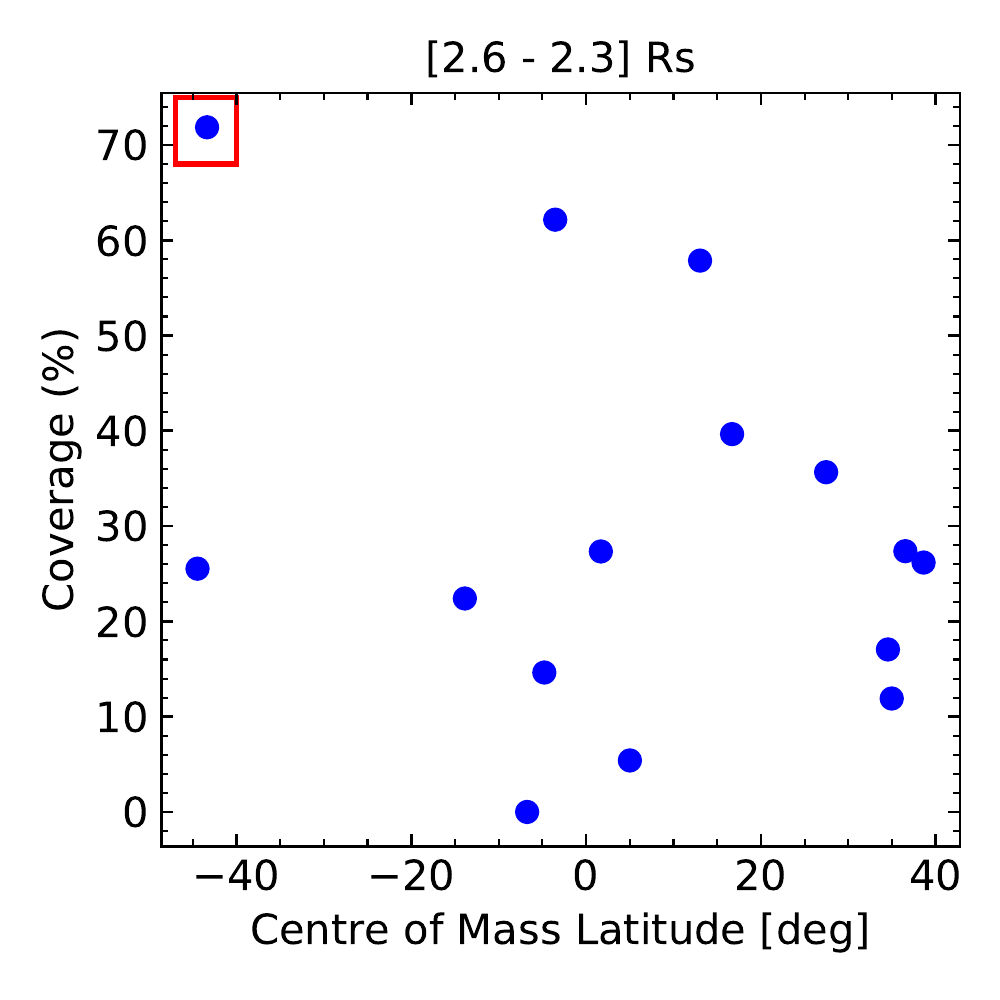}
        \end{overpic}
    \end{minipage}%
    \hfil
    \begin{minipage}{0.33\textwidth}
        \centering
        \begin{overpic}[width = 1\textwidth]{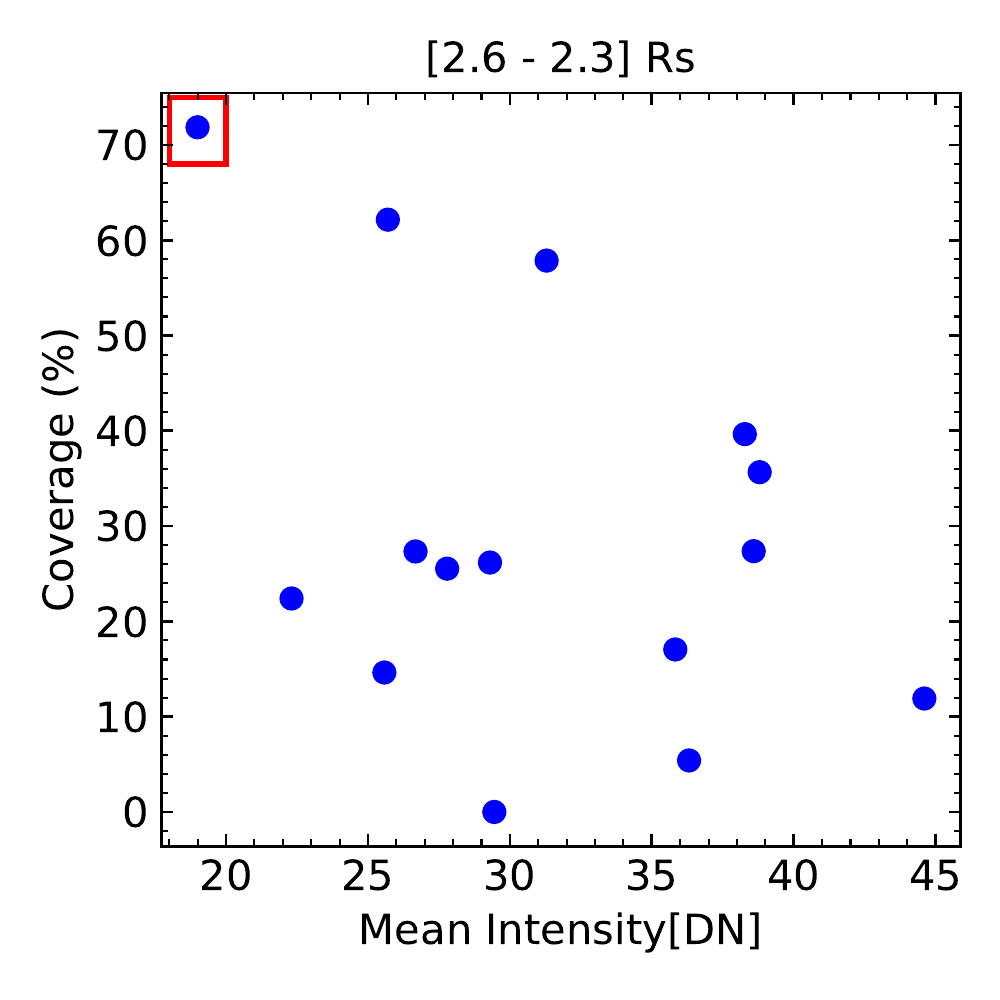}
        \end{overpic}
    \end{minipage}%
    \hfil
    \begin{minipage}{0.33\textwidth}
        \centering
        \begin{overpic}[width = 1\textwidth]{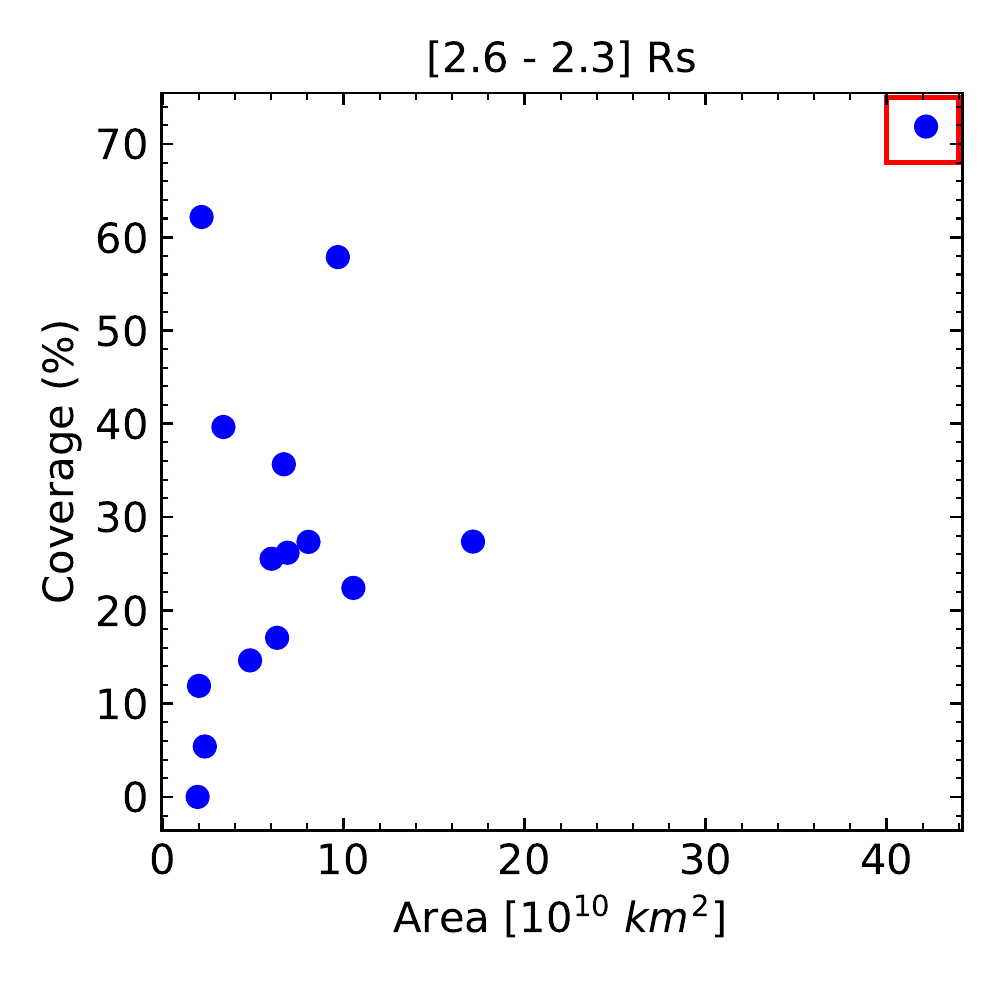}
        \end{overpic}
    \end{minipage}
    \caption{Scatter plots showing the impact on the coverage that the CH characteristics such as the centre of mass latitude, the mean intensity and the CH area have (left, middle and right panel respectively). The presented results are based on running the model for the default pair of heights [2.3, 2.6]$R_{\rm \odot}$ and indicate that neither CoM latitude, mean intensity or area had any effect in the performance of the model results. The red box encircles the south polar CH with the long extension to equatorial latitudes.}
 \label{CoverageVsAreaetc}
\end{figure}

In Figure \ref{CoverageVsAreaetc}, the left panel presents the coverage with respect to the CoM latitude. From the level of scatter in the points it is clear that the latitudinal position of the CH does not affect the performance of the modelled result. The same applies for the other two parameters shown in the middle panel (Mean Intensity) and the right most panel (Area). For the case of large CHs we have only one point so the relation between coverage and area remains inconclusive. We also checked whether the visual characteristics of elongation and patchiness of a CH have any effect. No such conclusion could be made; however, the number of CHs is too low for the result to be compelling \cite<for details see>[in preparation]{asvestari_parameters_2019}. We also accounted for the possibility of a coupling of different parameters, e.g. area or intensity and CoM latitude, which could have an effect in the modelled results. But even for this scenario no trend is apparent with respect to the coverage. One conclusion that could be made is that all CHs located in the south hemisphere of the Sun have low Mean Intensity, while those in the northern have high. This can possibly be a solar cycle related effect, since we studied CHs from one SC only (SC24), and indeed, from 2013-2015 southern hemisphere clearly dominated both in the number and in the area size of sunspots \cite{li_present_2019} (see also sunspot number information at \url{http://www.sidc.be/silso/ssngraphics}). Our sample size is rather small for an unambiguous conclusion to be made, so a more detailed analysis of a larger CH sample is necessary.

\subsection{Assessing for systematic shifting and solar cycle trends}
\label{subsec:results3}

Maps of open - closed flux indicated the possibility of shifting of the modelled CH areas with respect to the expected location. To assess this we investigated the likelihood of a systematic shifting. All directions (eastward/westward and northward/southward) were investigated. In addition, we considered both a 2 and a 4 degree shift in each direction. From this analysis no systematic effect could be identified. The role of CH characteristics, such as mean intensity, elongation, size, CoM latitude, and hemispheric position, in shifting effects was evaluated independently \cite[in preparation]{asvestari_parameters_2019}. Regardless of whether the sample is assessed as a whole or divided in groups based on the CH characteristics the conclusion for possible shifting remained negative for all cases.

In earlier studies \cite{lee_coronal_2011, arden_breathing_2014} the concept of a solar cycle dependent source surface height has been suggested, i.e. varying within the course of the solar cycle, as well as, from one solar cycle to another. We investigated this prospect at first for the whole CH sample and then by separating CHs to groups based on their apparent characteristics mentioned also in the shifting investigation. We found no solar cycle trends when assessing the result for the default heights, but also for our full set of [$R_{\rm i}$, $R_{\rm ss}$] heights and discussed below \cite<for details see>[in preparation]{asvestari_parameters_2019}. We note that the number of considered CHs is rather small and therefore, no conclusive results can be drawn about the systematic shift. In addition, the period over which the sample extends does not cover a complete solar cycle, so to understand solar cycle effects on the modelling output it is important for such a study to consider a sample that spans over the entire solar cycle.

\begin{figure}[h]
    \centering
    \begin{minipage}{0.43\textwidth}
        \centering
        \begin{overpic}[width = 1\textwidth,clip]{20140624_default_heights_148.pdf}
        \end{overpic}
    \end{minipage}%
    \hfil
    \begin{minipage}{0.43\textwidth}
        \centering
        \begin{overpic}[width = 1\textwidth]{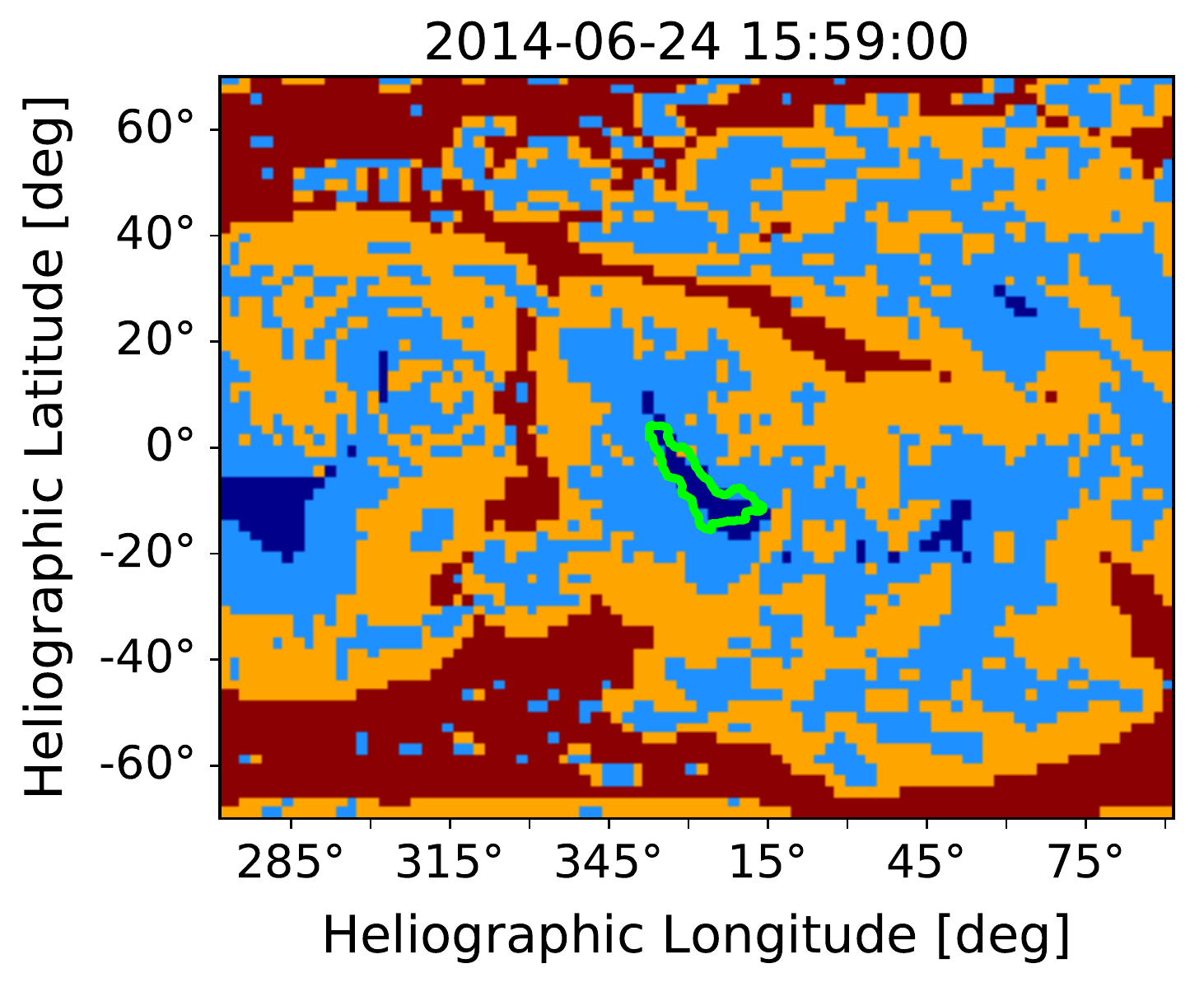}
        \end{overpic}
    \end{minipage}
    \caption{Maps of open-closed flux generated for the same CH by two different model runs. The one on the left resulted by EUHFORIA running for the default pair of heights [2.3, 2.6]$R_{\rm \odot}$, while the one on the right for the pair [1.3, 1.8]$R_{\rm \odot}$. It is clear that the CH is not present in the model result on the left, but is present and well captured by the run setup based on lower heights. This example highlights the impact the heights of the source surface and the inner boundary of the SCS model have in the modelling result.}
 \label{DefaultVsBetter}
\end{figure}

\subsection{Finding the optimum paired values for [$R_{\rm i}$, $R_{\rm ss}$]}
\label{subsec:results4}

After running the model for all 184 pairs of [$R_{\rm i}$, $R_{\rm ss}$] heights, it was made evident that lowering the source surface height significantly improved the modelled CHs. One very clear example is that of 2014-06-24, shown in Figure \ref{DefaultVsBetter}. This particular CH was invisible in the open - closed flux maps created with the boundaries placed at the default pair of heights. As can be seen from the right panel in Figure \ref{DefaultVsBetter} the model nicely maps the CH area when a lower pair of heights is considered. We specifically compare the default heights output to this one because the source surface height of 1.8$R_{\rm \odot}$ has been suggested by other studies as a better choice for particular periods and has been used in earlier studies for comparison purposes \cite<e.g.,>[]{lee_coronal_2011}.

\begin{figure}[h!]
    \centering
    \includegraphics[width = 1\textwidth]{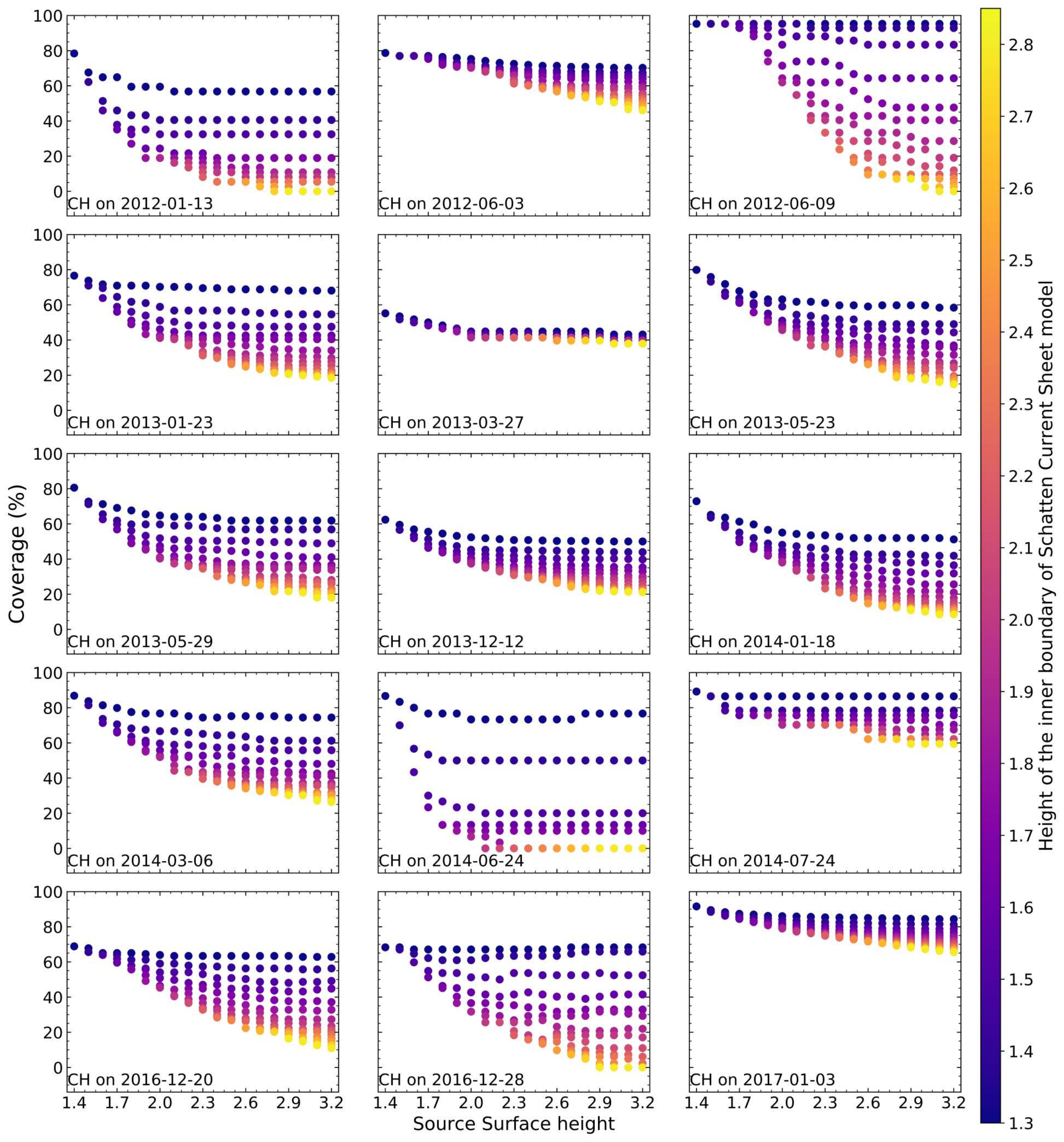}
    \caption{Collective results of model runs based on all the 184 pairs of heights for the source surface and the inner boundary of the SCS model. Each panel presents the results for one CH, and they are placed in chronological order. The x-axis is the source surface height and the colour bar is the height of the inner boundary of the SCS model increasing from dark to bright colours. The y-axis shows the coverage. As can be seen, for the majority of CHs best coverage is achieved for lower height values for the inner boundary of the SCS model.}
 \label{Coverage_collective}
 \end{figure}

Figure \ref{Coverage_collective} is a collective plot for the coverage resulting from the model runs for each of the 184 height pairs and the 15 CHs. Each panel represents the results for a particular CH, while the x-axis is the source surface height, the y-axis the coverage and the colour map the height of the inner boundary of the SCS model, $R_{\rm i}$. The striking feature in this figure is that for the vast majority of the CHs reducing $R_{\rm i}$ significantly improves the modelled coverage. For some cases the coverage even rises from 0 to almost 100$\%$. Overall, a lower source surface height has the same effect; however, for a fixed low value $R_{\rm i}$ a decrease in the source surface height does not necessarily have a high impact. The large CH that consisted of the southern polar one and its long extension to equatorial latitudes was very well modelled by EUHFORIA run with the default values. Lowering the values of the pair of heights below the default ones did result in improvement of the coverage; however as it was already high the improvement was not significant (within 10\%). The CH on 2013-03-27 is a very patchy CH that is not well modelled by EUHFORIA's adaptation of the WSA. The coverage is below 60\% regardless of the selected pair of heights considered for running EUHFORIA. Another patchy CH, but significantly less patchy than that on 2013-03-27, is the CH on 2016-12-20. This CH is a bit better modelled when lower values of the pair of heights are considered, but the coverage remained below 70\%. Reconstruction of these strongly (2013-03-27) and moderately (2016-12-20) patchy CHs suggest that the model might have difficulty in reconstructing patchy CHs. It will be interesting for this to be investigated further by considering a larger sample of patchy CHs. Also, it will be interesting to assess possible impact in modelling high speed streams originating from these CHs.

Even though, there is a clear indication that a pair of lower heights will result in a significantly improved coverage for all CHs not a single pair could be specified as the ideal one. In addition, lowering the two heights results in opening more flux to the heliosphere, not only within the CH areas, but also outside. This is illustrated in Figure \ref{DefaultVsBetter} where the modelled CH area extends beyond the EUV defined boundaries. This feature is present in all CH cases and rises the question of how low one can place the two heights, [$R_{\rm i}$, $R_{\rm ss}$], in order to best model the coronal hole area, but without overestimating the open flux outside the EUV defined CH boundaries. To answer that question we consider pixels of open flux that lie outside the EUV defined CH boundaries but have the same polarity as the CH and are connected to the open flux pixels enclosed by the EUV defined CH boundaries, i.e., neighbouring pixels. To identify these areas, we apply the region growth method by selecting seed pixels within the CH area that are assigned as open field. We calculated the total number of connected open flux pixels, $N_{\rm total}$, and divided them to those that lie within the boundaries, $N_{\rm in}$, and those that lie outside, $N_{\rm out}$. From these we define the percentage of connected area within the boundary, $CA_{\rm in}$ which is given by:

\begin{linenomath*}
\begin{equation}
CA_{in} = \frac{N_{in}}{N_{total}}*100 [\%]
\end{equation}
\end{linenomath*}

\begin{figure}[h!]
    \centering
    \includegraphics[width = 1\textwidth]{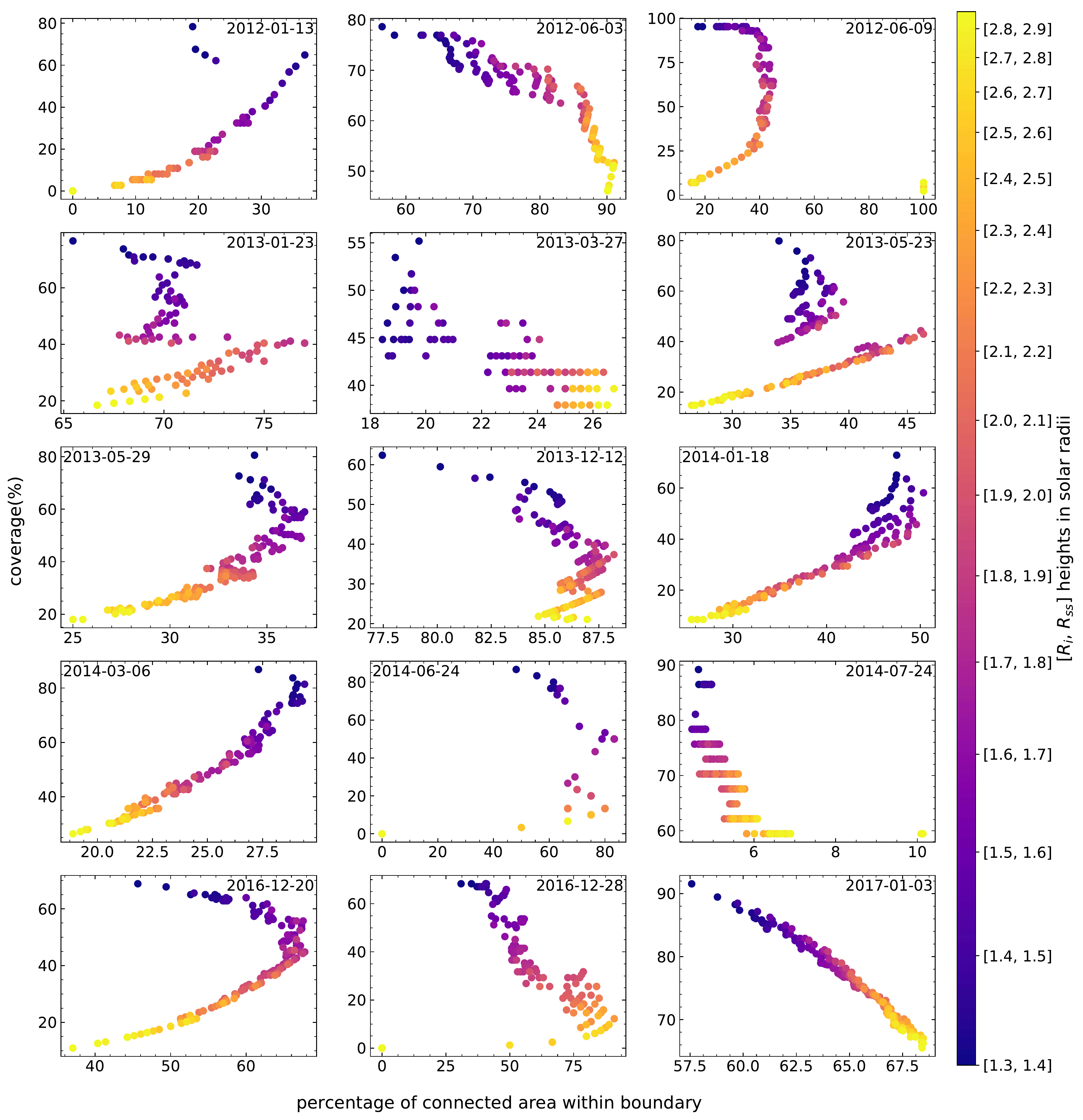}
    \caption{Each panel depicts how, for a particular CH, coverage and $CA_{\rm in}$ change when EUHFORIA runs are set up based on different pair of heights [$R_{\rm i}$, $R_{\rm ss}$]. For most CHs a saturation limit is reached beyond which the amount of open flux pixels lying inside the expected from EUV observations area is significantly lower than those lying outside.}
 \label{growth}
\end{figure}

In Figure \ref{growth} we plot this parameter as a function of the coverage. We can see that for the majority of the CHs both parameters at first increase but then the system reaches a point where the percentage of connected area that lies within the boundaries starts decreasing. This implies that, although, open flux pixels grow the majority lies outside the EUV defined boundaries. If flux was opening only inside the boundaries then the curves would be approximately linear. This does not happen though and a saturation point is reached from which further decreasing of the two heights opens more flux outside the boundaries than inside. Following the standard paradigm that the primary source of open flux are CH areas we can conclude that too low heights lead to nonphysical results due to cutting actually closed loops. This is an important finding since it limits how low the two heights can be. From the same figure one can notice that for some CHs there is a double saturation limit.

\begin{figure}[h!]
    \centering
    \begin{minipage}{0.96\textwidth}
        \centering
        \begin{overpic}[width = 0.96\textwidth]{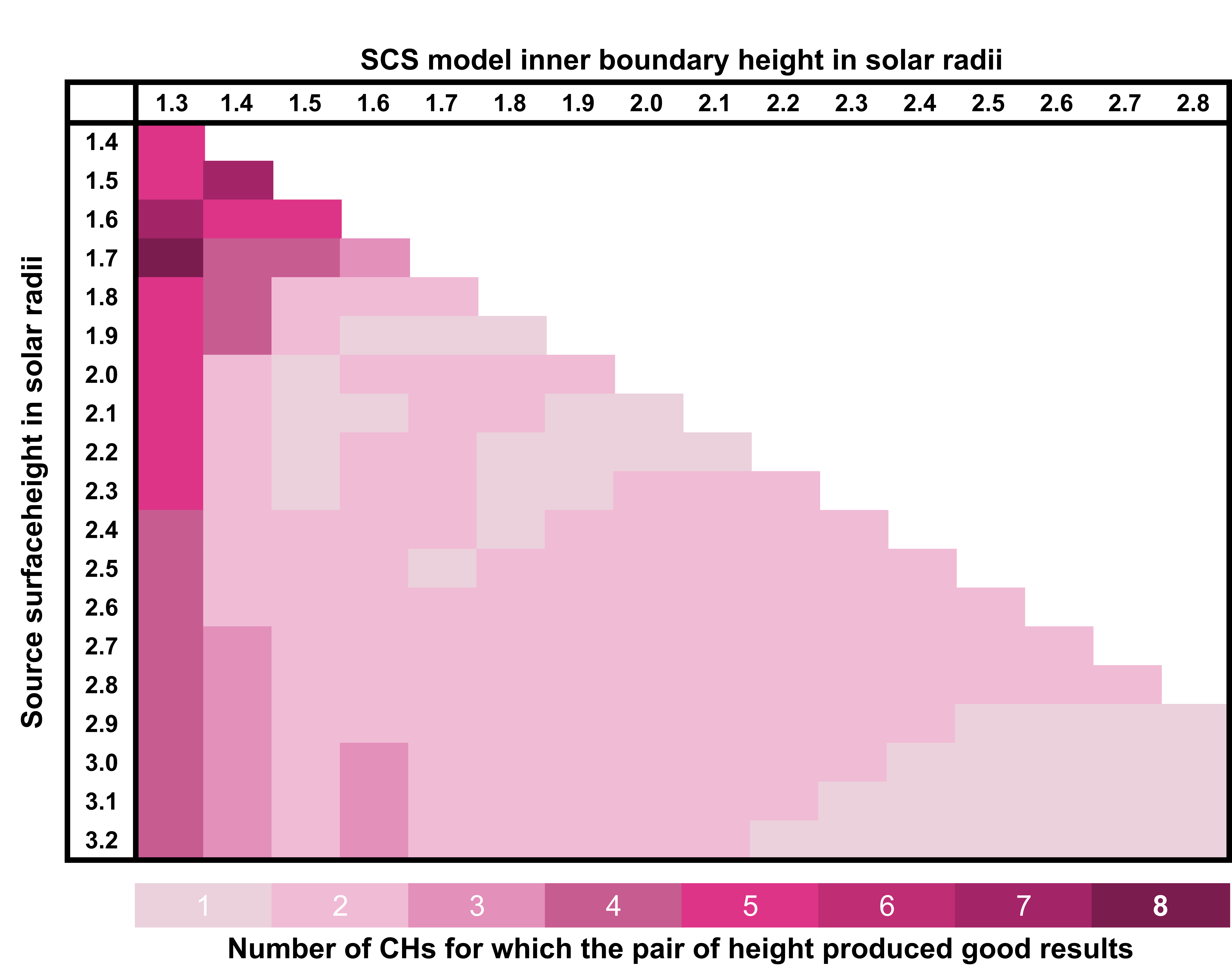}
        \end{overpic}
    \end{minipage}
    \caption{Collective table of all 184 pair of heights for which the model was tested. The colour in each shell indicates that this pair of heights produced good results both in coverage and $CA_{\rm in}$ for a certain number of CHs, as indicated in the colour map. The most successful pairs of heights, i.e., the ones that were successful for more CHs are collected on the top left corner of the table where lower height values can be found.}
 \label{succesS_table}
\end{figure}

These saturation limits identified in Figure \ref{growth} can be used in order to constrain the possible pairs of boundary heights to those that provide an improved result both in coverage and $CA_{\rm in}$.  We define two criteria with the first being that the coverage parameter needs to be above 60\% and the second is that $CA_{\rm in}$ should lay no less than 5\% from the saturation limit. Although, for most CHs the coverage is above 70\%, for four of them (i.e. 2013-03-27, 2013-12-12, 2016-12-20 and 2016-12-28) the maximum coverage is well below that value, which is the reason for this 60\% limit. Especially for the CH on 2013-03-27 the coverage parameter is below 55\%, thus, a special limit of minimum 50\% coverage is applied. For 8 out of the 15 CHs the value of $CA_{\rm in}$ is below 50\%. To this we include the CH on 2012-06-09 for which the high values of $CA_{\rm in}$ are only for when the coverage is below 10\% which indicates that the number of open flux pixels is extremely low. A $CA_{\rm in}<50\%$ is the result of having more open flux lying outside the CH area defined by the EUV boundaries for all the model runs performed. For some of the CH a secondary saturation limit can be identified less than 10\% from the first one, which, however, results in better coverage without affecting greatly $CA_{\rm in}$. For those cases we do consider the second saturation limit for the $CA_{\rm in}$ in investigating the optimal [$R_{\rm i}$, $R_{\rm ss}$] pair. After analysing the results for each CH we collected the pairs of heights that justify the criteria posed. The table is presented in Figure \ref{succesS_table}. The colour map indicate the number of CHs for which the pair fitted the criteria. Although, all the height pairs fulfill the criteria for at least one CH, some appear as good candidates more frequently. As can be seen in the table low heights of the SCS model inner boundary (below 1.5$R_{\rm \odot}$) are better options for successfully reconstructing CH areas ($cov > 60\%$), while the selection of the source surface height varies more strongly. Although, for the CH sample studied here these heights gave better results in terms of the coverage, it is important to highlight that there still can be open flux growth outside the expected from observations CH areas. So the use of the heights need to be done with caution.

\section{Conclusion and Discussion}
\label{subsec:discussion}

As already discussed in other works \cite<e.g.,>[]{arden_breathing_2014, linker_open_2017, wallace_estimating_2019} discrepancies exist between in situ measured at 1AU and model computed open flux using the PFSS model, and subsequently the WSA model. Considering the standard paradigm that CHs are primary sources of open flux \cite{mackay_suns_2012, linker_open_2017}, testing whether CHs are properly modelled is a logical step. \citeA{linker_open_2017} highlighted the importance for the model output to match CHs observed in EUV emission. They used automatically detected CHs to quantitatively make comparisons between the PFSS model output and CHs areas. Earlier, \citeA{lowder_measurements_2014} thoroughly computed open flux based on automatically detected CHs and compared the result to the PFSS estimations. For the period \citeA{linker_open_2017} investigated, placing the source surface height at 2.5$R_{\rm ss}$ resulted in significantly underestimated open field area and flux. They also concluded that lowering the source surface, although improved the open flux value, it overestimates the areas of open flux. 

In this work we presented a collective study assessing the capability of the PFSS+SCS part of the adopted WSA model by EUHFORIA forecasting tool in modelling CH areas. Although, values between 1.6 to 3.25 solar radii are considered allowed for the source surface height, a commonly used value is 2.5$R_{\rm \odot}$ \cite<e.g.,>[]{arge_improvement_2000, riley_role_2015, wallace_estimating_2019}. On the other hand, \citeA{mcgregor_analysis_2008} concluded that for the modified WSA model which they presented, and which is also adopted in EUHFORIA, the optimal heights of the source surface and the SCS model inner boundary are [2.3, 2.6]$R_{\rm \odot}$, which we considered as the default pair of heights. Other studies suggested an even lower or higher source surface height \cite{phillips_ulysses_1994, lee_coronal_2011} or even a solar cycle varying source surface \cite{arden_breathing_2014}. We considered 184 pairs of heights aiming to investigate whether the modelled result can be improved. We selected 15 CHs having CoM latitude within the central meridional zone of the Sun, and a variety of morphology defined by their elongation, area size, intensity and level of compactness. Using EUHFORIA we reconstructed the CH area based on model runs for the 184 pairs. The modelled areas were compared to those derived with \textsc{catch} using EUV imaging observations. 

Our main result is that the default heights employed so far, [2.3, 2.6]$R_{\rm \odot}$, fail to model the expected CH areas and/or their locations, with coverage below 60\%. This suggests that CH areas are not properly modelled using the default setup. This might provide some explanation for the discrepancy between the open flux output by the WSA model and that measured in-situ at Earth \cite{riley_role_2015, linker_open_2017, wallace_estimating_2019}. In some cases even the whole CH was missed. Our analysis clearly shows that lower heights ($R_{\rm i}$ below 1.5$R_{\rm \odot}$ and $R_{\rm ss}$ below 2.3$R_{\rm \odot}$) lead to an improved coverage ($cov > 60\%$) of the areas defined by the EUV extracted boundaries; however, it strongly results in opening flux outside those boundaries as well, which is opposing the standard paradigm for the sources of open field. This last point is in agreement with \citeA{linker_open_2017} result, i.e. that lowering the heights lead to overestimating open flux regions. Assuming that the standard paradigm for open flux origin is correct, this raises the question of how low the source surface height can be placed in order to properly model CH areas and at the same time not model excess open flux areas outside the EUV defined boundaries. Considering modelling the solar wind speed at 1AU it will be very interesting to see the effect of the modelled open flux areas outside the EUV defined boundaries can have. Our analysis also shows that it is rather the height of the inner boundary of the SCS model and not of the source surface that has a stronger effect on the modelled result. We also deduced that low values of this boundary lead to better results in terms or CH area reconstruction. It is noteworthy that this result is based on analysing a CH sample from only one solar cycle and considers CHs located in the central meridional zone. Increasing the size of the sample and also extending it over a larger period of time can potentially lead to different height selection.

Although, following different techniques \citeA{lowder_measurements_2014, linker_open_2017,wallace_estimating_2019} over-plotted automatically detected CHs on synoptic magnetic maps but still resulted in underestimated open flux. According to \citeA{linker_open_2017} the result could only be improved by considering areas larger than the detected CHs, leading them to the conclusion that either synoptic maps underestimate the solar magnetic flux or that CHs are not the sole source of open flux. In addition, there are studies that challenge the certainty that the open flux from in-situ measurements should be an exact match to that modelled close to the Sun, arguing that there might be other processes/factors affecting it, as suggested for example in \citeA{lockwood_excess_2009-1}, where they found that the excess open flux between models and in situ measurements can be explained by kinematic effects of variations of the solar wind speed. Thus, opening the question of whether one to one comparison is a sensible approach.

\citeA{lee_coronal_2011, arden_breathing_2014} investigated the possibility of a solar cycle varying source surface height and concluded that a lower source surface height during solar minimum periods is required. We investigated this possibility but we identified no solar cycle trends in the success of the model in reconstructing CH areas; however, capturing the whole cycle could improve the results. In addition, we investigated a possible latitude dependence; however, the latitudinal position of the CH did not have any effect on the modelled results. These are interesting results which, however, need to be considered with care due to rather small sample size. A more detailed analysis, considering a larger number of CHs, spanning over more than one solar cycle and over all solar cycle phases, and having more variety in their latitudinal positions, can potentially lead to a different outcome.

An interesting remark is that for the CH on 2014-07-24 a channeling to a polar CH was present regardless of the selected pair of heights. This was also appearing for other CHs in the sample when lower heights were selected for the source surface and the inner boundary of the SCS model. The idea of CHs being linked to polar ones via narrow passages in the photosphere has been discussed in \citeA<e.g.,>[]{antiochos_structure_2007, antiochos_model_2011}. It will be interesting to investigate this aspect on a larger sample of CHs.

It is necessary to mention that the outcome of this study, and other similar studies before it, is focused on how the two free parameters involved in the PFSS+SCS, and subsequently in the WSA model, can affect the modelled results. Saying that we aim to bring forward that the missing open flux in the model outputs might likewise require fine tuning of the other free parameters involved in the model, an idea also discussed in \citeA{wallace_estimating_2019}. Future work should also focus on improving the statistics and also on different aspects of the model outcomes, i.e. CH areas, open flux at Earth, velocity of high speed streams, and current sheet position.

\acknowledgments
E.A. would like to acknowledge the financial support by the Finnish Academy of Science and Letters via the Postdoc Pool funding. S.G.H. and M.T. acknowledge the support by
the FFG/ASAP Program under grant No. 859729 (SWAMI). EUHFORIA is developed as a joint effort between the University of Helsinki and KU Leuven. The validation of solar wind and CME modeling with EUHFORIA is being performed within the BRAIN-be project CCSOM (Constraining CMEs and Shocks by Observations and Modelling throughout the inner heliosphere; \url{http://www.sidc.be/ccsom/}).


%

\begin{thebibliography}{}

\bibitem [\protect \citeauthoryear {%
Altschuler%
\ \BBA {} Newkirk%
}{%
Altschuler%
\ \BBA {} Newkirk%
}{%
{\protect \APACyear {1969}}%
}]{%
altschuler_magnetic_1969}
\APACinsertmetastar {%
altschuler_magnetic_1969}%
\begin{APACrefauthors}%
Altschuler, M\BPBI D.%
\BCBT {}\ \BBA {} Newkirk, G.%
\end{APACrefauthors}%
\unskip\
\newblock
\APACrefYearMonthDay{1969}{{\APACmonth{09}}}{}.
\newblock
{\BBOQ}\APACrefatitle {Magnetic {Fields} and the {Structure} of the {Solar}
  {Corona}. {I}: {Methods} of {Calculating} {Coronal} {Fields}} {Magnetic
  {Fields} and the {Structure} of the {Solar} {Corona}. {I}: {Methods} of
  {Calculating} {Coronal} {Fields}}.{\BBCQ}
\newblock
\APACjournalVolNumPages{Solar Physics}{9}{}{131--149}.
\newblock
\begin{APACrefDOI} \doi{10.1007/BF00145734} \end{APACrefDOI}
\PrintBackRefs{\CurrentBib}

\bibitem [\protect \citeauthoryear {%
Antiochos%
, DeVore%
, Karpen%
\BCBL {}\ \BBA {} Mikić%
}{%
Antiochos%
\ \protect \BOthers {.}}{%
{\protect \APACyear {2007}}%
}]{%
antiochos_structure_2007}
\APACinsertmetastar {%
antiochos_structure_2007}%
\begin{APACrefauthors}%
Antiochos, S\BPBI K.%
, DeVore, C\BPBI R.%
, Karpen, J\BPBI T.%
\BCBL {}\ \BBA {} Mikić, Z.%
\end{APACrefauthors}%
\unskip\
\newblock
\APACrefYearMonthDay{2007}{{\APACmonth{12}}}{}.
\newblock
{\BBOQ}\APACrefatitle {Structure and {Dynamics} of the {Sun}'s {Open}
  {Magnetic} {Field}} {Structure and {Dynamics} of the {Sun}'s {Open}
  {Magnetic} {Field}}.{\BBCQ}
\newblock
\APACjournalVolNumPages{The Astrophysical Journal}{671}{1}{936}.
\newblock
\begin{APACrefDOI} \doi{10.1086/522489} \end{APACrefDOI}
\PrintBackRefs{\CurrentBib}

\bibitem [\protect \citeauthoryear {%
Antiochos%
, Mikić%
, Titov%
, Lionello%
\BCBL {}\ \BBA {} Linker%
}{%
Antiochos%
\ \protect \BOthers {.}}{%
{\protect \APACyear {2011}}%
}]{%
antiochos_model_2011}
\APACinsertmetastar {%
antiochos_model_2011}%
\begin{APACrefauthors}%
Antiochos, S\BPBI K.%
, Mikić, Z.%
, Titov, V\BPBI S.%
, Lionello, R.%
\BCBL {}\ \BBA {} Linker, J\BPBI A.%
\end{APACrefauthors}%
\unskip\
\newblock
\APACrefYearMonthDay{2011}{{\APACmonth{04}}}{}.
\newblock
{\BBOQ}\APACrefatitle {A {Model} for the {Sources} of the {Slow} {Solar}
  {Wind}} {A {Model} for the {Sources} of the {Slow} {Solar} {Wind}}.{\BBCQ}
\newblock
\APACjournalVolNumPages{The Astrophysical Journal}{731}{2}{112}.
\newblock
\begin{APACrefDOI} \doi{10.1088/0004-637X/731/2/112} \end{APACrefDOI}
\PrintBackRefs{\CurrentBib}

\bibitem [\protect \citeauthoryear {%
Arden%
, Norton%
\BCBL {}\ \BBA {} Sun%
}{%
Arden%
\ \protect \BOthers {.}}{%
{\protect \APACyear {2014}}%
}]{%
arden_breathing_2014}
\APACinsertmetastar {%
arden_breathing_2014}%
\begin{APACrefauthors}%
Arden, W\BPBI M.%
, Norton, A\BPBI A.%
\BCBL {}\ \BBA {} Sun, X.%
\end{APACrefauthors}%
\unskip\
\newblock
\APACrefYearMonthDay{2014}{{\APACmonth{03}}}{}.
\newblock
{\BBOQ}\APACrefatitle {A "breathing" source surface for cycles 23 and 24} {A
  "breathing" source surface for cycles 23 and 24}.{\BBCQ}
\newblock
\APACjournalVolNumPages{Journal of Geophysical Research (Space
  Physics)}{119}{3}{1476}.
\newblock
\begin{APACrefDOI} \doi{10.1002/2013JA019464} \end{APACrefDOI}
\PrintBackRefs{\CurrentBib}

\bibitem [\protect \citeauthoryear {%
Arge%
\ \protect \BOthers {.}}{%
Arge%
\ \protect \BOthers {.}}{%
{\protect \APACyear {2010}}%
}]{%
arge_air_2010}
\APACinsertmetastar {%
arge_air_2010}%
\begin{APACrefauthors}%
Arge, C\BPBI N.%
, Henney, C\BPBI J.%
, Koller, J.%
, Compeau, C\BPBI R.%
, Young, S.%
, MacKenzie, D.%
\BDBL {}Harvey, J\BPBI W.%
\end{APACrefauthors}%
\unskip\
\newblock
\APACrefYearMonthDay{2010}{{\APACmonth{03}}}{}.
\newblock
{\BBOQ}\APACrefatitle {Air {Force} {Data} {Assimilative} {Photospheric} {Flux}
  {Transport} ({ADAPT}) {Model}} {Air {Force} {Data} {Assimilative}
  {Photospheric} {Flux} {Transport} ({ADAPT}) {Model}}.{\BBCQ}
\newblock
\APACjournalVolNumPages{Twelfth International Solar Wind
  Conference}{1216}{}{343--346}.
\newblock
\begin{APACrefDOI} \doi{10.1063/1.3395870} \end{APACrefDOI}
\PrintBackRefs{\CurrentBib}

\bibitem [\protect \citeauthoryear {%
Arge%
, Hildner%
, Pizzo%
\BCBL {}\ \BBA {} Harvey%
}{%
Arge%
\ \protect \BOthers {.}}{%
{\protect \APACyear {2002}}%
}]{%
arge_two_2002}
\APACinsertmetastar {%
arge_two_2002}%
\begin{APACrefauthors}%
Arge, C\BPBI N.%
, Hildner, E.%
, Pizzo, V\BPBI J.%
\BCBL {}\ \BBA {} Harvey, J\BPBI W.%
\end{APACrefauthors}%
\unskip\
\newblock
\APACrefYearMonthDay{2002}{{\APACmonth{10}}}{}.
\newblock
{\BBOQ}\APACrefatitle {Two solar cycles of nonincreasing magnetic flux} {Two
  solar cycles of nonincreasing magnetic flux}.{\BBCQ}
\newblock
\APACjournalVolNumPages{Journal of Geophysical Research (Space
  Physics)}{107}{}{1319}.
\newblock
\begin{APACrefDOI} \doi{10.1029/2001JA000503} \end{APACrefDOI}
\PrintBackRefs{\CurrentBib}

\bibitem [\protect \citeauthoryear {%
Arge%
\ \BBA {} Pizzo%
}{%
Arge%
\ \BBA {} Pizzo%
}{%
{\protect \APACyear {2000}}%
}]{%
arge_improvement_2000}
\APACinsertmetastar {%
arge_improvement_2000}%
\begin{APACrefauthors}%
Arge, C\BPBI N.%
\BCBT {}\ \BBA {} Pizzo, V\BPBI J.%
\end{APACrefauthors}%
\unskip\
\newblock
\APACrefYearMonthDay{2000}{{\APACmonth{05}}}{}.
\newblock
{\BBOQ}\APACrefatitle {Improvement in the prediction of solar wind conditions
  using near-real time solar magnetic field updates} {Improvement in the
  prediction of solar wind conditions using near-real time solar magnetic field
  updates}.{\BBCQ}
\newblock
\APACjournalVolNumPages{Journal of Geophysical Research}{105}{}{10465--10480}.
\newblock
\begin{APACrefDOI} \doi{10.1029/1999JA000262} \end{APACrefDOI}
\PrintBackRefs{\CurrentBib}

\bibitem [\protect \citeauthoryear {%
Asvestari%
\ \protect \BOthers {.}}{%
Asvestari%
\ \protect \BOthers {.}}{%
{\protect \APACyear {submitted}}%
}]{%
asvestari_parameters_2019}
\APACinsertmetastar {%
asvestari_parameters_2019}%
\begin{APACrefauthors}%
Asvestari, E.%
, Heinemann, S\BPBI G.%
, Temmer, M.%
, Pomoell, J.%
, Kilpua, E.%
, Magdalenic, J.%
\BCBL {}\ \BBA {} Poedts, S.%
\end{APACrefauthors}%
\unskip\
\newblock
\APACrefYearMonthDay{submitted}{}{}.
\newblock
{\BBOQ}\APACrefatitle {The impact of coronal hole characteristics and solar
  cycle activity in reconstructing coronal holes with EUHFORIA} {The impact of
  coronal hole characteristics and solar cycle activity in reconstructing
  coronal holes with euhforia}.{\BBCQ}
\newblock
\APACjournalVolNumPages{10th Young Researchers Meeting}{}{}{}.
\PrintBackRefs{\CurrentBib}

\bibitem [\protect \citeauthoryear {%
Gressl%
\ \protect \BOthers {.}}{%
Gressl%
\ \protect \BOthers {.}}{%
{\protect \APACyear {2014}}%
}]{%
gressl_comparative_2014}
\APACinsertmetastar {%
gressl_comparative_2014}%
\begin{APACrefauthors}%
Gressl, C.%
, Veronig, A\BPBI M.%
, Temmer, M.%
, Odstrčil, D.%
, Linker, J\BPBI A.%
, Mikić, Z.%
\BCBL {}\ \BBA {} Riley, P.%
\end{APACrefauthors}%
\unskip\
\newblock
\APACrefYearMonthDay{2014}{{\APACmonth{05}}}{}.
\newblock
{\BBOQ}\APACrefatitle {Comparative {Study} of {MHD} {Modeling} of the
  {Background} {Solar} {Wind}} {Comparative {Study} of {MHD} {Modeling} of the
  {Background} {Solar} {Wind}}.{\BBCQ}
\newblock
\APACjournalVolNumPages{Solar Physics}{289}{5}{1783}.
\newblock
\begin{APACrefDOI} \doi{10.1007/s11207-013-0421-6} \end{APACrefDOI}
\PrintBackRefs{\CurrentBib}

\bibitem [\protect \citeauthoryear {%
{Heinemann}%
\ \protect \BOthers {.}}{%
{Heinemann}%
\ \protect \BOthers {.}}{%
{\protect \APACyear {2019}}%
}]{%
Heinemann_CATCH_2019}
\APACinsertmetastar {%
Heinemann_CATCH_2019}%
\begin{APACrefauthors}%
{Heinemann}, S\BPBI G.%
, {Temmer}, M.%
, {Heinemann}, N.%
, {Dissauer}, K.%
, {Samara}, E.%
, {Jer{\v{c}}i{\'c}}, V.%
, {Hofmeister}, S.%
\BDBL {}{Veronig}, A\BPBI M.%
\end{APACrefauthors}%
\unskip\
\newblock
\APACrefYearMonthDay{2019}{Jul}{}.
\newblock
{\BBOQ}\APACrefatitle {{Coronal Hole Statistical Analysis and Catalogue
  covering the SDO-era}} {{Coronal Hole Statistical Analysis and Catalogue
  covering the SDO-era}}.{\BBCQ}
\newblock
\APACjournalVolNumPages{arXiv e-prints}{}{}{arXiv:1907.01990}.
\PrintBackRefs{\CurrentBib}

\bibitem [\protect \citeauthoryear {%
Hickmann%
, Godinez%
, Henney%
\BCBL {}\ \BBA {} Arge%
}{%
Hickmann%
\ \protect \BOthers {.}}{%
{\protect \APACyear {2015}}%
}]{%
hickmann_data_2015}
\APACinsertmetastar {%
hickmann_data_2015}%
\begin{APACrefauthors}%
Hickmann, K\BPBI S.%
, Godinez, H\BPBI C.%
, Henney, C\BPBI J.%
\BCBL {}\ \BBA {} Arge, C\BPBI N.%
\end{APACrefauthors}%
\unskip\
\newblock
\APACrefYearMonthDay{2015}{{\APACmonth{04}}}{}.
\newblock
{\BBOQ}\APACrefatitle {Data {Assimilation} in the {ADAPT} {Photospheric} {Flux}
  {Transport} {Model}} {Data {Assimilation} in the {ADAPT} {Photospheric}
  {Flux} {Transport} {Model}}.{\BBCQ}
\newblock
\APACjournalVolNumPages{Solar Physics}{290}{}{1105--1118}.
\newblock
\begin{APACrefDOI} \doi{10.1007/s11207-015-0666-3} \end{APACrefDOI}
\PrintBackRefs{\CurrentBib}

\bibitem [\protect \citeauthoryear {%
Hinterreiter%
\ \protect \BOthers {.}}{%
Hinterreiter%
\ \protect \BOthers {.}}{%
{\protect \APACyear {to be submitted}}%
}]{%
hinterreiter_solar_wind_2019}
\APACinsertmetastar {%
hinterreiter_solar_wind_2019}%
\begin{APACrefauthors}%
Hinterreiter, J.%
, Magdalenic, J.%
, Temmer, M.%
, Verbeke, C.%
, Jerabaj, I\BPBI C.%
, Samara, E.%
\BDBL {}Isavnin, A.%
\end{APACrefauthors}%
\unskip\
\newblock
\APACrefYearMonthDay{to be submitted}{}{}.
\newblock
{\BBOQ}\APACrefatitle {Testing the background solar wind modelled by EUHFORIA}
  {Testing the background solar wind modelled by euhforia}.{\BBCQ}
\newblock
\APACjournalVolNumPages{Solar Physics}{}{}{}.
\PrintBackRefs{\CurrentBib}

\bibitem [\protect \citeauthoryear {%
Hoeksema%
\ \BBA {} Scherrer%
}{%
Hoeksema%
\ \BBA {} Scherrer%
}{%
{\protect \APACyear {1986}}%
}]{%
hoeksema_atlas_1986}
\APACinsertmetastar {%
hoeksema_atlas_1986}%
\begin{APACrefauthors}%
Hoeksema, J\BPBI T.%
\BCBT {}\ \BBA {} Scherrer, P\BPBI H.%
\end{APACrefauthors}%
\unskip\
\newblock
\APACrefYearMonthDay{1986}{{\APACmonth{05}}}{}.
\newblock
{\BBOQ}\APACrefatitle {An atlas of photospheric magnetic field observations and
  computed coronal magnetic fields: 1976-1985} {An atlas of photospheric
  magnetic field observations and computed coronal magnetic fields:
  1976-1985}.{\BBCQ}
\newblock
\APACjournalVolNumPages{Solar Physics}{105}{}{205--211}.
\newblock
\begin{APACrefDOI} \doi{10.1007/BF00156388} \end{APACrefDOI}
\PrintBackRefs{\CurrentBib}

\bibitem [\protect \citeauthoryear {%
Hoeksema%
, Wilcox%
\BCBL {}\ \BBA {} Scherrer%
}{%
Hoeksema%
\ \protect \BOthers {.}}{%
{\protect \APACyear {1982}}%
}]{%
hoeksema_structure_1982}
\APACinsertmetastar {%
hoeksema_structure_1982}%
\begin{APACrefauthors}%
Hoeksema, J\BPBI T.%
, Wilcox, J\BPBI M.%
\BCBL {}\ \BBA {} Scherrer, P\BPBI H.%
\end{APACrefauthors}%
\unskip\
\newblock
\APACrefYearMonthDay{1982}{{\APACmonth{12}}}{}.
\newblock
{\BBOQ}\APACrefatitle {Structure of the heliospheric current sheet in the early
  portion of sunspot cycle 21} {Structure of the heliospheric current sheet in
  the early portion of sunspot cycle 21}.{\BBCQ}
\newblock
\APACjournalVolNumPages{Journal of Geophysical Research}{87}{}{10331--10338}.
\newblock
\begin{APACrefDOI} \doi{10.1029/JA087iA12p10331} \end{APACrefDOI}
\PrintBackRefs{\CurrentBib}

\bibitem [\protect \citeauthoryear {%
Hoeksema%
, Wilcox%
\BCBL {}\ \BBA {} Scherrer%
}{%
Hoeksema%
\ \protect \BOthers {.}}{%
{\protect \APACyear {1983}}%
}]{%
hoeksema_structure_1983}
\APACinsertmetastar {%
hoeksema_structure_1983}%
\begin{APACrefauthors}%
Hoeksema, J\BPBI T.%
, Wilcox, J\BPBI M.%
\BCBL {}\ \BBA {} Scherrer, P\BPBI H.%
\end{APACrefauthors}%
\unskip\
\newblock
\APACrefYearMonthDay{1983}{{\APACmonth{12}}}{}.
\newblock
{\BBOQ}\APACrefatitle {The structure of the heliospheric current sheet -
  1978-1982} {The structure of the heliospheric current sheet -
  1978-1982}.{\BBCQ}
\newblock
\APACjournalVolNumPages{Journal of Geophysical Research}{88}{}{9910--9918}.
\newblock
\begin{APACrefDOI} \doi{10.1029/JA088iA12p09910} \end{APACrefDOI}
\PrintBackRefs{\CurrentBib}

\bibitem [\protect \citeauthoryear {%
Hofmeister%
\ \protect \BOthers {.}}{%
Hofmeister%
\ \protect \BOthers {.}}{%
{\protect \APACyear {2018}}%
}]{%
hofmeister_dependence_2018}
\APACinsertmetastar {%
hofmeister_dependence_2018}%
\begin{APACrefauthors}%
Hofmeister, S\BPBI J.%
, Veronig, A.%
, Temmer, M.%
, Vennerstrom, S.%
, Heber, B.%
\BCBL {}\ \BBA {} Vršnak, B.%
\end{APACrefauthors}%
\unskip\
\newblock
\APACrefYearMonthDay{2018}{{\APACmonth{03}}}{}.
\newblock
{\BBOQ}\APACrefatitle {The {Dependence} of the {Peak} {Velocity} of
  {High}-{Speed} {Solar} {Wind} {Streams} as {Measured} in the {Ecliptic} by
  {ACE} and the {STEREO} satellites on the {Area} and {Co}-latitude of {Their}
  {Solar} {Source} {Coronal} {Holes}} {The {Dependence} of the {Peak}
  {Velocity} of {High}-{Speed} {Solar} {Wind} {Streams} as {Measured} in the
  {Ecliptic} by {ACE} and the {STEREO} satellites on the {Area} and
  {Co}-latitude of {Their} {Solar} {Source} {Coronal} {Holes}}.{\BBCQ}
\newblock
\APACjournalVolNumPages{Journal of Geophysical Research (Space
  Physics)}{123}{3}{1738}.
\newblock
\begin{APACrefDOI} \doi{10.1002/2017JA024586} \end{APACrefDOI}
\PrintBackRefs{\CurrentBib}

\bibitem [\protect \citeauthoryear {%
Jian%
\ \protect \BOthers {.}}{%
Jian%
\ \protect \BOthers {.}}{%
{\protect \APACyear {2011}}%
}]{%
jian_comparison_2011}
\APACinsertmetastar {%
jian_comparison_2011}%
\begin{APACrefauthors}%
Jian, L\BPBI K.%
, Russell, C\BPBI T.%
, Luhmann, J\BPBI G.%
, MacNeice, P\BPBI J.%
, Odstrcil, D.%
, Riley, P.%
\BDBL {}Steinberg, J\BPBI T.%
\end{APACrefauthors}%
\unskip\
\newblock
\APACrefYearMonthDay{2011}{{\APACmonth{10}}}{}.
\newblock
{\BBOQ}\APACrefatitle {Comparison of {Observations} at {ACE} and {Ulysses} with
  {Enlil} {Model} {Results}: {Stream} {Interaction} {Regions} {During}
  {Carrington} {Rotations} 2016 - 2018} {Comparison of {Observations} at {ACE}
  and {Ulysses} with {Enlil} {Model} {Results}: {Stream} {Interaction}
  {Regions} {During} {Carrington} {Rotations} 2016 - 2018}.{\BBCQ}
\newblock
\APACjournalVolNumPages{Solar Physics}{273}{1}{179}.
\newblock
\begin{APACrefDOI} \doi{10.1007/s11207-011-9858-7} \end{APACrefDOI}
\PrintBackRefs{\CurrentBib}

\bibitem [\protect \citeauthoryear {%
Lee%
\ \protect \BOthers {.}}{%
Lee%
\ \protect \BOthers {.}}{%
{\protect \APACyear {2011}}%
}]{%
lee_coronal_2011}
\APACinsertmetastar {%
lee_coronal_2011}%
\begin{APACrefauthors}%
Lee, C\BPBI O.%
, Luhmann, J\BPBI G.%
, Hoeksema, J\BPBI T.%
, Sun, X.%
, Arge, C\BPBI N.%
\BCBL {}\ \BBA {} de Pater, I.%
\end{APACrefauthors}%
\unskip\
\newblock
\APACrefYearMonthDay{2011}{{\APACmonth{04}}}{}.
\newblock
{\BBOQ}\APACrefatitle {Coronal {Field} {Opens} at {Lower} {Height} {During} the
  {Solar} {Cycles} 22 and 23 {Minimum} {Periods}: {IMF} {Comparison} {Suggests}
  the {Source} {Surface} {Should} {Be} {Lowered}} {Coronal {Field} {Opens} at
  {Lower} {Height} {During} the {Solar} {Cycles} 22 and 23 {Minimum} {Periods}:
  {IMF} {Comparison} {Suggests} the {Source} {Surface} {Should} {Be}
  {Lowered}}.{\BBCQ}
\newblock
\APACjournalVolNumPages{Solar Physics}{269}{}{367--388}.
\newblock
\begin{APACrefDOI} \doi{10.1007/s11207-010-9699-9} \end{APACrefDOI}
\PrintBackRefs{\CurrentBib}

\bibitem [\protect \citeauthoryear {%
Lee%
\ \protect \BOthers {.}}{%
Lee%
\ \protect \BOthers {.}}{%
{\protect \APACyear {2009}}%
}]{%
lee_solar_2009}
\APACinsertmetastar {%
lee_solar_2009}%
\begin{APACrefauthors}%
Lee, C\BPBI O.%
, Luhmann, J\BPBI G.%
, Odstrcil, D.%
, MacNeice, P\BPBI J.%
, de Pater, I.%
, Riley, P.%
\BCBL {}\ \BBA {} Arge, C\BPBI N.%
\end{APACrefauthors}%
\unskip\
\newblock
\APACrefYearMonthDay{2009}{{\APACmonth{01}}}{}.
\newblock
{\BBOQ}\APACrefatitle {The {Solar} {Wind} at 1 {AU} {During} the {Declining}
  {Phase} of {Solar} {Cycle} 23: {Comparison} of 3D {Numerical} {Model}
  {Results} with {Observations}} {The {Solar} {Wind} at 1 {AU} {During} the
  {Declining} {Phase} of {Solar} {Cycle} 23: {Comparison} of 3d {Numerical}
  {Model} {Results} with {Observations}}.{\BBCQ}
\newblock
\APACjournalVolNumPages{Solar Physics}{254}{1}{155}.
\newblock
\begin{APACrefDOI} \doi{10.1007/s11207-008-9280-y} \end{APACrefDOI}
\PrintBackRefs{\CurrentBib}

\bibitem [\protect \citeauthoryear {%
Lemen%
\ \protect \BOthers {.}}{%
Lemen%
\ \protect \BOthers {.}}{%
{\protect \APACyear {2012}}%
}]{%
lemen_atmospheric_2012}
\APACinsertmetastar {%
lemen_atmospheric_2012}%
\begin{APACrefauthors}%
Lemen, J\BPBI R.%
, Title, A\BPBI M.%
, Akin, D\BPBI J.%
, Boerner, P\BPBI F.%
, Chou, C.%
, Drake, J\BPBI F.%
\BDBL {}Waltham, N.%
\end{APACrefauthors}%
\unskip\
\newblock
\APACrefYearMonthDay{2012}{{\APACmonth{01}}}{}.
\newblock
{\BBOQ}\APACrefatitle {The {Atmospheric} {Imaging} {Assembly} ({AIA}) on the
  {Solar} {Dynamics} {Observatory} ({SDO})} {The {Atmospheric} {Imaging}
  {Assembly} ({AIA}) on the {Solar} {Dynamics} {Observatory} ({SDO})}.{\BBCQ}
\newblock
\APACjournalVolNumPages{Solar Physics}{275}{}{17--40}.
\newblock
\begin{APACrefDOI} \doi{10.1007/s11207-011-9776-8} \end{APACrefDOI}
\PrintBackRefs{\CurrentBib}

\bibitem [\protect \citeauthoryear {%
Levine%
, Altschuler%
\BCBL {}\ \BBA {} Harvey%
}{%
Levine%
\ \protect \BOthers {.}}{%
{\protect \APACyear {1977}}%
}]{%
levine_solar_1977}
\APACinsertmetastar {%
levine_solar_1977}%
\begin{APACrefauthors}%
Levine, R\BPBI H.%
, Altschuler, M\BPBI D.%
\BCBL {}\ \BBA {} Harvey, J\BPBI W.%
\end{APACrefauthors}%
\unskip\
\newblock
\APACrefYearMonthDay{1977}{{\APACmonth{03}}}{}.
\newblock
{\BBOQ}\APACrefatitle {Solar sources of the interplanetary magnetic field and
  solar wind} {Solar sources of the interplanetary magnetic field and solar
  wind}.{\BBCQ}
\newblock
\APACjournalVolNumPages{Journal of Geophysical Research}{82}{}{1061--1065}.
\newblock
\begin{APACrefDOI} \doi{10.1029/JA082i007p01061} \end{APACrefDOI}
\PrintBackRefs{\CurrentBib}

\bibitem [\protect \citeauthoryear {%
Li%
, Xiang%
, Xie%
\BCBL {}\ \BBA {} Xu%
}{%
Li%
\ \protect \BOthers {.}}{%
{\protect \APACyear {2019}}%
}]{%
li_present_2019}
\APACinsertmetastar {%
li_present_2019}%
\begin{APACrefauthors}%
Li, F\BPBI Y.%
, Xiang, N\BPBI B.%
, Xie, J\BPBI L.%
\BCBL {}\ \BBA {} Xu, J\BPBI C.%
\end{APACrefauthors}%
\unskip\
\newblock
\APACrefYearMonthDay{2019}{{\APACmonth{03}}}{}.
\newblock
{\BBOQ}\APACrefatitle {The {Present} {Special} {Solar} {Cycle} 24: {Casting} a
  {Shadow} over {Periodicity} of the {North}-{South} {Hemispherical}
  {Asymmetry}} {The {Present} {Special} {Solar} {Cycle} 24: {Casting} a
  {Shadow} over {Periodicity} of the {North}-{South} {Hemispherical}
  {Asymmetry}}.{\BBCQ}
\newblock
\APACjournalVolNumPages{The Astrophysical Journal}{873}{2}{121}.
\newblock
\begin{APACrefDOI} \doi{10.3847/1538-4357/ab06bf} \end{APACrefDOI}
\PrintBackRefs{\CurrentBib}

\bibitem [\protect \citeauthoryear {%
Linker%
\ \protect \BOthers {.}}{%
Linker%
\ \protect \BOthers {.}}{%
{\protect \APACyear {2017}}%
}]{%
linker_open_2017}
\APACinsertmetastar {%
linker_open_2017}%
\begin{APACrefauthors}%
Linker, J\BPBI A.%
, Caplan, R\BPBI M.%
, Downs, C.%
, Riley, P.%
, Mikic, Z.%
, Lionello, R.%
\BDBL {}Owens, M\BPBI J.%
\end{APACrefauthors}%
\unskip\
\newblock
\APACrefYearMonthDay{2017}{{\APACmonth{10}}}{}.
\newblock
{\BBOQ}\APACrefatitle {The {Open} {Flux} {Problem}} {The {Open} {Flux}
  {Problem}}.{\BBCQ}
\newblock
\APACjournalVolNumPages{The Astrophysical Journal}{848}{}{70}.
\newblock
\begin{APACrefDOI} \doi{10.3847/1538-4357/aa8a70} \end{APACrefDOI}
\PrintBackRefs{\CurrentBib}

\bibitem [\protect \citeauthoryear {%
Lockwood%
, Owens%
\BCBL {}\ \BBA {} Rouillard%
}{%
Lockwood%
\ \protect \BOthers {.}}{%
{\protect \APACyear {2009}}%
}]{%
lockwood_excess_2009-1}
\APACinsertmetastar {%
lockwood_excess_2009-1}%
\begin{APACrefauthors}%
Lockwood, M.%
, Owens, M.%
\BCBL {}\ \BBA {} Rouillard, A\BPBI P.%
\end{APACrefauthors}%
\unskip\
\newblock
\APACrefYearMonthDay{2009}{{\APACmonth{11}}}{}.
\newblock
{\BBOQ}\APACrefatitle {Excess open solar magnetic flux from satellite data: 2.
  {A} survey of kinematic effects} {Excess open solar magnetic flux from
  satellite data: 2. {A} survey of kinematic effects}.{\BBCQ}
\newblock
\APACjournalVolNumPages{Journal of Geophysical Research (Space
  Physics)}{114}{A11}{A11104}.
\newblock
\begin{APACrefDOI} \doi{10.1029/2009JA014450} \end{APACrefDOI}
\PrintBackRefs{\CurrentBib}

\bibitem [\protect \citeauthoryear {%
Lowder%
, Qiu%
, Leamon%
\BCBL {}\ \BBA {} Liu%
}{%
Lowder%
\ \protect \BOthers {.}}{%
{\protect \APACyear {2014}}%
}]{%
lowder_measurements_2014}
\APACinsertmetastar {%
lowder_measurements_2014}%
\begin{APACrefauthors}%
Lowder, C.%
, Qiu, J.%
, Leamon, R.%
\BCBL {}\ \BBA {} Liu, Y.%
\end{APACrefauthors}%
\unskip\
\newblock
\APACrefYearMonthDay{2014}{{\APACmonth{03}}}{}.
\newblock
{\BBOQ}\APACrefatitle {Measurements of {EUV} {Coronal} {Holes} and {Open}
  {Magnetic} {Flux}} {Measurements of {EUV} {Coronal} {Holes} and {Open}
  {Magnetic} {Flux}}.{\BBCQ}
\newblock
\APACjournalVolNumPages{The Astrophysical Journal}{783}{2}{142}.
\newblock
\begin{APACrefDOI} \doi{10.1088/0004-637X/783/2/142} \end{APACrefDOI}
\PrintBackRefs{\CurrentBib}

\bibitem [\protect \citeauthoryear {%
Mackay%
\ \BBA {} Yeates%
}{%
Mackay%
\ \BBA {} Yeates%
}{%
{\protect \APACyear {2012}}%
}]{%
mackay_suns_2012}
\APACinsertmetastar {%
mackay_suns_2012}%
\begin{APACrefauthors}%
Mackay, D\BPBI H.%
\BCBT {}\ \BBA {} Yeates, A\BPBI R.%
\end{APACrefauthors}%
\unskip\
\newblock
\APACrefYearMonthDay{2012}{{\APACmonth{11}}}{}.
\newblock
{\BBOQ}\APACrefatitle {The {Sun}'s {Global} {Photospheric} and {Coronal}
  {Magnetic} {Fields}: {Observations} and {Models}} {The {Sun}'s {Global}
  {Photospheric} and {Coronal} {Magnetic} {Fields}: {Observations} and
  {Models}}.{\BBCQ}
\newblock
\APACjournalVolNumPages{Living Reviews in Solar Physics}{9}{1}{6}.
\newblock
\begin{APACrefDOI} \doi{10.12942/lrsp-2012-6} \end{APACrefDOI}
\PrintBackRefs{\CurrentBib}

\bibitem [\protect \citeauthoryear {%
MacNeice%
\ \protect \BOthers {.}}{%
MacNeice%
\ \protect \BOthers {.}}{%
{\protect \APACyear {2018}}%
}]{%
macneice_assessing_2018}
\APACinsertmetastar {%
macneice_assessing_2018}%
\begin{APACrefauthors}%
MacNeice, P.%
, Jian, L\BPBI K.%
, Antiochos, S\BPBI K.%
, Arge, C\BPBI N.%
, Bussy-Virat, C\BPBI D.%
, DeRosa, M\BPBI L.%
\BDBL {}Sokolov, I.%
\end{APACrefauthors}%
\unskip\
\newblock
\APACrefYearMonthDay{2018}{{\APACmonth{11}}}{}.
\newblock
{\BBOQ}\APACrefatitle {Assessing the {Quality} of {Models} of the {Ambient}
  {Solar} {Wind}} {Assessing the {Quality} of {Models} of the {Ambient} {Solar}
  {Wind}}.{\BBCQ}
\newblock
\APACjournalVolNumPages{Space Weather}{16}{}{1644--1667}.
\newblock
\begin{APACrefDOI} \doi{10.1029/2018SW002040} \end{APACrefDOI}
\PrintBackRefs{\CurrentBib}

\bibitem [\protect \citeauthoryear {%
McGregor%
, Hughes%
, Arge%
\BCBL {}\ \BBA {} Owens%
}{%
McGregor%
\ \protect \BOthers {.}}{%
{\protect \APACyear {2008}}%
}]{%
mcgregor_analysis_2008}
\APACinsertmetastar {%
mcgregor_analysis_2008}%
\begin{APACrefauthors}%
McGregor, S\BPBI L.%
, Hughes, W\BPBI J.%
, Arge, C\BPBI N.%
\BCBL {}\ \BBA {} Owens, M\BPBI J.%
\end{APACrefauthors}%
\unskip\
\newblock
\APACrefYearMonthDay{2008}{{\APACmonth{08}}}{}.
\newblock
{\BBOQ}\APACrefatitle {Analysis of the magnetic field discontinuity at the
  potential field source surface and {Schatten} {Current} {Sheet} interface in
  the {Wang}-{Sheeley}-{Arge} model} {Analysis of the magnetic field
  discontinuity at the potential field source surface and {Schatten} {Current}
  {Sheet} interface in the {Wang}-{Sheeley}-{Arge} model}.{\BBCQ}
\newblock
\APACjournalVolNumPages{Journal of Geophysical Research (Space
  Physics)}{113}{}{A08112}.
\newblock
\begin{APACrefDOI} \doi{10.1029/2007JA012330} \end{APACrefDOI}
\PrintBackRefs{\CurrentBib}

\bibitem [\protect \citeauthoryear {%
Nolte%
\ \protect \BOthers {.}}{%
Nolte%
\ \protect \BOthers {.}}{%
{\protect \APACyear {1976}}%
}]{%
nolte_coronal_1976}
\APACinsertmetastar {%
nolte_coronal_1976}%
\begin{APACrefauthors}%
Nolte, J\BPBI T.%
, Krieger, A\BPBI S.%
, Timothy, A\BPBI F.%
, Gold, R\BPBI E.%
, Roelof, E\BPBI C.%
, Vaiana, G.%
\BDBL {}McIntosh, P\BPBI S.%
\end{APACrefauthors}%
\unskip\
\newblock
\APACrefYearMonthDay{1976}{{\APACmonth{02}}}{}.
\newblock
{\BBOQ}\APACrefatitle {Coronal holes as sources of solar wind.} {Coronal holes
  as sources of solar wind.}{\BBCQ}
\newblock
\APACjournalVolNumPages{Solar Physics}{46}{2}{303}.
\newblock
\begin{APACrefDOI} \doi{10.1007/BF00149859} \end{APACrefDOI}
\PrintBackRefs{\CurrentBib}

\bibitem [\protect \citeauthoryear {%
Owens%
\ \protect \BOthers {.}}{%
Owens%
\ \protect \BOthers {.}}{%
{\protect \APACyear {2008}}%
}]{%
owens_metrics_2008}
\APACinsertmetastar {%
owens_metrics_2008}%
\begin{APACrefauthors}%
Owens, M\BPBI J.%
, Spence, H\BPBI E.%
, McGregor, S.%
, Hughes, W\BPBI J.%
, Quinn, J\BPBI M.%
, Arge, C\BPBI N.%
\BDBL {}Odstrcil, D.%
\end{APACrefauthors}%
\unskip\
\newblock
\APACrefYearMonthDay{2008}{{\APACmonth{08}}}{}.
\newblock
{\BBOQ}\APACrefatitle {Metrics for solar wind prediction models: {Comparison}
  of empirical, hybrid, and physics-based schemes with 8 years of {L}1
  observations} {Metrics for solar wind prediction models: {Comparison} of
  empirical, hybrid, and physics-based schemes with 8 years of {L}1
  observations}.{\BBCQ}
\newblock
\APACjournalVolNumPages{Space Weather}{6}{8}{S08001}.
\newblock
\begin{APACrefDOI} \doi{10.1029/2007SW000380} \end{APACrefDOI}
\PrintBackRefs{\CurrentBib}

\bibitem [\protect \citeauthoryear {%
Pesnell%
, Thompson%
\BCBL {}\ \BBA {} Chamberlin%
}{%
Pesnell%
\ \protect \BOthers {.}}{%
{\protect \APACyear {2012}}%
}]{%
pesnell_solar_2012}
\APACinsertmetastar {%
pesnell_solar_2012}%
\begin{APACrefauthors}%
Pesnell, W\BPBI D.%
, Thompson, B\BPBI J.%
\BCBL {}\ \BBA {} Chamberlin, P\BPBI C.%
\end{APACrefauthors}%
\unskip\
\newblock
\APACrefYearMonthDay{2012}{{\APACmonth{01}}}{}.
\newblock
{\BBOQ}\APACrefatitle {The {Solar} {Dynamics} {Observatory} ({SDO})} {The
  {Solar} {Dynamics} {Observatory} ({SDO})}.{\BBCQ}
\newblock
\APACjournalVolNumPages{Solar Physics}{275}{1-2}{3}.
\newblock
\begin{APACrefDOI} \doi{10.1007/s11207-011-9841-3} \end{APACrefDOI}
\PrintBackRefs{\CurrentBib}

\bibitem [\protect \citeauthoryear {%
Phillips%
\ \protect \BOthers {.}}{%
Phillips%
\ \protect \BOthers {.}}{%
{\protect \APACyear {1994}}%
}]{%
phillips_ulysses_1994}
\APACinsertmetastar {%
phillips_ulysses_1994}%
\begin{APACrefauthors}%
Phillips, J\BPBI L.%
, Balogh, A.%
, Bame, S\BPBI J.%
, Goldstein, B\BPBI E.%
, Gosling, J\BPBI T.%
, Hoeksema, J\BPBI T.%
\BDBL {}Wang, Y\BHBI M.%
\end{APACrefauthors}%
\unskip\
\newblock
\APACrefYearMonthDay{1994}{{\APACmonth{06}}}{}.
\newblock
{\BBOQ}\APACrefatitle {Ulysses at 50° south: constant immersion in the
  high-speed solar wind} {Ulysses at 50° south: constant immersion in the
  high-speed solar wind}.{\BBCQ}
\newblock
\APACjournalVolNumPages{Geophysical Research Letters}{21}{12}{1105}.
\newblock
\begin{APACrefDOI} \doi{10.1029/94GL01065} \end{APACrefDOI}
\PrintBackRefs{\CurrentBib}

\bibitem [\protect \citeauthoryear {%
Pomoell%
\ \BBA {} Poedts%
}{%
Pomoell%
\ \BBA {} Poedts%
}{%
{\protect \APACyear {2018}}%
}]{%
pomoell_euhforia:_2018}
\APACinsertmetastar {%
pomoell_euhforia:_2018}%
\begin{APACrefauthors}%
Pomoell, J.%
\BCBT {}\ \BBA {} Poedts, S.%
\end{APACrefauthors}%
\unskip\
\newblock
\APACrefYearMonthDay{2018}{{\APACmonth{06}}}{}.
\newblock
{\BBOQ}\APACrefatitle {{EUHFORIA}: {European} heliospheric forecasting
  information asset} {{EUHFORIA}: {European} heliospheric forecasting
  information asset}.{\BBCQ}
\newblock
\APACjournalVolNumPages{Journal of Space Weather and Space Climate}{8}{}{A35}.
\newblock
\begin{APACrefDOI} \doi{10.1051/swsc/2018020} \end{APACrefDOI}
\PrintBackRefs{\CurrentBib}

\bibitem [\protect \citeauthoryear {%
Riley%
, Linker%
\BCBL {}\ \BBA {} Arge%
}{%
Riley%
\ \protect \BOthers {.}}{%
{\protect \APACyear {2015}}%
}]{%
riley_role_2015}
\APACinsertmetastar {%
riley_role_2015}%
\begin{APACrefauthors}%
Riley, P.%
, Linker, J\BPBI A.%
\BCBL {}\ \BBA {} Arge, C\BPBI N.%
\end{APACrefauthors}%
\unskip\
\newblock
\APACrefYearMonthDay{2015}{{\APACmonth{03}}}{}.
\newblock
{\BBOQ}\APACrefatitle {On the role played by magnetic expansion factor in the
  prediction of solar wind speed} {On the role played by magnetic expansion
  factor in the prediction of solar wind speed}.{\BBCQ}
\newblock
\APACjournalVolNumPages{Space Weather}{13}{}{154--169}.
\newblock
\begin{APACrefDOI} \doi{10.1002/2014SW001144} \end{APACrefDOI}
\PrintBackRefs{\CurrentBib}

\bibitem [\protect \citeauthoryear {%
Riley%
\ \protect \BOthers {.}}{%
Riley%
\ \protect \BOthers {.}}{%
{\protect \APACyear {2006}}%
}]{%
riley_comparison_2006}
\APACinsertmetastar {%
riley_comparison_2006}%
\begin{APACrefauthors}%
Riley, P.%
, Linker, J\BPBI A.%
, Mikić, Z.%
, Lionello, R.%
, Ledvina, S\BPBI A.%
\BCBL {}\ \BBA {} Luhmann, J\BPBI G.%
\end{APACrefauthors}%
\unskip\
\newblock
\APACrefYearMonthDay{2006}{{\APACmonth{12}}}{}.
\newblock
{\BBOQ}\APACrefatitle {A {Comparison} between {Global} {Solar}
  {Magnetohydrodynamic} and {Potential} {Field} {Source} {Surface} {Model}
  {Results}} {A {Comparison} between {Global} {Solar} {Magnetohydrodynamic} and
  {Potential} {Field} {Source} {Surface} {Model} {Results}}.{\BBCQ}
\newblock
\APACjournalVolNumPages{The Astrophysical Journal}{653}{}{1510--1516}.
\newblock
\begin{APACrefDOI} \doi{10.1086/508565} \end{APACrefDOI}
\PrintBackRefs{\CurrentBib}

\bibitem [\protect \citeauthoryear {%
Rotter%
, Veronig%
, Temmer%
\BCBL {}\ \BBA {} Vršnak%
}{%
Rotter%
\ \protect \BOthers {.}}{%
{\protect \APACyear {2012}}%
}]{%
rotter_relation_2012}
\APACinsertmetastar {%
rotter_relation_2012}%
\begin{APACrefauthors}%
Rotter, T.%
, Veronig, A\BPBI M.%
, Temmer, M.%
\BCBL {}\ \BBA {} Vršnak, B.%
\end{APACrefauthors}%
\unskip\
\newblock
\APACrefYearMonthDay{2012}{{\APACmonth{12}}}{}.
\newblock
{\BBOQ}\APACrefatitle {Relation {Between} {Coronal} {Hole} {Areas} on the {Sun}
  and the {Solar} {Wind} {Parameters} at 1 {AU}} {Relation {Between} {Coronal}
  {Hole} {Areas} on the {Sun} and the {Solar} {Wind} {Parameters} at 1
  {AU}}.{\BBCQ}
\newblock
\APACjournalVolNumPages{Solar Physics}{281}{2}{793}.
\newblock
\begin{APACrefDOI} \doi{10.1007/s11207-012-0101-y} \end{APACrefDOI}
\PrintBackRefs{\CurrentBib}

\bibitem [\protect \citeauthoryear {%
Schatten%
}{%
Schatten%
}{%
{\protect \APACyear {1971}}%
}]{%
schatten_current_1971}
\APACinsertmetastar {%
schatten_current_1971}%
\begin{APACrefauthors}%
Schatten, K\BPBI H.%
\end{APACrefauthors}%
\unskip\
\newblock
\APACrefYearMonthDay{1971}{}{}.
\newblock
{\BBOQ}\APACrefatitle {Current sheet magnetic model for the solar corona.}
  {Current sheet magnetic model for the solar corona.}{\BBCQ}
\newblock
\APACjournalVolNumPages{Cosmic Electrodynamics}{2}{}{232--245}.
\newblock
\begin{APACrefURL}
  [{2019-05-09}]\url{http://adsabs.harvard.edu/abs/1971CosEl...2..232S}
  \end{APACrefURL}
\PrintBackRefs{\CurrentBib}

\bibitem [\protect \citeauthoryear {%
Schatten%
, Wilcox%
\BCBL {}\ \BBA {} Ness%
}{%
Schatten%
\ \protect \BOthers {.}}{%
{\protect \APACyear {1969}}%
}]{%
schatten_model_1969}
\APACinsertmetastar {%
schatten_model_1969}%
\begin{APACrefauthors}%
Schatten, K\BPBI H.%
, Wilcox, J\BPBI M.%
\BCBL {}\ \BBA {} Ness, N\BPBI F.%
\end{APACrefauthors}%
\unskip\
\newblock
\APACrefYearMonthDay{1969}{{\APACmonth{03}}}{}.
\newblock
{\BBOQ}\APACrefatitle {A model of interplanetary and coronal magnetic fields}
  {A model of interplanetary and coronal magnetic fields}.{\BBCQ}
\newblock
\APACjournalVolNumPages{Solar Physics}{6}{}{442--455}.
\newblock
\begin{APACrefDOI} \doi{10.1007/BF00146478} \end{APACrefDOI}
\PrintBackRefs{\CurrentBib}

\bibitem [\protect \citeauthoryear {%
Schulz%
}{%
Schulz%
}{%
{\protect \APACyear {1997}}%
}]{%
schulz_non-spherical_1997}
\APACinsertmetastar {%
schulz_non-spherical_1997}%
\begin{APACrefauthors}%
Schulz, M.%
\end{APACrefauthors}%
\unskip\
\newblock
\APACrefYearMonthDay{1997}{{\APACmonth{11}}}{}.
\newblock
{\BBOQ}\APACrefatitle {Non-spherical source-surface model of the heliosphere: a
  scalar formulation} {Non-spherical source-surface model of the heliosphere: a
  scalar formulation}.{\BBCQ}
\newblock
\APACjournalVolNumPages{Annales Geophysicae}{15}{}{1379--1387}.
\newblock
\begin{APACrefDOI} \doi{10.1007/s00585-997-1379-1} \end{APACrefDOI}
\PrintBackRefs{\CurrentBib}

\bibitem [\protect \citeauthoryear {%
Schulz%
}{%
Schulz%
}{%
{\protect \APACyear {2008}}%
}]{%
schulz_non-spherical_2008}
\APACinsertmetastar {%
schulz_non-spherical_2008}%
\begin{APACrefauthors}%
Schulz, M.%
\end{APACrefauthors}%
\unskip\
\newblock
\APACrefYearMonthDay{2008}{{\APACmonth{05}}}{}.
\newblock
{\BBOQ}\APACrefatitle {Non-{Spherical} {Source}-{Surface} {Model} of the
  {Corona} and {Heliosphere} for a {Quadrupolar} {Main} {Field} of the {Sun}}
  {Non-{Spherical} {Source}-{Surface} {Model} of the {Corona} and {Heliosphere}
  for a {Quadrupolar} {Main} {Field} of the {Sun}}.{\BBCQ}
\newblock
\APACjournalVolNumPages{AGU Spring Meeting Abstracts}{44}{}{SH44A--04}.
\PrintBackRefs{\CurrentBib}

\bibitem [\protect \citeauthoryear {%
Schwenn%
}{%
Schwenn%
}{%
{\protect \APACyear {2006}}%
{\protect \APACexlab {{\protect \BCnt {1}}}}}]{%
schwenn_solar_2006}
\APACinsertmetastar {%
schwenn_solar_2006}%
\begin{APACrefauthors}%
Schwenn, R.%
\end{APACrefauthors}%
\unskip\
\newblock
\APACrefYearMonthDay{2006{\protect \BCnt {1}}}{{\APACmonth{06}}}{}.
\newblock
{\BBOQ}\APACrefatitle {Solar {Wind} {Sources} and {Their} {Variations} {Over}
  the {Solar} {Cycle}} {Solar {Wind} {Sources} and {Their} {Variations} {Over}
  the {Solar} {Cycle}}.{\BBCQ}
\newblock
\APACjournalVolNumPages{Space Science Reviews}{124}{1-4}{51}.
\newblock
\begin{APACrefDOI} \doi{10.1007/s11214-006-9099-5} \end{APACrefDOI}
\PrintBackRefs{\CurrentBib}

\bibitem [\protect \citeauthoryear {%
Schwenn%
}{%
Schwenn%
}{%
{\protect \APACyear {2006}}%
{\protect \APACexlab {{\protect \BCnt {2}}}}}]{%
schwenn_space_2006}
\APACinsertmetastar {%
schwenn_space_2006}%
\begin{APACrefauthors}%
Schwenn, R.%
\end{APACrefauthors}%
\unskip\
\newblock
\APACrefYearMonthDay{2006{\protect \BCnt {2}}}{{\APACmonth{08}}}{}.
\newblock
{\BBOQ}\APACrefatitle {Space {Weather}: {The} {Solar} {Perspective}} {Space
  {Weather}: {The} {Solar} {Perspective}}.{\BBCQ}
\newblock
\APACjournalVolNumPages{Living Reviews in Solar Physics}{3}{1}{2}.
\newblock
\begin{APACrefDOI} \doi{10.12942/lrsp-2006-2} \end{APACrefDOI}
\PrintBackRefs{\CurrentBib}

\bibitem [\protect \citeauthoryear {%
Vršnak%
, Temmer%
\BCBL {}\ \BBA {} Veronig%
}{%
Vršnak%
\ \protect \BOthers {.}}{%
{\protect \APACyear {2007}}%
}]{%
vrsnak_coronal_2007}
\APACinsertmetastar {%
vrsnak_coronal_2007}%
\begin{APACrefauthors}%
Vršnak, B.%
, Temmer, M.%
\BCBL {}\ \BBA {} Veronig, A\BPBI M.%
\end{APACrefauthors}%
\unskip\
\newblock
\APACrefYearMonthDay{2007}{{\APACmonth{02}}}{}.
\newblock
{\BBOQ}\APACrefatitle {Coronal {Holes} and {Solar} {Wind} {High}-{Speed}
  {Streams}: {I}. {Forecasting} the {Solar} {Wind} {Parameters}} {Coronal
  {Holes} and {Solar} {Wind} {High}-{Speed} {Streams}: {I}. {Forecasting} the
  {Solar} {Wind} {Parameters}}.{\BBCQ}
\newblock
\APACjournalVolNumPages{Solar Physics}{240}{2}{315}.
\newblock
\begin{APACrefDOI} \doi{10.1007/s11207-007-0285-8} \end{APACrefDOI}
\PrintBackRefs{\CurrentBib}

\bibitem [\protect \citeauthoryear {%
Wallace%
, Arge%
, Pattichis%
, Hock-Mysliwiec%
\BCBL {}\ \BBA {} Henney%
}{%
Wallace%
\ \protect \BOthers {.}}{%
{\protect \APACyear {2019}}%
}]{%
wallace_estimating_2019}
\APACinsertmetastar {%
wallace_estimating_2019}%
\begin{APACrefauthors}%
Wallace, S.%
, Arge, C\BPBI N.%
, Pattichis, M.%
, Hock-Mysliwiec, R\BPBI A.%
\BCBL {}\ \BBA {} Henney, C\BPBI J.%
\end{APACrefauthors}%
\unskip\
\newblock
\APACrefYearMonthDay{2019}{{\APACmonth{02}}}{}.
\newblock
{\BBOQ}\APACrefatitle {Estimating {Total} {Open} {Heliospheric} {Magnetic}
  {Flux}} {Estimating {Total} {Open} {Heliospheric} {Magnetic} {Flux}}.{\BBCQ}
\newblock
\APACjournalVolNumPages{Solar Physics}{294}{}{19}.
\newblock
\begin{APACrefDOI} \doi{10.1007/s11207-019-1402-1} \end{APACrefDOI}
\PrintBackRefs{\CurrentBib}

\bibitem [\protect \citeauthoryear {%
Wang%
}{%
Wang%
}{%
{\protect \APACyear {2009}}%
}]{%
wang_coronal_2009}
\APACinsertmetastar {%
wang_coronal_2009}%
\begin{APACrefauthors}%
Wang, Y\BHBI M.%
\end{APACrefauthors}%
\unskip\
\newblock
\APACrefYearMonthDay{2009}{{\APACmonth{04}}}{}.
\newblock
{\BBOQ}\APACrefatitle {Coronal {Holes} and {Open} {Magnetic} {Flux}} {Coronal
  {Holes} and {Open} {Magnetic} {Flux}}.{\BBCQ}
\newblock
\APACjournalVolNumPages{Space Science Reviews}{144}{1-4}{383}.
\newblock
\begin{APACrefDOI} \doi{10.1007/s11214-008-9434-0} \end{APACrefDOI}
\PrintBackRefs{\CurrentBib}

\bibitem [\protect \citeauthoryear {%
Wang%
, Hawley%
\BCBL {}\ \BBA {} Sheeley%
}{%
Wang%
\ \protect \BOthers {.}}{%
{\protect \APACyear {1996}}%
}]{%
wang_magnetic_1996}
\APACinsertmetastar {%
wang_magnetic_1996}%
\begin{APACrefauthors}%
Wang, Y\BHBI M.%
, Hawley, S\BPBI H.%
\BCBL {}\ \BBA {} Sheeley, N\BPBI R., Jr.%
\end{APACrefauthors}%
\unskip\
\newblock
\APACrefYearMonthDay{1996}{{\APACmonth{01}}}{}.
\newblock
{\BBOQ}\APACrefatitle {The {Magnetic} {Nature} of {Coronal} {Holes}} {The
  {Magnetic} {Nature} of {Coronal} {Holes}}.{\BBCQ}
\newblock
\APACjournalVolNumPages{Science}{271}{}{464--469}.
\newblock
\begin{APACrefDOI} \doi{10.1126/science.271.5248.464} \end{APACrefDOI}
\PrintBackRefs{\CurrentBib}

\bibitem [\protect \citeauthoryear {%
Wang%
\ \BBA {} Sheeley%
}{%
Wang%
\ \BBA {} Sheeley%
}{%
{\protect \APACyear {1990}}%
}]{%
wang_solar_1990}
\APACinsertmetastar {%
wang_solar_1990}%
\begin{APACrefauthors}%
Wang, Y\BHBI M.%
\BCBT {}\ \BBA {} Sheeley, N\BPBI R., Jr.%
\end{APACrefauthors}%
\unskip\
\newblock
\APACrefYearMonthDay{1990}{{\APACmonth{06}}}{}.
\newblock
{\BBOQ}\APACrefatitle {Solar wind speed and coronal flux-tube expansion} {Solar
  wind speed and coronal flux-tube expansion}.{\BBCQ}
\newblock
\APACjournalVolNumPages{The Astrophysical Journal}{355}{}{}.
\newblock
\begin{APACrefDOI} \doi{10.1086/168805} \end{APACrefDOI}
\PrintBackRefs{\CurrentBib}

\bibitem [\protect \citeauthoryear {%
Wang%
\ \BBA {} Sheeley%
}{%
Wang%
\ \BBA {} Sheeley%
}{%
{\protect \APACyear {1992}}%
}]{%
wang_potential_1992}
\APACinsertmetastar {%
wang_potential_1992}%
\begin{APACrefauthors}%
Wang, Y\BHBI M.%
\BCBT {}\ \BBA {} Sheeley, N\BPBI R., Jr.%
\end{APACrefauthors}%
\unskip\
\newblock
\APACrefYearMonthDay{1992}{{\APACmonth{06}}}{}.
\newblock
{\BBOQ}\APACrefatitle {On potential field models of the solar corona} {On
  potential field models of the solar corona}.{\BBCQ}
\newblock
\APACjournalVolNumPages{The Astrophysical Journal}{392}{}{310--319}.
\newblock
\begin{APACrefDOI} \doi{10.1086/171430} \end{APACrefDOI}
\PrintBackRefs{\CurrentBib}

\bibitem [\protect \citeauthoryear {%
Wang%
\ \BBA {} Sheeley%
}{%
Wang%
\ \BBA {} Sheeley%
}{%
{\protect \APACyear {1995}}%
}]{%
wang_solar_1995}
\APACinsertmetastar {%
wang_solar_1995}%
\begin{APACrefauthors}%
Wang, Y\BHBI M.%
\BCBT {}\ \BBA {} Sheeley, N\BPBI R., Jr.%
\end{APACrefauthors}%
\unskip\
\newblock
\APACrefYearMonthDay{1995}{{\APACmonth{07}}}{}.
\newblock
{\BBOQ}\APACrefatitle {Solar {Implications} of {ULYSSES} {Interplanetary}
  {Field} {Measurements}} {Solar {Implications} of {ULYSSES} {Interplanetary}
  {Field} {Measurements}}.{\BBCQ}
\newblock
\APACjournalVolNumPages{The Astrophysical Journal Letters}{447}{}{L143}.
\newblock
\begin{APACrefDOI} \doi{10.1086/309578} \end{APACrefDOI}
\PrintBackRefs{\CurrentBib}

\bibitem [\protect \citeauthoryear {%
Wang%
\ \BBA {} Sheeley%
}{%
Wang%
\ \BBA {} Sheeley%
}{%
{\protect \APACyear {2003}}%
}]{%
wang_topological_2003}
\APACinsertmetastar {%
wang_topological_2003}%
\begin{APACrefauthors}%
Wang, Y\BHBI M.%
\BCBT {}\ \BBA {} Sheeley, N\BPBI R., Jr.%
\end{APACrefauthors}%
\unskip\
\newblock
\APACrefYearMonthDay{2003}{{\APACmonth{12}}}{}.
\newblock
{\BBOQ}\APACrefatitle {On the {Topological} {Evolution} of the {Coronal}
  {Magnetic} {Field} {During} the {Solar} {Cycle}} {On the {Topological}
  {Evolution} of the {Coronal} {Magnetic} {Field} {During} the {Solar}
  {Cycle}}.{\BBCQ}
\newblock
\APACjournalVolNumPages{The Astrophysical Journal}{599}{}{1404--1417}.
\newblock
\begin{APACrefDOI} \doi{10.1086/379348} \end{APACrefDOI}
\PrintBackRefs{\CurrentBib}

\end{thebibliography}

\end{document}